\documentclass[a4paper,11pt]{article}
\usepackage{jheppub}
\usepackage[utf8]{inputenc}
\usepackage{url}
\usepackage{mathrsfs}  
\usepackage{pdflscape}

\usepackage{refcount}
\usepackage{booktabs}
\usepackage{todonotes}
\usepackage{enumitem}
\usepackage{adjustbox}
\usepackage{afterpage}
\usepackage{dsfont}
\usepackage{nameref}
\usepackage{dsfont}

\makeatletter
\newcommand\level[1]{%
  \ifcase#1\relax\expandafter\chapter\or
    \expandafter\section\or
    \expandafter\subsection\or
    \expandafter\subsubsection\else
    \def\next{\@level{#1}}\expandafter\next
  \fi}
\newcommand{\@level}[1]{%
  \@startsection{level#1}
    {#1}
    {\z@}%
    {-3.25ex\@plus -1ex \@minus -.2ex}%
    {1.5ex \@plus .2ex}%
    {\normalfont\normalsize\bfseries}}

\newcounter{level4}[subsubsection]
\@namedef{thelevel4}{\thesubsubsection.\arabic{level4}}
\@namedef{level4mark}#1{}
\count@=4
\loop\ifnum\count@<100
  \begingroup\edef\x{\endgroup
    \noexpand\newcounter{level\number\numexpr\count@+1\relax}[level\number\count@]
    \noexpand\@namedef{thelevel\number\numexpr\count@+1\relax}{%
      \noexpand\@nameuse{thelevel\number\count@}.\noexpand\arabic{level\number\numexpr\count@+1\relax}}
    \noexpand\@namedef{level\number\numexpr\count@+1\relax mark}####1{}}
  \x
  \advance\count@\@ne
\repeat
\makeatother
\usepackage[utf8]{inputenc} 
\usepackage{amsmath}
\usepackage{amsthm}
\usepackage{amsfonts,mathrsfs} 
\usepackage{xcolor}
\usepackage[english]{babel} 
\usepackage[autostyle]{csquotes}
\usepackage{mathtools}

\newlist{inparaenum}{enumerate}{3}
\setlist[inparaenum,1]{label=\arabic*.}% labels for top level
\setlist[inparaenum,2]{label=\emph{\alph*})}% labels for second level
\setlist[inparaenum,3]{label=\emph{\roman*})}% labels for third level

\numberwithin{equation}{section}

\makeatletter
\newcommand\footnoteref[1]{\protected@xdef\@thefnmark{\ref{#1}}\@footnotemark}
\makeatother

\usepackage{xstring}
\usepackage{tikz} 
\usepackage{booktabs}
\usepackage{xcolor}
\usetikzlibrary{decorations.pathreplacing} 
\usetikzlibrary{decorations.pathmorphing} 
\usetikzlibrary{decorations.markings} 
\usetikzlibrary{arrows} 
\usetikzlibrary{shapes} 
\usetikzlibrary{matrix} 
\usetikzlibrary{positioning} 
\usetikzlibrary{shapes.geometric}
\usetikzlibrary{calc}

\tikzstyle{brane}=[draw]
\tikzset{D7/.style={circle, draw=black, inner sep=0pt, fill=white, minimum size=3mm}}
\tikzset{hasse/.style={circle, fill,inner sep=2pt}}
\tikzset{flavor/.style={regular polygon,regular polygon sides=4,inner sep=2.5pt, draw}}
\tikzset{gauge/.style={circle, draw,inner sep=2.5pt}}
\tikzset{gauger/.style={circle, draw=red, fill=red, inner sep=2.5pt}}
\tikzset{gaugeb/.style={circle, draw=lightblue, fill=lightblue, inner sep=2.5pt}}
\tikzset{emptygauge/.style={circle, draw=gray, dashed, inner sep=2.5pt}}
\tikzset{bd/.style={circle, draw=black, inner sep=0pt, fill=black, minimum size=2mm}}
\tikzset{wd/.style={circle, draw=black, inner sep=0pt, fill=white, minimum size=2mm}}
\tikzset{Dynkin/.style={circle, draw=black, inner sep=0pt, fill=white, minimum size=2mm}}
\tikzstyle{ligne}=[draw, thick] 
\tikzset{doublearrow/.style={ draw=black!75, color=black!75, thick, double distance=3pt, }}

\colorlet{lightblue}{blue!65}

\preprint{Imperial/TP/21/AH/06}

\title{Poisson Brackets for some Coulomb Branches}

\author{Kirsty Gledhill}
\author{and Amihay Hanany}
\affiliation{Theoretical Physics Group, The Blackett Laboratory, Imperial College London, Prince Consort Road
London, SW7 2AZ, UK}
\emailAdd{k.gledhill20@imperial.ac.uk}
\emailAdd{a.hanany@imperial.ac.uk}

\abstract{We construct Poisson bracket relations between the operators which generate the chiral ring of the Coulomb branch of certain $3d$ $\mathcal{N}=4$ quiver gauge theories. In the case where the Coulomb branch is a free space, $ADE$ Klein singularity, or the minimal $A_2$ nilpotent orbit, we explicitly compute the Poisson brackets between the generators using either inherited properties of the abstract Coulomb branch variety, or the expected charges of these operators under the global symmetry (known through use of the monopole formula). We also conjecture Poisson brackets for Higgs branches that originate from $6d$ theories with tensionless strings or $5d$ theories with massless instantons for which the HWG is known, based on representation theoretic and operator content constraints known from the study of their magnetic quiver.}

\begin{document}

\maketitle

\section{Introduction}
The moduli space $\mathcal{M}$ of a supersymmetric theory with $8$ supercharges is an algebraic variety $\mathcal{V}$; it is parameterised by the values of scalar fields in the theory, subject to the vacuum constraints and gauge invariance. A subset of the global symmetry of the theory is preserved on $\mathcal{M}$, and this induces a conserved charge \cite{Noether:1918} which distinguishes between scalar operators lying on the moduli space. In some cases, the exact variety $\mathcal{V}$ is known explicitly. However sometimes it is not known how to compute $\mathcal{V}$, and so other techniques are required to explore the moduli space.\\

\noindent In this paper we study $3d$ $\mathcal{N}=4$ unitary quiver gauge theories, which are good or ugly in the sense of \cite{Gaiotto:2008ak}, whose moduli spaces are symplectic singularities \cite{2000InMat.139..541B}. Such moduli spaces are equipped with a symplectic form, and can be split into the so-called Coulomb, Higgs and mixed branches, which are the parts parameterised by only vectormultiplet scalars, hypermultiplet scalars and both respectively. The Higgs branch $\mathcal{H}$ for a theory at finite coupling is classically exact, and thus the exact form of this variety (which will be a HyperK\"ahler quotient) can be computed explicitly in principle. The metric on the Coulomb branch $\mathcal{C}$ does however receive quantum corrections, and it is not yet known how to obtain this corrected metric for a generic theory. In recent years, methods for determining the dimension, global symmetry, representation content and singularity structure of $\mathcal{C}$ have been developed and provide us with great insight into the Coulomb branch. In this paper, we present various methods to learn of the symplectic form on $\mathcal{C}$ by determining the unordered Poisson brackets between its generators $\mathcal{G}_{\mathcal{C}}$, $\{\cdot,\cdot\}_{\mathcal{C}}$. The antisymmetry and Leibniz properties of the Poisson bracket and the ring structure of the Coulomb branch then fully fix the Poisson brackets between any two operators on $\mathcal{C}$. There has been previous progress on determining Poisson brackets for Coulomb branches in the literature (see for instance \cite{Bullimore:2015lsa,Nakajima:2015txa,Dedushenko:2017avn,2020AnHP...21.1235B}); in this note we add to this by providing explicit computations for a selected set of examples, including quivers with many unitary gauge nodes of high rank.\\

\noindent So, given a quiver, how does one compute the Poisson brackets between the generators of its Coulomb branch? We can answer this question in two ways, depending on how we choose to view $\mathcal{C}$. The first way will be the topic of Sections \ref{sec: pbs for monops} -- \ref{sec: nilp orbs}: we view the Coulomb branch as the space of dressed monopole operators,\footnote{This name is common in the literature but is a slight abuse of language; what we really mean is that the Coulomb branch, viewed as an affine algebraic variety, is the spectrum of the Coulomb branch chiral ring, which is generated by dressed monopole operators. The Coulomb branch variety, its chiral ring, and the VEVs of operators on $\mathcal{C}$ can be used interchangably in the literature, so it's important to clarify the distinction.} and study the Poisson brackets between those that generate the variety. This method gives more explicit information, but has the drawbacks of being more computationally intensive or requiring the exact description of the Coulomb branch as an algebraic variety to be known\footnote{Here, by ``an exact description" we mean an explicit realisation of the Coulomb branch as the space traced out by some known set of generators which are expressed as holomorphic polynomials in some complex coordinates, and any relations they are subject to.} (via mirror symmetry \cite{Intriligator:1996ex, Hanany:1996ie}). The second viewpoint we can take, if the representation content of $\mathcal{C}$ at low orders is known, is slightly more reductionist: we can characterise the Coulomb branch simply as a space of representations, isolate those which generate the space and then conjecture the Poisson brackets between them based on representation theoretic and operator content constraints. This viewpoint is discussed in Sections \ref{sec: pbs for cb as reps} -- \ref{sec:eg pbs from reps}; it is a more abstract approach, but allows us to say something about the Poisson structure of a wider set of Coulomb branches.\\

\noindent The outline of this paper is as follows. In Sections \ref{sec: pbs for monops} -- \ref{sec: nilp orbs} we detail the computation of $\{\mathcal{G}_{\mathcal{C}},\mathcal{G}_{\mathcal{C}}\}_{\mathcal{C}}$ for $\mathcal{G}_{\mathcal{C}}$ explicit dressed monopole operators of the physical theory. In Section \ref{sec: pbs for monops} we explain the method used to do this for the cases where either the explicit variety $\mathcal{C}=\mathcal{V}$ is known, or the dimension of the Coulomb branch is sufficiently small. In Sections \ref{sec: free spaces}, \ref{sec: klein sings} and \ref{sec: nilp orbs}, we report the results for the examples of Coulomb branches which are free spaces, $A$ and $D$ Klein singularities and small dimension nilpotent orbits respectively. Sections \ref{sec: pbs for cb as reps} -- \ref{sec:eg pbs from reps} cover the alternative method of computing $\{\mathcal{G}_{\mathcal{C}},\mathcal{G}_{\mathcal{C}}\}_{\mathcal{C}}$ for generators $\mathcal{G}_{\mathcal{C}}$ viewed abstractly as representations of the global symmetry: Section \ref{sec: pbs for cb as reps} details the method, and Section \ref{sec:eg pbs from reps} details the results for several families of quivers whose Coulomb branch describes the Higgs branch of $5$ or $6d$ theories at infinite coupling (such quivers are called magnetic quivers for these higher dimensional theories). The appendices attempt to cover some basics of the necessary content for unfamiliar readers: Appendices \ref{app: ss} describes symplectic singularities and the Poisson bracket, and Appendix \ref{app:tens symm antisymm prods} reviews tensor, symmetric and antisymmetric products of representations respectively. In the final part of this introduction, we briefly list some notation that will be used throughout:
\begin{itemize}
    \item[--] $\mathcal{G}_{\mathcal{C}}$, $\mathcal{G}_{\mathcal{V}}$ and $\mathcal{G}_{\mathcal{H}}$ are used to denote the generators for a Coulomb branch $\mathcal{C}$, variety $V$ and Higgs branch $\mathcal{H}$ respectively.
    \item[--] $\{\cdot,\cdot\}_{\mathcal{C}}$ and $\{\cdot,\cdot\}_{\mathcal{V}}$ denote the Poisson bracket between any two Coulomb branch operators written in terms of the Coulomb branch degrees of freedom and written in terms of the holomorphic functions on the abstract variety we know the Coulomb branch to be respectively. 
    \item[--] Throughout, $\{\cdots\}$ is used to denote both a set and a Poisson bracket (the latter is only applicable when there are two arguments), but the context should make clear in which sense it is being used.
    \item[--] We use $\boldsymbol{m}(\cdot)$, $\boldsymbol{J}(\cdot)$ and $\Delta(\cdot)$ to denote the magnetic, topological and conformal dimension of an operator respectively.
    \item[--] In Sections \ref{sec: pbs for cb as reps} and \ref{sec:eg pbs from reps}, $\mu$, $\nu$ and $\mu_i$ will be used both as an index labelling representation generators that takes on certain specified values, and as the highest weight fugacities to denote a representation. Its meaning in a given situation should be clear from context (i.e. whether it lies in an exponent/subscript or not).
\end{itemize}

\section{Poisson Brackets for Dressed Monopole Operators}\label{sec: pbs for monops}
In this section we take the viewpoint of seeing the Coulomb branch as the space of dressed monopole operators, and explain how to compute $\{\mathcal{G}_{\mathcal{C}},\mathcal{G}_{\mathcal{C}}\}_{\mathcal{C}}$ in this case. \\

\noindent Before we get into how to do this however, let's first make concrete the notion of a Poisson bracket by giving an explicit example: the Poisson bracket on $\mathbb{C}^2 \cong \mathbb{C}[z_1,z_2]$. This variety is generated by the two complex coordinates over which it is formed, which here we have called $z_1$ and $z_2$. As a result, we can find the Poisson bracket between any two functions on $\mathbb{C}^2$ by postulating a relation between the generators on this space, $z_1$ and $z_2$, and then invoking the Leibniz, bilinear and antisymmetric properties of $\{\cdot,\cdot\}$. We postulate that $z_1$ and $z_2$ satisfy
\begin{equation}\label{defining PB on C2}
    \{z_1,z_2\} = 1,
\end{equation}
a direct analogue of that of phase space coordinates in classical mechanics. Leibniz can then be used to see that the Poisson bracket between any two functions $f$ and $g$ on $\mathbb{C}^2$ is
\begin{equation}\label{generic PB on C2}
    \{f(z_1, \, z_2), \, g(z_1, \, z_2) \} = \frac{\partial f}{\partial z_1} \, \frac{\partial g}{\partial z_2} - \frac{\partial f}{\partial z_2} \, \frac{\partial g}{\partial z_1}.
\end{equation}

\noindent We are now ready to present the method for finding Poisson brackets between dressed monopole operators on $\mathcal{C}$. In words, the method we use is fairly simple:
\level{4}*{Method}
\label{Alg: MPBA}
\begin{inparaenum}
    \item Explicitly write down the generators of the Coulomb branch in terms of its basic degrees of freedom (bare monopoles and adjoint scalars).
    \item Constrain the Poisson brackets by demanding the result have the expected conserved charges under the global symmetries of the Coulomb branch. This will fix the results up to constant factors.
    \item 
    \begin{inparaenum}
        \item If the Coulomb branch is known as a variety with coordinates on which there exists a canonical Poisson bracket,\footnote{We use the term canonical Poisson bracket to mean the one associated to a known algebraic symplectic variety.} use the results of Step $2$ to identify which dressed monopoles correspond to which holomorphic functions on this variety, and declare that their Poisson brackets are equal.
        \item Otherwise, fix the constant factors by demanding that the Poisson bracket of the Cartan elements of the Coulomb branch generators with any operator on the Coulomb branch yield the correct weight under the global symmetry of the Coulomb branch.
    \end{inparaenum}
\end{inparaenum}

\noindent As already alluded to, these steps are easier said than done: the number of quivers whose Coulomb branch satisfies the condition of $3.a)$ is rather small; and even for quivers with fairly low-dimension Coulomb branches there quickly become too many variables to actually find an explicit solution following the idea of $3.b)$ (at least using the methods we have so far). Essentially, there are many quivers for which Step $3$ becomes too hard to do and we cannot ascertain the Poisson relations between the explicit dressed monopole operators of $\mathcal{C}$. It is in these cases that we turn to an alternate viewpoint, as discussed in Section \ref{sec: pbs for cb as reps} and executed in Section \ref{sec:eg pbs from reps}. In Section \ref{sec: cb ops}, \ref{sec: rough form pbs from charge cons}, \ref{sec: explicit pbs from var} and \ref{sec: explicit pbs from gs charges} respectively we elaborate on each of the steps $1$, $2$, $3.a)$ and $3.b)$ in the method above (assuming that the Coulomb branches we consider satisfy the above conditions of either being known exactly as a variety or of low enough dimension), detailing exactly how to perform the relevant computations and illsutrating them in the example of the Coulomb branch of SQED with $2$ electrons.\\

\subsection{Writing down Coulomb branch generators}\label{sec: cb ops}
Recall that the Coulomb branch is the space of gauge invariant bare and dressed monopole operators labelled by magnetic charges in the weight lattice of the GNO dual of the gauge group. Before we see explicitly how to construct these, we make a brief remark on terminology. We refer to (gauge invariant) monopole operators which are not dressed by any adjoint scalars as (physical) bare monopoles and the supersymmetry-preserving dressing factors as adjoint scalars or Casimirs (these are naturally physical). Together these form the physical Coulomb branch degrees of freedom $\mathcal{D}_{\mathcal{C}}$. These degrees of freedom can then be used to construct the Coulomb branch operators $\mathcal{O}_{\mathcal{C}}$, which we define as linear combinations of products of $\mathcal{D}_{\mathcal{C}}$ that have matching conserved charges (topological charge and conformal dimension).\\

\paragraph{Physical bare monopoles} By the work of \cite{Goddard:1976qe}, bare monopole operators are parameterised by magnetic charges lying in the weight lattice of the GNO dual group of the gauge group $G$, $\Lambda_{G^{\vee}}$. Suppose we have just a single gauge group $G$ with $\text{rank}(G)=n$, then a bare monopole operator has magnetic charge $\boldsymbol{m} = (m_1,...,m_n) \in \Lambda_{G^{\vee}}$. We denote this bare monopole operator $v_{\boldsymbol{m}}=v_{m_1 \, \cdots \, m_n}$ correspondingly. However this is not necessarily a physical degree of freedom $\mathcal{D}_{\mathcal{C}}$, as it may not be gauge invariant. The gauge group acts on the weights of $G^{\vee}$ via its Weyl group $\mathcal{W}_{G^{\vee}}$, and if $\boldsymbol{m}$ is not invariant under $\mathcal{W}_{G^{\vee}}$ then this bare monopole operator $v_{\boldsymbol{m}}$ is not physical. To rectify this, for a given magnetic charge $\boldsymbol{m}$ we need to sum up the unphysical bare monopole operators with magnetic charges in the $\mathcal{W}_{G^{\vee}}$ orbit of $\boldsymbol{m}$. We will call the physical bare monopole corresponding to $v_{\boldsymbol{m}}$ by the name $V_{\boldsymbol{m}}$:
\begin{equation}\label{gauge inv monops}
   V_{\boldsymbol{m}} = \sum_{\sigma \in \mathcal{W}_{G^{\vee}}} v_{\sigma(\boldsymbol{m}) }.
\end{equation}
Note that in this notation we clearly have $V_{\boldsymbol{m}}=V_{\sigma (\boldsymbol{m})}$ for any $\sigma \in \mathcal{W}_{G^{\vee}}$, and so if we write down all the $V_{\boldsymbol{m}}$ for every $\boldsymbol{m} \in \Lambda_{G^{\vee}}$, many will be identical and the duplicates will need deleting so as not to overcount. To avoid these duplicates, we can just compute (\ref{gauge inv monops}) for each $\boldsymbol{m}$ in a single Weyl chamber of $\Lambda_{G^{\vee}}$:
\begin{equation}\label{gauge inv monops not overcount}
    \{ V_{\boldsymbol{m}} \} = \left\{  \sum_{\sigma \in \mathcal{W}_{G^{\vee}}} v_{\sigma(\boldsymbol{m})} \ \Bigg| \ \boldsymbol{m} \in \Lambda_{G^{\vee}} / \mathcal{W}_{G^{\vee}} \right\}.
\end{equation}

\noindent In this paper we only study Coulomb branches of gauge theories involving unitary gauge groups. In the case of $G=U(n)$, since $U(n)$ is GNO self-dual we have that
\begin{equation}
    \Lambda_{G^{\vee}} = \Lambda_{U(n)} = \mathbb{Z}^n,
\end{equation}
and so a bare monopole operator $v_{\boldsymbol{m}}=v_{m_1 \, \cdots \, m_n}$ is labelled by $n$ integers. The Weyl group here is $\mathcal{W}_{G^{\vee}}=S_n$, and so if $m_1,...,m_n$ are not all equal then $v_{\boldsymbol{m}}$ will not be gauge invariant. The corresponding physical degree of freedom $V_{\boldsymbol{m}}$ would be $\sum_{\sigma \in S_n} v_{\sigma(\boldsymbol{m})}$. As mentioned in the remark above, to reduce the number of duplicate monopole operators we'll need to remove, we restrict to a single Weyl chamber of the GNO dual weight lattice of $G=U(n)$:
\begin{equation}
    \boldsymbol{m} \in \mathbb{Z}^n / S_n = \{m_1,...,m_n \, \in \, \mathbb{Z}^n \ | \ m_1 \geq m_2 \geq \cdots \geq m_n \},
\end{equation}
and find the physical bare monopoles for each magnetic charge in this chamber:
\begin{equation}
\begin{split}
    \{ V_{\boldsymbol{m} \,\in \, \mathbb{Z}^n / S_n }\} = \left\{ \sum_{\sigma \in S_n} v_{\sigma(\boldsymbol{m})} \ \Bigg| \ \boldsymbol{m}\in \mathbb{Z}^n / S_n\right\}.
\end{split}
\end{equation}

\noindent By extension, the unphysical bare monopoles in a quiver with $p$ unitary gauge nodes $U(n_i)$ $i=1,...,p$ are labelled by vectors $\boldsymbol{m}=(\boldsymbol{m_{1}},...,\boldsymbol{m_{p}})=(m_{1,1},...,m_{1,n_1},...,m_{p,1},...,m_{p,n_p})$
of length $\sum_{i=1}^{p} n_i$ with integer entries: $\{v_{\boldsymbol{m}} \ | \ \boldsymbol{m} \in \prod_{i=1}^{p}\mathbb{Z}^{n_i}\}$. The set of all physical bare monopole degrees of freedom is then
\begin{equation}\label{bare monop cb ops}
    \begin{split}
    \left\{ V_{\boldsymbol{m}}  \right\} = \left\{ \sum_{\sigma \, \in \, \prod_{i=1}^{p} S_{n_i}} v_{\sigma(\boldsymbol{m})} \ \Big| \  \boldsymbol{m}\in \prod_{i=1}^{p} \, \mathbb{Z}^{n_i} / S_{n_i} \right\}.
    \end{split}
\end{equation}

\paragraph{Adjoint valued complex scalars}
A complex scalar parameterising the Coulomb branch lives in the adjoint representation of the gauge group $G$, and therefore can be thought of as a matrix valued in the Lie algebra $\mathfrak{g}$ of $G$, with $\text{dim}(G)$ degrees of freedom. This can be diagonalised, which boils down this Lie-algebra-valued matrix to a diagonal matrix with just $\text{rank}(G)$ entries on the diagonal. These adjoint scalars have magnetic charge $\boldsymbol{m}=\boldsymbol{0}$ as they are not monopoles. As with the monopoles, the physical adjoint scalars that can dress monopoles on the Coulomb branch will be gauge invariant combinations of these $\text{rank}(G)$ Eigenvalues.\\

\noindent For example, in $G=U(n)$, a generic $\mathfrak{g}$ (adjoint) valued scalar would take the form
\begin{equation}
   \phi =  \begin{pmatrix}
        \phi_{11} & \cdots & \phi_{1n}\\
        \cdots & \cdots & \cdots \\
        \phi_{n1} & \cdots &\phi_{nn}\\
        \end{pmatrix},
\end{equation}
with $\phi_{ij}\in\mathbb{C}$ for $i,j=1,...,n$ and $\phi=-\phi^\dagger$. This can be diagonalised to an equivalent matrix
\begin{equation}
    \begin{pmatrix}
        \lambda_{1} & 0 & \cdots & 0 \\
        0 & \lambda_2 &  \cdots & 0\\
        \cdots & \cdots & \cdots & \cdots\\
        0 & \cdots & 0 & \lambda_n
        \end{pmatrix},
\end{equation}
where $\lambda_i$, $i=1,...,n$ are the Eigenvalues of $\phi$. The adjoint valued scalars on the Coulomb branch are then fully parameterised by gauge invariant combinations of $\{\lambda_1, \ \cdots \ , \ \lambda_n\}$. The Weyl group $\mathcal{W}(U(n))=S_n$ acts to permute these eigenvalues, hence the physical adjoint scalar degree of freedom is
\begin{equation}
    \lambda_1+ \cdots + \lambda_n.
\end{equation}

\noindent For a generic unitary quiver with $p$ nodes with ranks $n_i$ for $i=1,...,p$, the above logically extends and we find that the physical adjoint scalar degrees of freedom are
\begin{equation}\label{adj valued scalars}
    \left\{ \sum_{j=1}^{n_i} \lambda_{i,j}  \,| \, i=1,...,p \right\}.
\end{equation}

\vspace{3mm}

\paragraph{All degrees of freedom and operators}
We have seen that the complete list of physical Coulomb branch degrees of freedom for a quiver with $p$ unitary gauge nodes of ranks $n_i$ for $i=1,...,p$ is comprised of (\ref{bare monop cb ops}) and (\ref{adj valued scalars}):
\begin{equation}\label{all cb dof}
    \mathcal{D}_{\mathcal{C}} = \left\{ \ V_{\boldsymbol{m}} \ | \ \boldsymbol{m}\in \prod_{i=1}^{p} \, \mathbb{Z}^{n_i} / S_{n_i}\right\} \ \cup \   \left\{ \sum_{j=1}^{n_i} \lambda_{i,j}  \,| \, i=1,...,p \right\},
\end{equation}
and so the Coulomb branch operators are products of the two
\begin{equation}\label{all physical cb ops}
    \mathcal{O}_{\mathcal{C}} = \Bigg\{ \prod_{i=1}^{p} \Big(\sum_{j=1}^{n_i} \lambda_{i,j}\Big)^{q_i}  \prod_{\boldsymbol{m} \, \in \, \prod_{i=1}^{p} \mathbb{Z}^{n_i} / S_{n_i} } {(V_{\boldsymbol{m}}})^{\tilde{q}_{\boldsymbol{m}}} \Bigg\}
\end{equation}
and linear combinations thereof, for any choice of non-negative integers $q_i$ and $\tilde{q}_{\boldsymbol{m}}$ such that each term in the linear combination has the same values for the conserved charges (from here on out we will refer to such linear combinations as ``consistent"). \\

\paragraph{Generators}
To find which linear combinations of (\ref{all physical cb ops}) have the correct charges to be generators of the Coulomb branch in question, we first consult the plethystic logarithm of its Hilbert series. This tells us the conformal dimension(s) at which the generators lie, which we'll call $\Delta^{\text{gen}}_i$ for $i=1,...,d$ where $d$ is the number of distinct conformal dimensions of generators (which is less than or equal to the number of generators). We then solve the $d$ equations 
\begin{equation}
    \Delta^{\text{gen}}_i = \Delta(\mathcal{G}_{\mathcal{C}}),
\end{equation}
to find which of the Coulomb branch operators $\mathcal{O}_{\mathcal{C}}$ (\ref{all physical cb ops}) are the generators of the Coulomb branch, $\mathcal{G}_{\mathcal{C}}$. To do this we need to know how to compute the conformal dimension of a Coulomb branch operator. For a bare monopole operator with magnetic charge $\boldsymbol{m}$ this is given for example in ($2.4$) of \cite{Cremonesi:2013lqa}, and since the adjoint valued scalars lie in the vectormultiplet, we always have
\begin{equation}
    \Delta(\text{adjoint scalar})=1.
\end{equation}

\paragraph{Example}
Consider the theory of $U(1)$ with $2$ electrons, described by the following quiver:
\begin{equation}\label{u1 2f}
    \mathcal{Q}_{A_1} = \raisebox{-0.7\height}{\begin{tikzpicture}[x=1cm,y=.8cm]
    \node (g1) at (0,0) [gauge,label=below:{$1$}] {};
    \node (g2) at (1,0) [flavor,label=below:{$2$}] {};
    \draw (g1)--(g2);
    \end{tikzpicture}}\ .
\end{equation}
Since there is just one gauge group and it is Abelian, here $\boldsymbol{m} = m \in \mathbb{Z}$ is just a vector of length one, and there is just one adjoint eigenvalue $\lambda \in \mathbb{C}$. The physical degrees of freedom on the Coulomb branch here are then 
\begin{equation}\label{u1 2f dof}
    \mathcal{D}_{\mathcal{C}} = \{\lambda, \ V_{m} \}.
\end{equation}
Note that since here the only Weyl group is the trivial group $S_1$, the $V_m$ are equal to the $v_m$, as $v_m$ are already physical. The Coulomb branch operators then are
\begin{equation}\label{u12f oc}
    \mathcal{O}_{\mathcal{C}} = \{ \lambda^q \, \prod_{m \, \in \, \mathbb{Z}} (V_{m})^{\tilde{q}_{m}} \}
\end{equation}
for non-negative integers $q$ and $\tilde{q}_{m}$, and consistent linear combinations thereof.\\

\noindent Now let's find the operators among $\mathcal{O}_{\mathcal{C}}$ which are generators. The plethystic logarithm of the Hilbert series is 
\begin{equation}\label{pl hs u1 2f}
    PL(HS(\mathcal{C}(\mathcal{Q}_{A_1}))) = [2]_{SU(2)} t^2 - t^4,
\end{equation}
hence every generator has $\Delta^{\text{gen}}=1$. This means that the Coulomb branch generators $\mathcal{G}_{\mathcal{C}}$ are the operators $\mathcal{O}_{\mathcal{C}}$ (\ref{u12f oc}) which have $\Delta(\mathcal{O}_{\mathcal{C}})=1$. For this quiver theory, the conformal dimension of a bare monopole is
\begin{equation}
    \Delta_{\mathcal{Q}_{A_1}}(m) = |m|,
\end{equation}
and we know
\begin{equation}
    \Delta_{\mathcal{Q}_{A_1}}(\lambda)=1,
\end{equation}
thus the generators of $\mathcal{C}(\mathcal{Q}_{A_1})$ are the bare monopoles $V_{\pm 1}$ and the complex valued adjoint scalar $\lambda$. So the generating operators of the Coulomb branch of (\ref{u1 2f}) are 
\begin{equation}\label{u1 2f cb gens}
    \mathcal{G}_{\mathcal{C}} = \{\lambda, \ V_{+1}, \ V_{-1} \},
\end{equation}
up to constants. More complex linear combinations are not allowed because (\ref{u1 2f cb gens}) all have distinct topological charges: $0$, $+1$ and $-1$ respectively. All other physical Coulomb branch operators $\mathcal{O}_{\mathcal{C}}$ (\ref{u12f oc}) can be formed from products of (\ref{u1 2f cb gens}) and consistent linear combinations thereof.
\hfill$\square$\\

\subsection{Fixing the form of Poisson bracket relations using charge conservation}\label{sec: rough form pbs from charge cons}
In the refined Hilbert series we grade the Coulomb branch operators by their charges under two global symmetries of the Coulomb branch: the conformal dimension $\Delta$ under $SU(2)_R$, and the topological charge $\boldsymbol{J}$ under the topological symmetry. For a unitary quiver comprised of $p$ gauge nodes, as described in Section \ref{sec: cb ops}, the topological charge is given by
\begin{equation}\label{top charge for unit quiv}
    \boldsymbol{J(m)}=(J(m_1),\, ...\, ,J(m_p))=(\sum_{k=1}^{n_1} m_{1,k}\, , \, ... \, , \sum_{k=1}^{n_p} m_{p,k}).
\end{equation} 
These charges must be logically conserved under action with the Poisson bracket.\\

\noindent We know the Poisson bracket acts like two derivatives with respect to the coordinates of the variety on which it acts: it has ``weight" $-2$ with respect to the degree of a monomial on this variety (see \cite{2020AnHP...21.1235B} for an intuitive physical reasoning for this). The Hilbert series counts the holomorphic functions (chiral operators) on the Coulomb branch, graded by a power of $t$ equal to twice their conformal dimension, so the conformal dimension of the Poisson bracket between two Coulomb branch operators $\mathcal{O}_1$ and $\mathcal{O}_2$ should be
\begin{equation}\label{delta for PB}
    \Delta(\{\mathcal{O}_1,\mathcal{O}_2\}) = \Delta(\mathcal{O}_1) + \Delta(\mathcal{O}_2) - 1.
\end{equation}

\noindent The Poisson bracket is a structure defined on the variety, independent of any gauge theory construction, and thus the magnetic charge $\boldsymbol{m}$ of the Poisson bracket itself should be zero. The magnetic charge of the result of a Poisson bracket between two Coulomb branch operators $\mathcal{O}_1$ and $\mathcal{O}_2$ should therefore be
\begin{equation}\label{top charge for PB}
    \boldsymbol{J}(\{\mathcal{O}_1,\mathcal{O}_2\}) = \boldsymbol{J}(\mathcal{O}_1) + \boldsymbol{J}(\mathcal{O}_2).
\end{equation}

\noindent Using (\ref{delta for PB}) and (\ref{top charge for PB}), we can constrain which operators can lie in the result of our Poisson bracket, up to linear combinations.

\paragraph{Example} As in Section \ref{sec: cb ops}, consider the example of the Coulomb branch of $U(1)$ with $2$ flavours (\ref{u1 2f}). Recall that here all the generators have conformal dimension $1$
\begin{equation}
\begin{gathered}
     \Delta(\lambda) = 1,\\[5pt]
     \Delta(V_{+1}) = 1,\\[5pt]
     \Delta(V_{-1})=1,\\
\end{gathered}
\end{equation}
and therefore, using (\ref{delta for PB}), the Poisson bracket between any two generators from (\ref{u1 2f cb gens}) must have conformal dimension
\begin{equation}\label{delta on pbs u1 2f}
    \Delta(\{ \mathcal{G}_{\mathcal{C}}, \mathcal{G}_{\mathcal{C}} \}) = 1.
\end{equation}
The topological charges of the generators (\ref{u1 2f cb gens}) are
\begin{equation}
\begin{gathered}
     J(\lambda) = 0,\\[5pt]
     J(V_{+1}) = +1,\\[5pt]
     J(V_{-1})=-1,\\
\end{gathered}
\end{equation}
and so, using (\ref{top charge for PB}), the topological charges for all possible Poisson brackets between them are\footnote{Remember that by antisymmetry, the Poisson bracket of anything with itself is zero and the only relevant Poisson brackets to consider are those between all non-ordered pairs of distinct generators.}
\begin{equation}\label{mag charges on pbs u1 2f}
\begin{gathered}
     J(\{ \lambda, V_{+1}\}) = +1,\\[5pt]
     J(\{ \lambda, V_{-1} \}) = -1,\\[5pt]
     J(\{ V_{+1}, V_{-1}\})=0.\\
\end{gathered}
\end{equation}
Hence, using the constraints of (\ref{delta on pbs u1 2f}) and (\ref{mag charges on pbs u1 2f}), the Poisson brackets themselves must take the form 
\begin{equation}\label{rough pbs u1 2f}
\begin{gathered}
     \{ \lambda, V_{+1}\} = c_1 \, V_{+1},\\[5pt]
     \{ \lambda, V_{-1} \} = c_2 \, V_{-1},\\[5pt]
     \{ V_{+1}, V_{-1}\}= c_3 \, \lambda,\\
\end{gathered}
\end{equation}
for some constants $c_1,c_2,c_3$.\footnote{These Poisson brackets indeed form a closed algebra as we would expect: $\{\cdot,\cdot\}$ acts as the Lie bracket of the global symmetry.} \hfill $\square$\\

\subsection{Fixing the remaining constants}\label{sec: fix rem const}
Thus far we've discussed how to fix the general form of the Poisson brackets between generators using charge conservation, but now we'd like to find the explicit constants of proportionality (e.g. the $c_1,c_2,c_3$ in (\ref{rough pbs u1 2f})). We can do this in one of two ways depending on the situation, as detailed in the \nameref{Alg: MPBA} at the start of Section \ref{sec: pbs for monops}: using the Poisson relations of the exact variety (if known), or demanding that each operator has the correct weight under the global symmetry.\\

\subsubsection{Using the Poisson relations of the variety}\label{sec: explicit pbs from var}
We start with determining the Poisson brackets in the situation where the Coulomb branch of the quiver in question is a known variety $\mathcal{V}$ (i.e. $\mathcal{C}(Q) \equiv \mathcal{V}$) that is equipped with a canonical Poisson bracket. To establish the Poisson brackets between the generating monopoles in this case, we adhere to the following steps:
\begin{enumerate}
    \item Explicitly write down the generators $\mathcal{G}_{\mathcal{V}}$ of $\mathcal{V}$ as holomorphic polynomials in its coordinates.
    \item Calculate the Poisson bracket between these generators using the canonical Poisson bracket on the coordinates of $\mathcal{V}$.
    \item Compare these with the ``rough form" of the Poisson brackets between the generating monopoles found using the constraints of Section \ref{sec: rough form pbs from charge cons}: identify which monopoles in $\mathcal{G}_{\mathcal{C}}$ correspond to which holomorphic polynomials in $\mathcal{G}_{\mathcal{V}}$, and declare their Poisson brackets to be those found in Step $2$ above.
\end{enumerate}

\paragraph{Example} 
Recall that for the Coulomb branch of (\ref{u1 2f}), $\mathcal{G}_{\mathcal{C}}$ were simply (\ref{u1 2f cb gens}) or mulitples thereof. In this case however, we also know (via mirror symmetry and the Higgs branch construction \cite{Intriligator:1996ex, Hanany:1996ie}) that
\begin{equation}
    \mathcal{C}(\mathcal{Q}_{A_1}) = \mathbb{C}^2/\mathbb{Z}_2,
\end{equation}
i.e. here $\mathcal{V}=\mathbb{C}^2/\mathbb{Z}_2$.
The coordinates on this variety are just inherited from those on $\mathbb{C}^2$, $(z_1,z_2)$, as is the canonical Poisson bracket:
\begin{equation}\label{C2 pb}
    \{z_1,z_2\} = 1.
\end{equation}
Let's proceed with the steps above to find the Poisson brackets between $\mathcal{G}_{\mathcal{C}}$ by finding those between $\mathcal{G}_{\mathcal{V}}$:
\begin{enumerate}
    \item The variety $\mathcal{V}=\mathbb{C}^2/\mathbb{Z}_2$ is simply comprised of the elements of $\mathbb{C}^2=\mathbb{C}[z_1,z_2]$ which are invariant under the $\mathbb{Z}_2$ action
    \begin{equation}
       \begin{split}
           \mathbb{Z}_2 &: z_1 \rightarrow -z_1\\
                        &: z_2 \rightarrow -z_2\\
       \end{split} 
    \end{equation}
    Thus we can clearly see that $\mathcal{V}$ is generated by 
    \begin{equation}
        \mathcal{G}_{\mathcal{V}} = \{ A = z_1 z_2 \,,\, B = z_2^2 \,,\, C = z_1^2 \}
    \end{equation}
    up to constant factors.
    
    \item We can then use the Leibniz property of the Poisson bracket and the canonical relation (\ref{C2 pb}) to find the Poisson brackets between any pair of generators:
    \begin{equation}\label{c2z2 pbs}
        \begin{gathered}
            \{A, B\} = 2 \, B,\\[5pt]
            \{A,C\} = -2 \, C,\\[5pt]
            \{B,C\} = -4 \, A.
        \end{gathered}
    \end{equation}
    
    \item Comparing (\ref{c2z2 pbs}) with (\ref{rough pbs u1 2f}), we can match the members of $\mathcal{G}_{\mathcal{V}}$ and $\mathcal{G}_{\mathcal{C}}$ as follows:
    \begin{equation}
        \begin{gathered}
            A = \lambda,\\[5pt]
            B = V_{+1},\\[5pt]
            C = V_{-1},
        \end{gathered}
    \end{equation}
    and therefore we could conclude that 
    \begin{equation}\label{pbs c2 z2}
       \begin{gathered}
            \{\lambda,V_{+1}\} = 2 \, V_{+1},\\[5pt]
            \{\lambda,V_{-1}\} = -2 \, V_{-1},\\[5pt]
            \{V_{+1},V_{-1}\} = -4 \, \lambda.
       \end{gathered} 
    \end{equation}
    Note that the constant factors here can be rescaled by redefining the generators by multiplication by a constant: we will do this in Section \ref{sec: explicit pbs from gs charges} to make the equivalence between the methods presented in that section and this one clear.
\end{enumerate}
\hfill $\square$

\subsubsection{Demanding correct global symmetry weights}\label{sec: explicit pbs from gs charges}

\noindent We now turn to Coulomb branches for which the exact variety $\mathcal{V}$, and canonical Poisson bracket thereof, is not known. In these cases we make headway with constructing the Poisson brackets between $\mathcal{G}_{\mathcal{C}}$ by inspecting the Hilbert series. We exploit the fact that the Poisson bracket acts on representations of the global symmetry as the Lie bracket: a Cartan element of the global symmetry is an Eigenoperator of the Poisson bracket, with Eigenvalue equal to the weight under the global symmetry of the operator it acts on.\footnote{We thank the referee for pointing out that this also follows from the fact that the Casimir elements are the complex moment maps for the topological symmetry.} Note that since it's an Eigenoperator of $\{\cdot,\cdot\}$, a Cartan element must have $\Delta=1$ and $\boldsymbol{J}=0$.\\

\noindent Suppose that after completing Step $2$ of the \nameref{Alg: MPBA} (as explained in Section \ref{sec: rough form pbs from charge cons}), we found that a Poisson bracket between two generators was proportional to the Coulomb branch operator $\mathcal{O}$ (for example, in the case of $U(1)$ with $2$ flavours we found that $\{\lambda, V_{+1}\}$ was proportional to $\mathcal{O}=V_{+1}$). Then the method to constrain the constant of proportionality is as follows:
\begin{enumerate}
    \item First, find which monomial in the refined Hilbert series corresponds to $\mathcal{O}$.
    \item Using an appropriate fugacity map, find which monomial in the character of the Coulomb branch global topological symmetry (GS) that this corresponds to, and hence deduce the weight of $\mathcal{O}$ under GS.
    \item Define $r=\text{rank}(\text{GS})$ Cartan elements, $C_1,...,C_{r}$, as linear combinations of all possible Coulomb branch operators with $\Delta=1$ and $\boldsymbol{J}=\boldsymbol{0}$. 
    \item Demand that the Poisson brackets $\{C_1,\mathcal{O} \} \,, \, \dots \,, \, \{C_{rank(GS)},\mathcal{O}\}$ yield the correct weight of $\mathcal{O}$ under GS (as found in Step $2$), and solve for the constants in the problem (the constants of proportionality in the postulated Poisson brackets based on charge conservation, and the constants appearing in the linear combinations in $C_1,...,C_r$). 
\end{enumerate}

\paragraph{Example} Recall the ``rough" (i.e. up to constants) Poisson relations (\ref{rough pbs u1 2f}) for the Coulomb branch generators (\ref{u1 2f cb gens}) of SQED with $2$ electrons (\ref{u1 2f}) considered in the previous two subsections. To fix the constants of proportionality $c_1,c_2,c_3$ here, let's follow the steps above.

\begin{enumerate}
    \item First, we find the monomials in the refined Hilbert series corresponding to each of $\lambda, V_{+1},V_{-1}$. Recall the monopole formula \cite{Cremonesi:2013lqa}:
        \begin{equation}
            \sum_{\boldsymbol{m}\in \Gamma_{G^\vee}/\mathcal{W}_{G^\vee}}  P_{G}(t,\boldsymbol{m}) \, t^{2\Delta(\boldsymbol{m})} \, z^{J(\boldsymbol{m})}.
        \end{equation}
    Then the monomials at order $t^2$ (i.e. $\Delta=1$) coming from $m=0,+1,-1$, respectively are
        \begin{equation}
            1 ,\  z, \ z^{-1}
        \end{equation}
    so $\lambda, V_{+1},V_{-1}$ (up to constants) correspond to monomials $1,z,z^{-1}$ respectively in the refined Hilbert series.
    
    \item This term $(1+z+z^{-1}) \,t^2$ in the Hilbert series can be put into the form of the adjoint character of $SU(2)$ via the fugacity map 
        \begin{equation}\label{fug map c2z2}
            z \rightarrow x^2,
        \end{equation}
    hence the global symmetry is $GS=SU(2)$. Note that $SU(2)$ is actually the local form of GS; its global form is actually $SU(2)/\mathbb{Z}_2$, as can be seen from the fact that no fermionic representations show up in the Hilbert series (\ref{pl hs u1 2f}). As we do not make use of this global form, we refer to GS simply as its local form $SU(2)$ in the subsequent discussion.\footnote{Throughout the rest of the paper, we will not specify the global form of the global symmetry $GS\times SU(2)_R$. We will always provide the terminating PL(HS) or PL(HWG) of all quivers we study, and this inherently contains the information of the global form of $GS\times SU(2)_R$, as it tells us exactly what representations show up. To write down the global form from this does not take much work, and since it will not be of use to us in our study of Poisson brackets we opt not to include it in this note.} Under such a map (\ref{fug map c2z2}), the physical Coulomb branch generators $\lambda$, $V_{+1}$ and $V_{-1}$ have charges $0$, $+2$ and $-2$ respectively under GS.
    
    \item The rank of the global symmetry here is $\text{rank}(SU(2))=1$, and so we need to just define one Cartan element $C_1$ in the Lie algebra of GS, $\mathfrak{su(2)}$. Only one Coulomb branch operator satisfies $\Delta=1$ and $m=0$: $\lambda$. So the Cartan operator here must be 
        \begin{equation}\label{cartan u1 2f}
            C_1 = \alpha \, \lambda
        \end{equation}
    for some constant $\alpha \in \mathbb{C}$.
    
    \item Finally we need to demand that the Poisson bracket, which takes the role of the Lie bracket of the complexification of the Lie algebra of the $SU(2)$ global symmetry on the Coulomb branch $\mathfrak{su(2)}_{\mathbb{C}}=\mathfrak{sl(2;\mathbb{C})}$, yields the correct charges of our Coulomb branch generators under this global symmetry (as determined in Step $2$):
        \begin{equation}\label{conditions on consts u1 2f}
            \begin{gathered}
                 \{C_1,\lambda\}\equiv0,  \\[5pt]
                 \{C_1, V_{+1} \} \equiv +2 \, V_{+1},\\[5pt]
                 \{C_1, V_{-1} \} \equiv -2 \, V_{-1}.\\
            \end{gathered}
        \end{equation}
    Substituting in our expression for the Cartan element (\ref{cartan u1 2f}) and using the solved Poisson bracket results fixed up to constants found in Section \ref{sec: rough form pbs from charge cons} (\ref{rough pbs u1 2f}), we find that the first of (\ref{conditions on consts u1 2f}) is automatically satisfied, and solving the second two amounts to
        \begin{equation}
            \begin{gathered}
                \alpha \, \{\lambda,V_{+1}\} = +2 \, V_{+1} = \alpha \, c_1 \, V_{+1},\\[5pt]
                \alpha \, \{\lambda,V_{-1}\} = -2 \, V_{-1} = \alpha \, c_2 \, V_{-1},\\
            \end{gathered}
        \end{equation}
    i.e. 
        \begin{equation}
            c_1=-c_2.
        \end{equation}
    Several $\alpha,c_1,c_2,c_3$ satisfy this, but we will pick values so that we can make the following identification of $\lambda$, $V_{+1}$, $V_{-1}$ with the canonical symmetric generators of $\mathfrak{sl(2;\mathbb{C})}$:
        \begin{equation}\label{sl2c gens}
            \lambda=\begin{pmatrix}
            0 & 1\\
            1 & 0\\
            \end{pmatrix},\ \ \ 
            V_{+1}=\begin{pmatrix}
            0 & 0\\
            0 & 1\\
            \end{pmatrix},\ \ \ 
            V_{-1}=\begin{pmatrix}
            1 & 0\\
            0 & 0\\
            \end{pmatrix}.
        \end{equation}
    The Poisson bracket acts on these $2 \times 2$ matrices as a commutator (see the discussion preceeding (\ref{e matrix mult}) for more details), and so we can compute:
    \begin{equation}\label{full pb u1 2f}
        \begin{gathered}
                \{\lambda,V_{+1}\} = 2 \, V_{+1},\\[5pt]
                \{\lambda,V_{-1}\} = -2 \, V_{-1},\\[5pt]
                \{V_{+1},V_{-1}\} = -\lambda.\\
            \end{gathered}
    \end{equation}
    Thus we pick $\alpha,c_1,c_2,c_3$ to be
        \begin{equation}\label{consts u1 2f}
            \alpha = 1, \ c_1=+2, \ c_2=-2, \ c_3 = -1,
        \end{equation}
    so that the Poisson brackets between the generators of our Coulomb branch (\ref{rough pbs u1 2f}) match those between the generating symmetric matrices of $\mathfrak{sl(2;\mathbb{C})}$ (\ref{full pb u1 2f}).
\end{enumerate} 
\hfill $\square$\\

\noindent The Poisson brackets we have computed in this section (\ref{full pb u1 2f}) may look different from those in Section \ref{sec: explicit pbs from var} (\ref{pbs c2 z2}), but infact they are equivalent by a redefinition of generators: if in (\ref{pbs c2 z2}) we rescale the generators as 
\begin{equation}
    \begin{gathered}
        \lambda \rightarrow \lambda,\\[5pt]
        V_{+1} \rightarrow \frac{1}{2}\, V_{+1},\\[5pt]
        V_{-1} \rightarrow \frac{1}{2} \, V_{-1},\\
    \end{gathered}
\end{equation}
then we recover the relations (\ref{full pb u1 2f}).\\

\noindent The form of the generators of $\mathfrak{sl(2;\mathbb{C})}$ used in (\ref{sl2c gens}) is worth a comment. Often $\mathfrak{sl(2;\mathbb{C})}$ is viewed as the $n=2$ version of $\mathfrak{sl(n;\mathbb{C})}$; generated by traceless $n\times n$ matrices, which (\ref{sl2c gens}) obviously are not. However, since the fundamental of $\mathfrak{sl(2;\mathbb{C})}$ is pseudo-real, not complex, and the adjoint is the second rank symmetric of the fundamental, not the product of the fundamental with the anti-fundamental minus the trace (in $\mathfrak{sl(2;\mathbb{C})}$ there is no concept of anti-fundamental, or ``up vs down" indices), it is more natural to think of $\mathfrak{sl(2;\mathbb{C})}$ matrices as being $2 \times 2$ symmetric matrices instead. The obvious generators for such matrices are (\ref{sl2c gens}). The only invariant of $\mathfrak{sl(2;\mathbb{C})}$ when viewed in this way is $\epsilon_{\alpha \beta}$ (the $\delta$ invariant that exists for all $\mathfrak{sl(n;\mathbb{C})}$ can be expressed in terms of the $\epsilon$ for $n=2$), and so matrix multiplication is done with contraction by $\epsilon$. For example, the Poisson bracket between the $\mathfrak{sl(2;\mathbb{C})}$ matrices identified with $\lambda$ and $V_{+1}$ is:
\begin{equation}\label{e matrix mult}
    \{\lambda, V_{+1} \}_{\alpha \beta} = [\lambda,V_{+1}]_{\alpha \beta} = \lambda_{\alpha \gamma} \epsilon_{\gamma \delta} (V_{+1})_{\delta \beta} - (V_{+1})_{\alpha \gamma} \epsilon_{\gamma \delta} \lambda_{\delta \beta}.
\end{equation}

\noindent We now move on to detail the results of such calculations for Coulomb branches that are free, Klein singularities and Nilpotent orbits in Sections \ref{sec: free spaces}, \ref{sec: klein sings} and \ref{sec: nilp orbs} respectively. The former two use Step $3.a)$ in the \nameref{Alg: MPBA} at the top of Section \ref{sec: pbs for monops} (the topic of Section \ref{sec: explicit pbs from var}), while the latter uses Step $3.b)$ (the topic of this section).\\

\section{Free Spaces}\label{sec: free spaces}
The first type of Coulomb branches we consider are those which are free, in the sense that they are some number of copies of the quaternionic plane: $\mathcal{C}=\mathbb{H}^k$. These are the simplest cases to consider, as the generators of $\mathbb{H}^k=(\mathbb{C}^2)^k$ are in the fundamental representation of the global symmetry $Sp(k)$, and the Poisson brackets between them are trivially inherited from those on each copy of $\mathbb{C}^2$ (\ref{defining PB on C2}). We can logically extend this structure for the $k=1$ case to $k > 1$ by fixing a complex structure
\begin{equation}
    \mathbb{H}^k \cong (\mathbb{C}^2)^k \cong \prod_{i=1}^{d} \mathbb{C}[z_{i,1},z_{i,2}],
\end{equation}
and equipping its $2k$ complex generators $z_{i,a}$ for $i=1,...,k$ and $a=1,2$ with Poisson brackets
\begin{equation}\label{defining PBs on Hk c2k}
    \{z_{i,a},z_{j,b} \} = \delta_{ij} \epsilon_{ab}.
\end{equation}
This notation is less useful however, as the $i$ and $j$ indices are not antisymmetrised. It is more convenient to label the $2k$ generators in the standard way for the fundamental representation of $Sp(k)$: with a single index taking values from $1,...,2k$. The Poisson bracket between two such generators $z_\alpha$ and $z_\beta$ is then just
\begin{equation}\label{defining PBs on Hk spk}
    \{z_\alpha,z_\beta\} = \Omega_{\alpha \beta},
\end{equation}
where $\alpha,\beta=1,...,2k$, and the $2k \, \times \, 2k$ matrix $\Omega_{\alpha \beta}$ is the invariant skew-symmetric two-form of $Sp(k)$, which we take to be
\begin{equation}\label{omega spk}
    \Omega_{\alpha \beta} = 
    \begin{pmatrix}
        \boldsymbol{0} & \mathds{1}\\
        -\mathds{1} & \boldsymbol{0}\\
    \end{pmatrix},
\end{equation}
where each entry in (\ref{omega spk}) is a $k \, \times \, k$ block matrix.

\subsection{Coulomb quivers for free spaces}\label{sec: Dynkin quivs}

\noindent A family of simple unframed quivers whose Coulomb branches\footnote{Note that because these quivers are unframed (i.e. they contain no flavours), there is a diagonal $U(1)$ which acts trivially on the Coulomb branch. Before computing the Coulomb branch of such a quiver, this $U(1)$ needs to be ungauged to avoid overcounting divergences \cite{Hanany:2020jzl}.} are $\mathbb{H}^k$ can be formed by removing the affine node from the balanced affine Dynkin quiver of any Lie group $G$ for which $h^{\vee}_G=k+2$.\footnote{If $G$ is a Lie group, $h^{\vee}_G$ is called the dual coxeter number of $G$. It is equal to the sum of the entries of the vector whose dot product with the simple roots of the corresponding Lie algebra $\mathfrak{g}$ gives the highest root of $\mathfrak{g}$. Equivalently $h^{\vee}_G$ is the sum of the node ranks on the affine quiver of $G$. The affine quiver of $G$ is the quiver in the shape of the affine Dynkin diagram of $G$, with node ranks equal to the smallest possible positive integers which render it balanced. The affine quivers are listed for example in Table 1 of \cite{Gledhill:2021cbe}, where also the notion of balance is explained in Section $2.1$.} We call such quivers \textit{Dynkin quivers of finite type}, and denote the one corresponding to $G$ with $\mathsf{D}_{G}$. They are drawn explicitly in Table \ref{tab: non affine Dds}.\\

\begin{table}[hbtp!]
    \centering
    
 \hspace*{-0.5cm}\begin{tabular}{|c|c|c|} \hline
 
Lie Group $G$ & Dynkin Quiver $\mathsf{D}_{G}$ & Coulomb Branch\\

\hline

$\begin{array}{c}
	A_{k}\\
	\end{array}$ &
$\raisebox{-.5\height}{\begin{tikzpicture}[x=1cm,y=.8cm]
\node (g8) at (-2,0) [gauger,label=below:{$1$}] {};
\node (g7) at (-1,0) [gauge,label=below:{$1$}] {};
\node (g5) at (0,0) {$\cdots$};
\node (g4) at (1,0) [gauge,label=below:{$1$}] {};
\node (g6) at (2,0) [gauger,label=below:{$1$}] {};
\draw (g8)--(g7)--(g5)--(g4)--(g6);
\draw [decorate,decoration={brace,mirror,amplitude=6pt}] (-2,-0.8) --node[below=6pt] {$k$} (2,-0.8);
\end{tikzpicture}}$ 
&
$\mathbb{H}^{k-1}$
\\

\hline

$\begin{array}{c}
	B_{k}\\
	\\
	k \geq 3
	\end{array}$ &
$\raisebox{-.5\height}{\begin{tikzpicture}[x=1cm,y=.8cm]
\node (g2) at (-3,0) [gauge,label=below:{$1$}] {};
\node (g3) at (-2,0) [gauger,label=below:{$2$}] {};
\node (g0) at (-1,0) [gauge,label=below:{$2$}] {};
\node (g4) at (0,0) {$\cdots$};
\node (g5) at (1,0) [gauge,label=below:{$2$}] {};
\node (g12) at (2,0) [gauge,label=below:{$1$}] {};
\draw (g2)--(g3)--(g0)--(g4)--(g5);
\draw[transform canvas={yshift=-1.5pt}] (g5)--(g12);
\draw[transform canvas={yshift=1.5pt}] (g5)--(g12);
\draw (1.4,-0.2)--(1.6,0)--(1.4,0.2);
\draw [decorate,decoration={brace,mirror,amplitude=6pt}] (-2,-0.8) --node[below=6pt] {$k-2$} (1,-0.8);
\end{tikzpicture}}$ 
&
$\mathbb{H}^{2k-3}$
\\

\hline

$\begin{array}{c}
	C_{k}\\
	\\
	k \geq 2
	\end{array}$ &
$\raisebox{-.5\height}{\begin{tikzpicture}[x=1cm,y=.8cm]
\node (g3) at (-2,0) [gauger,label=below:{$1$}] {};
\node (g2) at (-1,0) [gauge,label=below:{$1$}] {};
\node (g4) at (0,0) {$\cdots$};
\node (g5) at (1,0) [gauge,label=below:{$1$}] {};
\node (g12) at (2,0) [gauge,label=below:{$1$}] {};
\draw (g3)--(g2)--(g4)--(g5);
\draw[transform canvas={yshift=-1.5pt}] (g5)--(g12);
\draw[transform canvas={yshift=1.5pt}] (g5)--(g12);
\draw (1.6,-0.2)--(1.4,0)--(1.6,0.2);
\draw [decorate,decoration={brace,mirror,amplitude=6pt}] (-2,-0.8) --node[below=6pt] {$k-1$} (1,-0.8);
\end{tikzpicture}}$ 
&
$\mathbb{H}^{k-1}$
\\

\hline

$\begin{array}{c}
	D_{k}\\
	\\
	k \geq 4
	\end{array}$ &
$\raisebox{-.5\height}{\begin{tikzpicture}[x=1cm,y=.8cm]
\node (g2) at (-3,0) [gauge,label=below:{$1$}] {};
\node (g3) at (-2,0) [gauger,label=below:{$2$}] {};
\node (g1) at (-1,0) [gauge,label=below:{$2$}] {};
\node (g4) at (0,0) {$\cdots$};
\node (g5) at (1,0) [gauge,label=below:{$2$}] {};
\node (g12) at (2,0) [gauge,label=below:{$1$}] {};
\node (g13) at (1,1) [gauge,label=above:{$1$}] {};
\draw (g2)--(g3)--(g1)--(g4)--(g5)--(g12);
\draw (g13)--(g5);
\draw [decorate,decoration={brace,mirror,amplitude=6pt}] (-2,-0.8) --node[below=6pt] {$k-3$} (1,-0.8);
\end{tikzpicture}}$ 
&
$\mathbb{H}^{2k-4}$\\

\hline

$\begin{array}{c}
	E_6
	\end{array}$ &
$\raisebox{-.5\height}{\begin{tikzpicture}[x=1cm,y=.8cm]
\node (g2) at (-2,0) [gauge,label=below:{$1$}] {};
\node (g3) at (-1,0) [gauge,label=below:{$2$}] {};
\node (g4) at (0,0) [gauge,label=below:{$3$}] {};
\node (g5) at (1,0) [gauge,label=below:{$2$}] {};
\node (g12) at (2,0) [gauge,label=below:{$1$}] {};
\node (g10) at (0,1) [gauger,label=right:{$2$}] {};
\draw (g2)--(g3)--(g4)--(g5)--(g12);
\draw (g4)--(g10);
\end{tikzpicture}}$ 
&
$\mathbb{H}^{10}$
\\

\hline

$\begin{array}{c}
	E_7
	\end{array}$ &
$\raisebox{-.5\height}{\begin{tikzpicture}[x=1cm,y=.8cm]
\node (g3) at (-3,0) [gauge,label=below:{$1$}] {};
\node (g4) at (-2,0) [gauge,label=below:{$2$}] {};
\node (g13) at (-1,0) [gauge,label=below:{$3$}] {};
\node (g5) at (0,0) [gauge,label=below:{$4$}] {};
\node (g12) at (1,0) [gauge,label=below:{$3$}] {};
\node (g6) at (2,0) [gauger,label=below:{$2$}] {};
\node (g7) at (0,1) [gauge,label=above:{$2$}] {};
\draw (g3)--(g4)--(g13)--(g5)--(g12)--(g6);
\draw (g7)--(g5);
\end{tikzpicture}}$ 
&
$\mathbb{H}^{16}$
\\

\hline

$\begin{array}{c}
	E_8
	\end{array}$ &
$\raisebox{-.5\height}{\begin{tikzpicture}[x=1cm,y=.8cm]
\node (g4) at (-2.5,0) [gauger,label=below:{$2$}] {};
\node (g5) at (-1.5,0) [gauge,label=below:{$3$}] {};
\node (g6) at (-0.5,0) [gauge,label=below:{$4$}] {};
\node (g7) at (0.5,0) [gauge,label=below:{$5$}] {};
\node (g8) at (1.5,0) [gauge,label=below:{$6$}] {};
\node (g9) at (2.5,0) [gauge,label=below:{$4$}] {};
\node (g10) at (3.5,0) [gauge,label=below:{$2$}] {};
\node (g11) at (1.5,1) [gauge,label=above:{$3$}] {};
\draw (g4)--(g5)--(g6)--(g7)--(g8)--(g9)--(g10);
\draw (g8)--(g11);
\end{tikzpicture}}$
&
$\mathbb{H}^{28}$\\

\hline

$\begin{array}{c}
    \\
	F_4\\
	\\
	\end{array}$
	&
$\raisebox{-.5\height}{\begin{tikzpicture}[x=1cm,y=.8cm]
\node (g3) at (-1,0) [gauger,label=below:{$2$}] {};
\node (g4) at (0,0) [gauge,label=below:{$3$}] {};
\node (g5) at (1,0) [gauge,label=below:{$2$}] {};
\node (g12) at (2,0) [gauge,label=below:{$1$}] {};
\draw (g3)--(g4);
\draw (g12)--(g5);
\draw[transform canvas={yshift=-1.5pt}] (g4)--(g5);
\draw[transform canvas={yshift=1.5pt}] (g4)--(g5);
\draw (0.4,-0.2)--(0.6,0)--(0.4,0.2);
\end{tikzpicture}}$
&
$\mathbb{H}^7$\\

\hline

$\begin{array}{c}
    \\
    G_2\\
    \\
    \end{array}$ 
&
$\raisebox{-.5\height}{\begin{tikzpicture}[x=1cm,y=.8cm]
\node (g4) at (0,0) [gauger,label=below:{$2$}] {};
\node (g5) at (1,0) [gauge,label=below:{$1$}] {};
\draw[transform canvas = {yshift=-2pt}] (g4)--(g5);
\draw[transform canvas = {yshift=0pt}] (g4)--(g5);
\draw[transform canvas = {yshift =2pt}] (g4)--(g5);
\draw (0.4,0.2)--(0.6,0)--(0.4,-0.2);
\end{tikzpicture}}$ 
&
$\mathbb{H}^2$\\

\hline

\end{tabular}
 \caption{Dynkin quivers of finite type. The quiver in the central column is the Dynkin quiver for the Lie group in the left-hand column. The Coulomb branch variety of this quiver, after ungauging the diagonal $U(1)$, is listed in the right hand column. Red nodes are those which are unbalanced. All unbalanced nodes in this table have an excess of $-1$.}
    \label{tab: non affine Dds}
\end{table}

\noindent The method to compute the Poisson brackets for the dressed monopole operators generating the Coulomb branch $\mathcal{G}_{\mathcal{C}}$ is the same for all quivers in Table \ref{tab: non affine Dds} and is fairly trivial, so we will just illustrate the process with one example, $\mathsf{D}_{D_5}$, to introduce and walk through some of the ideas about constructing generating dressed monopole operators and imposing constraints on Poisson brackets that are used heavily in this paper.\\

\subsection{\texorpdfstring{Example: $\mathbb{H}^6 = \mathcal{C}(\mathsf{D}_{D_5})$}{H6=C(DD5)}}\label{sec: free space eg dd5}
The $\mathsf{D}_{D_5}$ quiver is given by\footnote{A red node indicates an unabalanced node. Here the excess is $-1$.}
\begin{equation}\label{D5 quiv}
     \mathsf{D}_{D_5} = \raisebox{-0.35\height}{\begin{tikzpicture}[x=1cm,y=.8cm]
    \node (g1) at (0,0) [gauge,label=below:{$1$},label={[blue]above:{$1$}}] {};
    \node (g2) at (1,0) [gauger,label=below:{$2$},label={[blue]above:{$2$}}] {};
    \node (g3) at (2,0) [gauge,label=below:{$2$},label={[blue]above:{$3$}}] {};
    \node (g4) at (2.8,-0.8) [gauge,label=right:{$1$},label={[blue]above:{$4$}}] {};
    \node (g5) at (2.8,0.8) [gauge,label=right:{$1$},label={[blue]above:{$5$}}] {};
    \draw (g1)--(g2)--(g3)--(g4);
    \draw (g3)--(g5);
    \end{tikzpicture}},
\end{equation}
where the black labels are the ranks of the unitary gauge nodes and the blue labels are names to distinguish between the nodes (we will make use of these shortly). The Coulomb branch here is entirely free ($\mathbb{H}^6$), as showcased by its Hilbert series,
\begin{equation}\label{D5 HS}
    HS(\mathcal{C}(\ref{D5 quiv}))=PE[\, [1,0,0,0,0,0]_{Sp(6)} \, t\,].
\end{equation}
In particular, its only generators lie in the fundamental representation of $Sp(6)$. The global symmetry group expected from the balance of the quiver \cite{Gledhill:2021cbe} is embedded inside the observed global symmetry $SU(2) \times SU(4) \hookrightarrow Sp(6)$. More generally, for $\mathsf{D}_{D_k}$ (shown in the fourth row of Table \ref{tab: non affine Dds}), there is an embedding of the balance global symmetry $SU(2) \times SO(2k-4)$ inside the observed global symmetry $Sp(2k-4)$. The Coulomb branch of $\mathsf{D}_{D_k}$ is $\mathbb{H}^{2k-4}$; we can deduce this quaternionic dimension $2k-4$ as half the complex dimension, given by the degrees of freedom in the unbalanced node. This node is connected to the only node of the balanced $SU(2)$ Dynkin diagram to its left and the vector node of the balanced $SO(2k-4)$ Dynkin diagram to its right; it lies in the representation formed from the product of the fundamental of $SU(2)$ and the vector of $SO(2k-4)$, and thus has $2 \cdot (2k-4) = 4k-8$ complex degrees of freedom. The quaternionic dimension of the corresponding space is then half of this: $2k-4$. In this case, $k=5$ and we see from the above that the quaternionic dimension of the Coulomb branch is $2\cdot 5-4=6$, which matches that of $\mathbb{H}^6$ as expected.\\

\noindent We will concretely demonstrate how to compute the Poisson brackets for the generating Coulomb branch operators of these theories, following steps $1,2,3.a)$ of the \nameref{Alg: MPBA}, as outlined in Sections \ref{sec: cb ops} - \ref{sec: explicit pbs from var}.\\

\subsubsection{\texorpdfstring{Writing down $\mathcal{G}_{\mathcal{C}}$}{Writing down GC}}\label{sec: free monops}
We start with Step $1$ of the \nameref{Alg: MPBA}: writing down the dressed monopole operators that generate the Coulomb branch explicitly.\\

\noindent Any operator on the Coulomb branch must be built out of the basic (non-gauge invariant) degrees of freedom, so let's write these down explicitly. They will just be the $p=5$, $\{n_1,...,n_5\}=\{1,2,2,1,1\}$ case of (\ref{all cb dof}), i.e.
\begin{equation}\label{D5 quiv cb dof}
\begin{split}
    \mathcal{D}_{\mathcal{C}} = \{v_{\boldsymbol{m}},\ \lambda_{1,1}, \ \lambda_{2,1}, \ \lambda_{2,2}, \ \lambda_{3,1}, \  \lambda_{3,2}, \  \lambda_{4,1}, \ \lambda_{5,1}\}
\end{split}
\end{equation}
where $v_{\boldsymbol{m}}= v_{\boldsymbol{m_1} \boldsymbol{m_2} \boldsymbol{m_3} \boldsymbol{m_4} \boldsymbol{m_5}} = v_{m_{1,1} m_{2,1} m_{2,2} m_{3,1} m_{3,2} m_{4,1} m_{5,1}}$. The $\lambda_{i,j} \in \mathbb{C}$ and $m_{i,j} \in \mathbb{Z}$ are the $j^{th}$ adjoint scalar and magnetic charge respectively for the $i^{th}$ node (the labeling of the nodes is given in blue in (\ref{D5 quiv})).\\

\noindent Now recall from (\ref{D5 HS}) that the generators all lie at order $t$ in the Hilbert series; they have conformal dimension $\Delta=\frac{1}{2}$. The $\lambda_{i,j}$ cannot generate the Coulomb branch as they all have $\Delta=1$, hence only bare monopoles with magnetic charges $\boldsymbol{m}=( \boldsymbol{m_1},\boldsymbol{m_2},\boldsymbol{m_3},\boldsymbol{m_4},\boldsymbol{m_5} )=(m_{1,1}, m_{2,1}, m_{2,2}, m_{3,1}, m_{3,2}, m_{4,1}, m_{5,1})$ which solve
\begin{equation}\label{D5 conf dim}
    \begin{split}
        \frac{1}{2} = \Delta = &-|m_{2,1}-m_{2,2}| - |m_{3,1}-m_{3,2}|+\frac{1}{2}(|m_{1,1}-m_{2,1}|+|m_{1,1}-m_{2,2}|+|m_{2,1}-m_{3,1}|\\
     &+|m_{2,1}-m_{3,2}|
    +|m_{2,2}-m_{3,1}|+|m_{2,2}-m_{3,2}|+|m_{3,1}-m_{4,1}|+|m_{3,2}-m_{4,1}|\\[3pt]
    &+|m_{3,1}-m_{5,1}|+|m_{3,2}-m_{5,1}|)
    \end{split} 
\end{equation} 
can be the generators. If $\boldsymbol{m}$ is not invariant under action by the Weyl group, then a single monopole operator $v_{\boldsymbol{m}}$ with this charge would be unphysical. Recall the physical counterpart $V_{\boldsymbol{m}}$ is then given by the sum of all the unphysical monopoles $v_{\boldsymbol{\tilde{m}}}$ with magnetic charge $\boldsymbol{\tilde{m}}$ in the Weyl orbit of $\boldsymbol{m}$ (\ref{gauge inv monops}). As mentioned before in the discussion around (\ref{gauge inv monops not overcount}), to simplify computational difficulty in solving (\ref{D5 conf dim}) and eliminate overcounting physical solutions $V_{\boldsymbol{m}}$, we will choose to restrict the possible solutions of (\ref{D5 conf dim}) to the dominant Weyl chambers: 
\begin{equation}\label{D5 weyl constraints}
    \begin{split}
        &m_{1,1} \geq 0,\\
        &m_{2,1} \geq m_{2,2} \geq 0,\\
        &m_{3,1} \geq m_{3,2} \geq 0,\\
        &m_{4,1} \geq 0,\\
        &m_{5,1} \geq 0,\\
    \end{split}
\end{equation}
and then act upon solutions to (\ref{D5 conf dim}) that satisfy (\ref{D5 weyl constraints}) with the Weyl group. Since (\ref{D5 quiv}) is an unframed quiver, in order to compute the Hilbert series we need to ungauge,\footnote{There is an overall $U(1)$ which acts trivially on the Coulomb branch, which needs to be ungauged to avoid overcounting divergences \cite{Hanany:2020jzl}.} which for us will manifest as setting one of the $m_{i,j}$ to be zero. The choice of where to ungauge will not change the Hilbert series or the gauge invariant generating monopole operators, but will change the solutions $( \boldsymbol{m_1},\boldsymbol{m_2},\boldsymbol{m_3},\boldsymbol{m_4},\boldsymbol{m_5} )$ to (\ref{D5 conf dim}). Here we'll choose to ungauge on the node labelled $1$ (the unbalanced node), as it turns out to be the simplest\footnote{Simplest in the sense of computational power: it yields the fewest solutions to (\ref{D5 conf dim}) meaning the solutions are easier to find and there are less duplicates generated (that need to be deleted) from action by the Weyl group.} choice: the solutions to (\ref{D5 conf dim}) for $m_{1,1}=0$ are
\begin{equation}\label{D5 conf dim sols}
    (m_{2,1},m_{2,2},m_{3,1},m_{3,2},m_{4,1},m_{5,1}) =
   \begin{cases}
     (0,-1,-1,-1,-1,-1) \\
     (0,-1,0,-1,-1,-1)\\
     (0,-1,0,-1,-1,0)\\
     (0,-1,0,-1,0,-1)\\
     (0,-1,0,-1,0,0)\\
     (0,-1,0,0,0,0)\\
     (1,0,0,0,0,0)\\
     (1,0,1,0,0,0)\\
     (1,0,1,0,0,1)\\
     (1,0,1,0,1,0)\\
     (1,0,1,0,1,1)\\
     (1,0,1,1,1,1)
   \end{cases}.
\end{equation}
Acting with the Weyl group $S_2 \times S_2 \times S_1 \times S_1$ (where $S_n$ is the finite symmetric group of order $n$)\footnote{Note that there are exactly $12$ solutions (\ref{D5 conf dim sols}) to (\ref{D5 conf dim}) when we ungauge on node $1$, which is exactly the number of generators we have (the coefficient of $t$ in the unrefined HS is indeed $12$): in this case there are no duplicate solutions to delete. If we had chosen to ungauge on a different node however, there would have been duplicates to delete.} to find the physical bare monopoles, we see that the generators for $\mathcal{C}$(\ref{D5 quiv}) (up to constants\footnote{No two operators in (\ref{D5 CB gens}) have matching topological chargers, so the only valid linear combinations are multiples of (\ref{D5 CB gens}) by a constant.}) are
\begin{equation}\label{D5 CB gens}
    \mathcal{G}_{\mathcal{C}} = \begin{cases}
        V_{\pm 1 0 0 0 0 0} = v_{\pm 1 0 0 0 0 0} + v_{0 \pm 1 0 0 0 0}\\
        V_{\pm 1 0 \pm 1 0 0 0} = v_{\pm 1 0 \pm 1 0 0 0} + v_{0 \pm 1 \pm 1 0 0 0} + v_{\pm 1 0 0 \pm 1 0 0} + v_{0 \pm 1 0 \pm 1 0 0}\\
        V_{\pm 1 0 \pm 1 0 0 \pm1} = v_{\pm 1 0 \pm 1 0 0 \pm1} + v_{0 \pm 1 \pm 1 0 0 \pm1} + v_{\pm 1 0 0 \pm 1 0 \pm1} + v_{0 \pm 1 0 \pm 1 0 \pm1}\\
        V_{\pm 1 0 \pm 1 0 \pm1 0} = v_{\pm 1 0 \pm 1 0 \pm1 0} + v_{0 \pm 1 \pm 1 0 0 \pm1 0} + v_{\pm 1 0 0 \pm 1 \pm1 0} + v_{0 \pm 1 0 \pm 1 \pm1 0}\\
        V_{\pm 1 0 \pm 1 0 \pm1 \pm1} = v_{\pm 1 0 \pm 1 0 \pm1 \pm1} + v_{0 \pm 1 \pm 1 0 \pm1 \pm1} + v_{\pm 1 0 0 \pm 1 \pm1 \pm1} + v_{0 \pm 1 0 \pm 1 \pm1 \pm1}\\
        V_{\pm 1 0 \pm 1 \pm 1 \pm 1 \pm 1} = v_{\pm 1 0 \pm 1 \pm 1 \pm 1 \pm 1} + v_{0 \pm 1 \pm 1 \pm 1 \pm 1 \pm 1}
    \end{cases}.
\end{equation}
The Casimir invariants $\lambda_{i,j}$ of the gauge group will be formed as bilinears in these generators (\ref{D5 CB gens}).

\subsubsection{Constraining Poisson brackets by charge conservation}
Next we will constrain the results of the Poisson brackets between unordered pairs of these generators using conservation of conformal dimension and magnetic charge, as discussed in Section \ref{sec: rough form pbs from charge cons}. \\

\noindent Using (\ref{delta for PB}), we see that imposing $\Delta$-conservation tells us that for any $V_1,V_2 \in \mathcal{G}_{\mathcal{C}}$,
\begin{equation}
    \Delta(\{V_1,V_2\})=0.
\end{equation}
This tells us that the Poisson brackets between any two of the $\mathcal{G}_{\mathcal{C}}$ must be proportional to operators with charge $\Delta=0$ under the $R$-symmetry. The Hilbert series tells us that there is only one such operator -- the identity -- and hence $\{V_1,V_2\}$ must be proportional to the identity. This agrees with (\ref{defining PBs on Hk spk}) which says that the brackets should have value either $\pm1$ or $0$.\\

\noindent To determine which Poisson brackets vanish and which do not, we invoke the conservation of topological charge. Recalling the topological charge for a physical monopole operator (\ref{top charge for unit quiv}), we can see that the topological charges of our $\mathcal{G}_{\mathcal{C}}$ are
\begin{equation}
    \begin{gathered}
        \boldsymbol{J}(V_{\pm 1 0 0 0 0 0}) = (\pm1,0,0,0),\\[5pt]
        \boldsymbol{J}(V_{\pm 1 0 \pm 1 0 0 0}) = (\pm1,\pm1,0,0),\\[5pt]
        \boldsymbol{J}(V_{\pm 1 0 \pm 1 0 0 \pm1}) = (\pm1,\pm1,0,\pm1),\\[5pt]
        \boldsymbol{J}(V_{\pm 1 0 \pm 1 0 \pm1 0}) = (\pm1,\pm1,\pm1,0),\\[5pt]
        \boldsymbol{J}(V_{\pm 1 0 \pm 1 0 \pm1 \pm1}) = (\pm1,\pm1,\pm1,\pm1),\\[5pt]
        \boldsymbol{J}(V_{\pm 1 0 \pm 1 \pm 1 \pm 1 \pm 1}) = (\pm1,\pm2,\pm1,\pm1).
    \end{gathered}
\end{equation}
Above, we saw that any non-zero result result of a Poisson bracket between two generators $V_1$ and $V_2$ should be proportional to the identity operator and thus have zero topological charge. Thus we can see from (\ref{top charge for PB}) that the only pairs of operators that can have non-zero Poisson pairing are those with equal and opposite magnetic charges. That is
\begin{equation}\label{rough D5 pbs}
    \{V_{\boldsymbol{m}},V_{\boldsymbol{\tilde{m}}}\} = 
    \begin{cases}
      c_{\boldsymbol{m},\boldsymbol{\tilde{m}}}=-c_{\boldsymbol{\tilde{m}},\boldsymbol{m}} & \text{if } \boldsymbol{m}=-\boldsymbol{\tilde{m}}\\
      0 & \text{if } \boldsymbol{m} \neq -\boldsymbol{\tilde{m}}
    \end{cases}
\end{equation}
for some antisymmetric constant $c_{\boldsymbol{m},\boldsymbol{\tilde{m}}}$.

\subsubsection{\texorpdfstring{Fixing the constants using the Poisson brackets of $(\mathbb{C}^2)^6$}{Fixing the Constants Using The Poisson Relations of C26}}

We now compare the relations (\ref{rough D5 pbs}) to those we expect of $\mathbb{H}^6$, which recall were (\ref{defining PBs on Hk spk}) for $k=6$. It is clear that $|c_{\boldsymbol{m},-\boldsymbol{m}}|=1$, but it is up to us which operator of $V_{\boldsymbol{m}}$ and $V_{-\boldsymbol{m}}$ to associate to with $z_{\alpha}$ for some $\alpha=1,...,6$, and which to associate with the corresponding $z_{\alpha+6}$. We'll choose to assign each $\boldsymbol{m} \in \mathbb{Z}_{\geq 0}^{6}$ in (\ref{D5 conf dim sols}) a different $\alpha=1,...,6$. This fixes the corresponding $-\boldsymbol{m} \in \mathbb{Z}_{\leq 0}^{6}$ to be assigned to $z_{\alpha+6}$, and gives the following Poisson brackets for the generating dressed monopoles $V_{\boldsymbol{m}}$ of $\mathsf{D}_{D_5}$:
\begin{equation}
    \{V_{\boldsymbol{m}},V_{\boldsymbol{\tilde{m}}}\} = 
     \begin{cases}
       1 & \text{if } \boldsymbol{m} = - \boldsymbol{\tilde{m}} \in \mathbb{Z}^6_{\geq 0}\\
       -1 & \text{if } \boldsymbol{m} = - \boldsymbol{\tilde{m}} \in \mathbb{Z}^6_{\leq 0}\\
       0 & \text{otherwise}
     \end{cases}\ ,
\end{equation}
recovering the exact structure of $\mathbb{H}^6$ (\ref{defining PBs on Hk spk}).

\section{Klein Singularities}\label{sec: klein sings}
We now turn our heads to Klein singularities. The results we arrive at here have previously been deduced with alternative methods, see for example \cite{lerman:1994,mcmillan:2011,mcmillan:2013}. We rederive them here from our perspective, using the method of Section \ref{sec: pbs for monops}.\\

\noindent Klein singularities are varieties $\mathcal{V}=\mathbb{C}^2/\Gamma$ which are quotients of $\mathbb{C}^2$ by a finite subgroup of its global symmetry $\Gamma \subset SU(2)$. An exhaustive list of such subgroups are in one to one correspondence with the affine $ADE$ Dynkin diagrams \cite{McKay:1980,McKay:1982} and as such we often denote $\mathbb{C}^2/\Gamma$ instead by the capital letter of the $ADE$ diagram it corresponds to, with its rank as subscript. In particular, the full list of Klein singularities can be denoted as follows:
\begin{equation}\label{klein sings}
    A_k, \ \ \ D_k, \ \ \ E_6, \ \ \ E_7, \ \ \ E_8.
\end{equation}

\noindent The generators $\mathcal{G}_{\Gamma}$ of the subgroups $\Gamma \subset SU(2)$ are listed on page $14$ of \cite{Benvenuti:2006qr} as $2\times 2$ matrices, which we can think of as acting on the vector containing the complex coordinates of $\mathbb{C}^2$:
\begin{equation}\label{c2 coords}
    \begin{pmatrix}
        z_1\\
        z_2
    \end{pmatrix}.
\end{equation}
The elements of the corresponding Klein singualrity $\mathbb{C}^2/\Gamma$ (\ref{klein sings}) will then just be the elements of $\mathbb{C}^2$ (i.e. polynomials in $z_1,z_2$) which are invariant under action by any product of these $\Gamma$ generators.\\

\noindent The Hilbert series of these Klein singularities is also listed in \cite{Benvenuti:2006qr}; the degree of their generators can be found by taking the plethystic logarithm. We can then explicitly construct these generators $\mathcal{G}_{\mathbb{C}^2/\Gamma}$ by finding the monomials of $\mathbb{C}^2$ which are: invariant under the action of $\Gamma$; of the correct polynomial degree; and irreducible (i.e. not generated by a product of elements of $\mathbb{C}^2/\Gamma$ of lower degrees). Each Klein singularity has $3$ such generators
\begin{equation}
    |\mathcal{G}_{\mathbb{C}^2/\Gamma}| = 3,
\end{equation}
and they satisfy a relation which is termed the \textit{defining relation} of the singularity, which is also given in \cite{Benvenuti:2006qr}.\footnote{Note that by a change of variables the precise form of this equation can be modified.} The Poisson brackets between these generators can be computed using $\{\cdot, \cdot \}_{\mathbb{C}^2}$, the inherited canonical Poisson bracket of $\mathbb{C}^2$ (\ref{C2 pb}). This can then be used to identify the Poisson brackets between members of $\mathcal{G}_{\mathcal{C}}$ for the Coulomb quiver\footnote{Recall the Coulomb quiver for a variety $V$ is the quiver $Q$ whose Coulomb branch is $V$: $\mathcal{C}(Q)=V$.} of $\mathbb{C}^2/\Gamma$, as in Step $3.a)$ of the \nameref{Alg: MPBA}.\\

\noindent We found that the Poisson brackets of $\mathcal{G}_{\mathbb{C}^2/\Gamma}$ for all Klein singularities (\ref{klein sings}), including the $\Gamma=E_{6,7,8}$ cases, can be summed up by the same succinct formula: if we call $g_1,g_2,g_3$ the generators and $E(g_1,g_2,g_3)$ the defining equation for $\mathbb{C}^2/\Gamma$, then
\begin{equation}\label{pbs for klein sings}
    \{g_i,g_j\} = \epsilon_{ijk} \frac{\partial E}{\partial g_k}.
\end{equation}
It is worth noting that the explicit form of a defining relation is not fixed inherently: a change of variables would alter it. However, imposing the Poisson brackets (\ref{pbs for klein sings}) fully fixes the defining equation and vice versa. \\

\noindent Since it is only the cases of $A_k$ and $D_k$ for which Coulomb quivers are known, it is just these that we will restrict our study to in the remainder of this chapter. The Higgs quivers for $E_{6,7,8}$ are known, and the Poisson brackets of the generators of the Higgs branch of these quivers can be found, but we do not cover this here as our goal is to focus on the Poisson brackets of Coulomb branch generators. In Section \ref{sec:A klein sing}, we spell out the full steps of this process and derive the $\mathcal{G}_{\mathcal{C}}$ Poisson brackets for the Coulomb quiver of $A_k$. In Section \ref{sec:D klein sing}, we jump straight to the result for the Coulomb quiver of $D_k$.\\

\subsection{\texorpdfstring{$A$ type}{A type}}\label{sec:A klein sing}
The cyclic subgroup $\Gamma=\mathbb{Z}_{k+1} \subset SU(2)$ corresponds to the affine $A_k$ Dynkin diagram, and hence $\mathbb{C}^2/\mathbb{Z}_{k+1} = A_k$ is referred to as the $A$ type Klein singularity. Analogously to how the Coulomb quiver for $\mathbb{C}^2/\mathbb{Z}_2 = A_1$ was (\ref{u1 2f}), the Coulomb quiver of $A_k$ ($k \geq 1$) is given by SQED with $k+1$ electrons:\footnote{You may also see this quiver with the flavour node labelled as $SU(k+1)$: this is because there is an $SU(k+1)$ symmetry rotating the $k+1$ hypermultiplets. This is a symmetry of the Higgs branch, so it does not play a part in our discussion.}
\begin{equation}\label{u1 kplus1 flav}
    \mathcal{Q}_{A_k}= \raisebox{-0.75\height}{\begin{tikzpicture}[x=1cm,y=.8cm]
    \node (g1) at (0,0) [gauge,label=below:{$1$}] {};
    \node (g2) at (1,0) [flavor,label=below:{$k+1$}] {};
    \draw (g1)--(g2);
    \end{tikzpicture}}.
\end{equation}
We will first derive the Poisson brackets of the generators of the abstract variety $\mathcal{G}_{\mathcal{V}=A_k}$ as described at the top of Section \ref{sec: klein sings}, before using this to do the same for the generating dressed monopoles $\mathcal{G}_{\mathcal{C}(\ref{u1 kplus1 flav})}$ of the Coulomb quiver.\\

\subsubsection{Poisson brackets for the abstract variety}\label{sec:A klein sing PBs}
Here we simply view $A_k$ as the set of polynomials in two complex variables invariant under action by $\mathbb{Z}_{k+1}$.\\

\noindent The first step is to find the generators of $A_k$. By taking the plethystic logarithm of the Hilbert series listed in \cite{Benvenuti:2006qr}, we can see that the three generators are of degrees $2$, $k+1$ and $k+1$:
\begin{equation}\label{pl u1 kplus1 f}
    PL(HS(A_k)) = t^2 + (q + \frac{1}{q})t^{k+1} - t^{2k+2}.
\end{equation}

\noindent In general to find the holomorphic functions corresponding to the degree $d$ generators of $\mathbb{C}^2/\Gamma$, we take all monomials in our complex coordinates $z_1,z_2$ of degree $d$, and for each one we total the results of the action on them by each $\Gamma$ group element. This construct invariants of $\Gamma$ in $\mathbb{C}^2$, and thus of $\mathbb{C}^2/\Gamma$. A vanishing result tells us the starting monomial was not an invariant.\\

\noindent To find the holomorphic functions corresponding to the degree $d$ generators of $A_k$, we construct the degree $d$ monomials in $z_1,z_2$ which are invariant under $\mathbb{Z}_{k+1}$. The generator of $\mathbb{Z}_{k+1}$ is
\begin{equation}
    C = \begin{pmatrix}
        \omega_{k+1} & 0\\
        0 & \omega_{k+1}^{-1}
    \end{pmatrix},
\end{equation}
where $\omega_{k}=e^{\frac{2\pi i}{k}}$, and thus under the $\mathbb{Z}_{k+1}$ action,
\begin{equation}\label{Zkplus1 action on C2}
\begin{gathered}
     z_1 \rightarrow e^{\frac{2\pi i}{k+1}} z_1,\\[5pt]
     z_2 \rightarrow e^{-\frac{2\pi i}{k+1}} z_2.
\end{gathered}
\end{equation}
The irreducible invariants under this action at degree $2$ and $k+1$ are
\begin{equation}
    \begin{split}
        \mathcal{G}_{A_k} &= \{ z_1 \, z_2,  \ z_1^{k+1}, \ z_2^{k+1} \}\\
        &\equiv \{ g_1', \ g_2', \ g_3' \},
    \end{split}
\end{equation}
and so these are our generators. We use primed variables here as we will have to redefine them by a constant to achieve the relation (\ref{pbs for klein sings}), and we will use the corresponding unprimed variable for these to match with those used in that relation. The generators $g_1', \, g_2', \, g_3'$ clearly satisfy the equation
\begin{equation}
    E'(g_1',g_2',g_3') = {g_1'}^{k+1} - g_2' \, g_3' = 0.
\end{equation}
The Poisson brackets between these generators can easily be calculated using (\ref{C2 pb}):
\begin{equation}
    \begin{gathered}
        \{g_1',g_2'\} = - (k+1) \,g_2',\\[5pt]
        \{g_1',g_3'\} = (k+1) \, g_3',\\[5pt]
        \{g_2',g_3'\} = (k+1)^2 \, {g_1'}^{k}.
    \end{gathered}
\end{equation}
We can see that if we rescale our generators to the following $g_1,g_2,g_3$
\begin{equation}\label{gens of c2 mod ak}
\begin{gathered}
    g_1 = \frac{1}{k+1} \, g_1' = \frac{1}{k+1} \, z_1 \, z_2,\\[5pt]
    g_2 = \frac{1}{(k+1)^{\frac{k+1}{2}}} g_2' = \frac{1}{(k+1)^{\frac{k+1}{2}}} z_1^{k+1},\\[5pt]
    g_3 = \frac{1}{(k+1)^{\frac{k+1}{2}}} g_3' = \frac{1}{(k+1)^{\frac{k+1}{2}}} z_2^{k+1},
\end{gathered}
\end{equation}
then we find they satisfy the same defining equation as the $g_i'$:
\begin{equation}\label{klein A E}
    E(g_1,g_2,g_3)=E'(g_1,g_2,g_3)= g_1^{k+1}-g_2g_3=0,
\end{equation}
and indeed that
\begin{equation}
    \{g_i,g_j\} = \epsilon_{ijk} \frac{\partial}{\partial g_k} E(g_1,g_2,g_3)
\end{equation}
as claimed in (\ref{pbs for klein sings}).\\

\subsubsection{Poisson brackets for the dressed monopoles of the Coulomb quiver}\label{sec:A klein sing CQ PBs}
We now make a connection with physics: we view $A_k$ as the Coulomb branch of (\ref{u1 kplus1 flav}) and find the Poisson brackets $\{\cdot,\cdot \}_{\mathcal{C}}$ of $\mathcal{G}_{\mathcal{C}}$ following Steps $1-3.a)$ of the \nameref{Alg: MPBA}.\\

\noindent For Step $1$, we need the generating dressed monopoles $\mathcal{G}_{\mathcal{C}}$. Recall the Hilbert series for $\mathcal{C}(\ref{u1 kplus1 flav})$ has plethystic logarithm given by (\ref{pl u1 kplus1 f}). In particular, the generating dressed monopoles $\mathcal{G}_{\mathcal{C}}$ have conformal dimensions $1$, $\frac{k+1}{2}$ and $\frac{k+1}{2}$. Following the same method as in Section \ref{sec: free monops}, we find that the gauge invariant dressed monopoles generating the Coulomb branch are\footnote{As in the example of Section \ref{sec: cb ops}, since the only gauge group is $U(1)$ there is no Weyl group action, and so all Coulomb branch degrees of freedom are gauge invariant: the $v$ and $V$ of Section \ref{sec: cb ops} are the same here.}
\begin{equation}\label{gc for u1 kplus1 f}
    \mathcal{G}_{\mathcal{C}} = \{\lambda, \ V_{+1}, \ V_{-1}\}
\end{equation}
up to linear combinations consistent with topological charge. In this case, since all $\mathcal{G}_{\mathcal{C}}$ have different topological charge, only constant multiples of each element are allowed to be considered as generators.\\

\noindent Step $2$ then tells us to constrain the $\{ \mathcal{G}_{\mathcal{C}},\mathcal{G}_{\mathcal{C}} \}_{\mathcal{C}}$ via charge conservation. Following the same ideas as illustrated in previous examples, this tells us
\begin{equation}
    \begin{gathered}
        \{\lambda,V_{+1}\} \propto V_{+1},\\[5pt]
        \{\lambda,V_{-1}\} \propto V_{-1},\\[5pt]
        \{V_{+1},V_{-1}\} \propto \lambda^k.\\[5pt]
    \end{gathered}
\end{equation}

\noindent Finally, we fix $\{\cdot,\cdot\}_{\mathcal{C}}$ by comparing $\mathcal{G}_{\mathcal{C}}$ (\ref{gc for u1 kplus1 f}) with the generators of the abstract variety $\mathcal{G}_{A_k}$ (\ref{gens of c2 mod ak}). We can see that if we declare that
\begin{equation}\label{GC PBs for Ak}
    \begin{gathered}
        \{\lambda,V_{+1}\} = V_{+1},\\[5pt]
        \{\lambda,V_{-1}\} = -V_{-1},\\[5pt]
        \{V_{+1},V_{-1}\} = \lambda^k,\\[5pt]
        V_{+1} V_{-1} = -\frac{\lambda^{n+1}}{n+1}\\
    \end{gathered}
\end{equation}
(which are consistent with (\ref{delta for PB}) and (\ref{top charge for PB})), then the following linear combinations of $\mathcal{G}_{\mathcal{C}}$
\begin{equation}
    \begin{gathered}
        g_1 = -\lambda ,\\[5pt]
        g_2 = (-1)^{\frac{k}{2}} \sqrt{k+1} V_{+1},\\[5pt]
        g_3 = (-1)^{\frac{k}{2}} \sqrt{k+1} V_{-1},\\[5pt]
    \end{gathered}
\end{equation}
satisfy the same defining equation (\ref{klein A E}) and Poisson bracket relations (\ref{pbs for klein sings}) of $\mathcal{G}_{A_k}$: our declaration of the Poisson brackets between generators (\ref{GC PBs for Ak}) for this Coulomb branch is valid. The Poisson bracket between any two operators on $\mathcal{C}(\ref{u1 kplus1 flav})$ can then be deduced from (\ref{GC PBs for Ak}).

\subsection{\texorpdfstring{$D$ type}{D type}}\label{sec:D klein sing}
The Coulomb quiver for the $D$ type Klein singularity $D_k$ is:\footnote{You may also see this quiver with the flavour node labelled as $SO(2k)$: this is because there is an $SO(2k)$ symmetry rotating the $k$ hypermultiplets. This is a symmetry of the Higgs branch, so it does not play a part in our discussion.}
\begin{equation}\label{su2 k flav}
    \mathcal{Q}_{D_k}= \raisebox{-0.75\height}{\begin{tikzpicture}[x=1cm,y=.8cm]
    \node (g1) at (0,0) [gauge,label=below:{$SU(2)$}] {};
    \node (g2) at (1,0) [flavor,label=below:{$k$}] {};
    \draw (g1)--(g2);
    \end{tikzpicture}}\ ,
\end{equation}
for $k\geq 4$. The plethystic logarithm of the Hilbert series is given by
\begin{equation}\label{Dk HS}
    PL(HS(\mathcal{C}(D_k))) = t^4 + t^{2k-4} + t^{2k-2} - t^{4k-4}.
\end{equation}
Note that here there is no topological $U(1)^{r}$ symmetry, unlike in the case where the gauge group $G$ is a product of $r$ unitary groups. As a result, in Step $2$ of the \nameref{Alg: MPBA} the only conservation of charge we need to impose under the Poisson bracket is that of the conformal dimension. \\

\noindent We won't go through the details here as they are very similar to those of Section \ref{sec:A klein sing}, but the generators of the Klein $D$ type singularity are
\begin{equation}\label{klein D sing gens}
\begin{split}
    \mathcal{G}_{D_k} &= \{ -\, \frac{z_1^2 \, z_2^2}{(2k-4)^2}\ , \ \frac{1}{2(2k-4)^{k-2}}(z_1^{2k-4} + (-1)^k z_2^{2k-4})\ , \\ 
    & \ \ \ \ \ \ \ \ \ \ \ \  \ \ \ \ \ \ \ \ \ \ \ \  \frac{z_1 \, z_2}{2(2k-4)^{k-1}}(z_1^{2k-4} + (-1)^{k-1} z_2^{2k-4}) \}\\
    &\equiv \{g_1, \ g_2, \ g_3 \},
\end{split}
\end{equation}
and they satisfy the defining equation
\begin{equation}\label{Dk def eq}
    E(g_1,g_2,g_3) = g_1 \, g_2^2 + g_3^2 - g_1^{k-1} = 0,
\end{equation}
with Poisson bracket relations as stated before in (\ref{pbs for klein sings}). \\

\noindent As in the case of the $A_k$ Coulomb quiver, since there is just one gauge group of rank one, a (un)physical bare monopole is labelled by a single integer $v_m$. There is also a single adjoint valued complex scalar $\lambda$. This time the Weyl group $\mathcal{W}(SU(2))=\mathbb{Z}_2$ is non-trivial; it acts on $m$ and $\lambda$ as
\begin{equation}
    \begin{split}
    \mathbb{Z}_2 : \  m &\rightarrow -m,\\
    \lambda &\rightarrow -\lambda,
\end{split}
\end{equation}
meaning that unlike before the basic degrees of freedom $v_m$ and $\lambda$ are not physical. From the Hilbert series (\ref{Dk HS}), we see that the Coulomb branch generators have conformal dimensions $\Delta_1=2$, $\Delta_2=k-2$ and $\Delta_3=k-1$. One can check that the physical Coulomb branch operators with these dimensions are 
\begin{equation}\label{Dk cb gens}
    \lambda^2, \ \ \ V^+_1, \ \ \ \lambda \, V^-_1
\end{equation}
respectively, where
\begin{equation}
    \begin{split}
        &V^+_1 = v_{+1} + v_{-1},\\
        &V^-_1 = v_{+1} - v_{-1}.
    \end{split}
\end{equation}
Note that under action by the Weyl group, $V^+_1$ is invariant and $V^-_1 \rightarrow - V^-_1$.\\

\noindent Step $2$ of the \nameref{Alg: MPBA} tells us that
\begin{equation}
    \begin{gathered}
    \{\lambda^2,V^+_1\} \propto \lambda \, V^-_1,\\[5pt]
    \{\lambda^2,\lambda \, V^-_1\} \propto \lambda \,  V_-,\\[5pt]
    \{V^+_1,\lambda \, V^-_1\} = c_1 \, \lambda^{2k-4} + c_2 \, (V^+_1)^2 + c_3 \, (V^-_1)^2,
    \end{gathered}
\end{equation}
for some constants $c_i$. If we declare that the constants of proportionality are as follows
\begin{equation}\label{Dk cb pbs}
    \begin{gathered}
        \{\lambda^2,V^+_1\} = 2 \lambda \,  V^-_1,\\[5pt]
        \{\lambda^2,\lambda \, V^-_1\} = -2 \lambda^2 \, V^+_1,\\[5pt]
        \{V^+_1, \lambda \, V^-_1\} = (V^+_1)^2 -(k-1)\, \lambda^{2k-4}
\end{gathered}
\end{equation}
and demand that the three generators (\ref{Dk cb gens}) satisfy
\begin{equation}\label{Dk cb def eq}
    (V^-_1)^2 = \lambda^{2k-4} - (V^+_1)^2,
\end{equation}
then can we can see that if we identify 
\begin{equation}
    \begin{gathered}
        g_1 = \lambda^2,\\[5pt]
        g_2 = V^+_1,\\[5pt]
        g_3 = \lambda \, V^-_1,
    \end{gathered}
\end{equation}
then the declared brackets (\ref{Dk cb pbs}) and defining equation (\ref{Dk cb def eq}) between the generators reproduce the brackets and defining equation for the $D_k$ Klein singularity (\ref{pbs for klein sings}) and (\ref{Dk def eq}) respectively.

\section{Nilpotent Orbits}\label{sec: nilp orbs}
Closures of nilpotent orbits of Lie algebras $\mathfrak{g}$ as moduli spaces of quiver gauge theories have been studied intensively due to their nice properties: they are classifiable, and entirely generated by the adjoint representation of $\mathfrak{g}$. Since the explicit construction of these moduli spaces is fully known in the math literature, it makes their Higgs and Coulomb quivers (the respective quivers whose Higgs and Coulomb branch is the nilpotent orbit in question) good candidates on which to explore new techniques or properties of interest. This is precisely our intention here. Since all generators in these moduli spaces are at order $t^2$ in the Hilbert series, and since operators at weight $2$ form a closed algebra under the symplectic form \cite{1992math......4227B}, we know a-priori that the results of the Poisson brackets between generators must be the structure constants of the global symmetry algebra. This has been previously discussed, see for example \cite{gaiotto:2010}.\\

\noindent Actually computing these constants in practise however can be quite tricky. Since there is no $\mathbb{C}^2$ (or equivalent) Poisson bracket to be inherited on these spaces, like there was in the cases of free spaces and Klein singularities (as shown in Sections \ref{sec: free spaces} and \ref{sec: klein sings} respectively), we must turn to Step $3.b)$ in the \nameref{Alg: MPBA} and use the refined Coulomb branch Hilbert series to find the Poisson brackets between $\mathcal{G}_{\mathcal{C}}$, as outlined in Section \ref{sec: explicit pbs from gs charges}. We demonstrate our successful execution of this method in the case of the closure of the minimal nilpotent orbit of $A_2$, denoted $a_2$, below. We have not provided Poisson brackets for closures of other nilpotent orbits in this way because the excess of unconstrained constants quickly becomes too many variables to deal with when the complexity\footnote{Here we consider a quiver more complex if its Coulomb branch has higher dimension (which scales with the sum of squares of the gauge group ranks in a unitary quiver).} of the quiver is increased.\\

\subsection{\texorpdfstring{Minimal $A_2$}{Minimal A2}}\label{sec: min A2}
The Coulomb quiver for $a_2$ is
\begin{equation}\label{a2 quiv}
    \mathcal{Q}_{a_2}= \raisebox{-0.75cm}{\begin{tikzpicture}[x=1cm,y=.8cm]
    \node (g1) at (0,0) [gauge,label=below:{$1$}] {};
    \node (g2) at (1,0) [gauge,label=below:{$1$}] {};
    \node (g3) at (0,1) [flavor,label=left:{$1$}] {};
    \node (g4) at (1,1) [flavor,label=right:{$1$}] {};
    \draw (g3)--(g1)--(g2)--(g4);
    \end{tikzpicture}}.
\end{equation}
To find the Poisson brackets we'll follow Steps $1,2$ and $3.b)$ of the \nameref{Alg: MPBA}.

\paragraph{Step 1} The global symmetry of $\mathcal{C}(\mathcal{Q}_{a_2})$ is $SU(3)$ and all generators lie in the adjoint representation, at $\Delta=1$. The adjoint of $SU(3)$ has dimension $8$, and thus we have $8$ generators:\footnote{As with previous cases where the only gauge groups were Abelian (e.g. Section \ref{sec:A klein sing}), the Coulomb branch degrees of freedom $v$ are automatically gauge invariant, and so the $v$ and $V$ of Section \ref{sec: cb ops} are interchangeable.} 
\begin{equation}\label{a2 gens}
    \mathcal{G}_{\mathcal{C}} = \{ {c^1}_1\, \lambda_1 + {c^2}_1 \, \lambda_2 \ , \ {c^1}_2\, \lambda_1 + {c^2}_2 \, \lambda_2  \ , \ V_{-1-1}\ , \ V_{-10} \ , \ V_{0-1}\ , \ V_{01} \ , \ V_{10} \ , \ V_{11} \},
\end{equation}
where $V_{m_1 m_2}$ are the bare monopole operators with magnetic charge\footnote{For monopole operators in an Abelian theory, the magnetic charge coincides with the topological charge.} $m_1 \in \mathbb{Z}$ under the first $U(1)$ gauge group and $m_2 \in \mathbb{Z}$ under the second; $\lambda_1,\lambda_2 \in \mathbb{C}$ are the adjoint scalars of the first and second gauged $U(1)$ respectively; and ${c^i}_j\in \mathbb{C}$ are constants.\\

\paragraph{Step 2} (\ref{delta for PB}) then tells us that the Poisson brackets between $\mathcal{G}_{\mathcal{C}}$ will give another operator with $\Delta=1$; another element in the adjoint of $SU(3)$. This is what we expect, as the Poisson bracket acts on representations of the global symmetry $SU(3)$ as the Lie bracket of $SU(3)$, under the action of which the adjoint representation is closed. Schematically, 
\begin{equation}\label{pb for delta 1 closedness}
    \{\mathcal{G}_{\mathcal{C}},\mathcal{G}_{\mathcal{C}}\} = \mathcal{G}_{\mathcal{C}}.
\end{equation}
On top of this, the conservation of topological charge (\ref{top charge for PB}) implies that the Poisson bracket between any two $\mathcal{G}_{\mathcal{C}}$, for which the componentwise sum of the vectors of their magnetic charges is not the magnetic charge of any other $\mathcal{G}_{\mathcal{C}}$, must vanish:
\begin{equation}\label{zero pbs a2}
    \begin{gathered}
        \{V_{-1-1},V_{-10}\}=0,\\[5pt]
        \{V_{-1-1},V_{0-1}\}=0,\\[5pt]
        \{V_{11},V_{10}\}=0,\\[5pt]
        \{V_{11},V_{01}\}=0,\\[5pt]
        \{V_{10},V_{0-1}\}=0,\\[5pt]
        \{V_{01},V_{-10}\}=0,\\
    \end{gathered}
\end{equation}
and fully fixes the others up to constants. For example,\footnote{The horrible looking subscript on the constants of proportionality ${\alpha^{(i)}}_{abcdefgh}=-{\alpha^{(i)}}_{efghabcd} \in \mathbb{C}$ (the bracket around the upper index is to indicate that it is not always present) was chosen to reflect the arguments in the Poisson bracket in question: with no upper index, it is the constant of proportionality for $\{ V_{ab} \, \lambda_1^c \, \lambda_2^d \ , \ V_{ef} \, \lambda_1^g \, \lambda_2^h \}$. Upper indices $1$ and $2$ are included in the cases where the result must have topological charge zero, because there are two operators ($\lambda_1$ and $\lambda_2$) which have this charge and they do not necessarily have the same coefficient.}
\begin{equation}\label{eg pbs a2}
\begin{gathered}
    \{V_{-1-1},V_{10}\} =\alpha_{-1-1001000} \, V_{0-1},\\[5pt]
    \{V_{01},V_{0-1}\} = {\alpha^1}_{01000-100} \, \lambda_1 + {\alpha^2}_{01000-100} \, \lambda_2.
\end{gathered}
\end{equation}

\paragraph{Step 3} We then fix the constants of proportionality ${\alpha^{(i)}}_{abcdefgh}$ in the relations found in Step $2$ by demanding consistency with global symmetry charges. To do this, we first find what global symmetry charge each generator $\mathcal{G}_{\mathcal{C}}$ has. We can see that the topological charges $\boldsymbol{J}(\boldsymbol{m})= (m_1,m_2)$ of $\mathcal{G}_{\mathcal{C}}$ (\ref{a2 gens}) respectively are:
\begin{equation}
    \{ (0,0) \ , \ (0,0) \ , \ (-1,-1) \ , \ (-1,0) \ , \ (0,-1) \ , \ (0,1) \ , \ (1,0) \ , \ (1,1) \}.
\end{equation}
Under the appropriate fugacity map,\footnote{A fugacity map is some linear map acting on the topological charges in the monopole formula to map the monomials formed into canonical characters of the global symmetry, making the representation content easier to read. See Appendix $C$ of \cite{Gledhill:2021cbe} for more information.} which one can find to be
\begin{equation}
    \boldsymbol{J}(\boldsymbol{m})
    =
    \begin{pmatrix}
        m_1\\
        m_2\\
    \end{pmatrix}
    \rightarrow
    \boldsymbol{\tilde{J}}(\boldsymbol{\tilde{m}})
    =
    \begin{pmatrix}
        \tilde{m}_1\\
        \tilde{m}_2\\
    \end{pmatrix}
    =
    \begin{pmatrix}
        2 & -1\\
        -1 & 2\\
    \end{pmatrix}
    \cdot 
    \begin{pmatrix}
        m_1\\
        m_2\\
    \end{pmatrix},
\end{equation}
we see that (\ref{a2 gens}) have charges $\boldsymbol{\tilde{J}}$ under the global symmetry  given respectively by:
\begin{equation}\label{a2 gs charges}
    \{ (0,0) \ , \ (0,0) \ , \ (-1,-1) \ , \ (-2,1) \ , \ (1,-2) \ , \  (-1,2) \ , \ (2,-1) \ , \ (1,1) \}.
\end{equation}
Since $\{\cdot,\cdot\}$ acts as the Lie bracket of $SU(3)$, which has rank $2$, there are $2$ Cartan elements $C_1$ and $C_2$ which act as Eigenoperators of the Poisson bracket with Eigenvalue equal to the weight under $SU(3)$ of the adjoint operator it acts on. For $\mathcal{G}_{\mathcal{C}}$, the weights are $\boldsymbol{\tilde{J}}$:
\begin{equation}\label{gs charges condition a2}
    \{C_i,\mathcal{G}_{\mathcal{C}} \} = \boldsymbol{\tilde{J}}_i \, \mathcal{G}_{\mathcal{C}} 
\end{equation}
for $i=1,2$. Thus we must ensure the brackets between each generator and the Cartan elements give the correct global symmetry charges (\ref{a2 gs charges}). The Cartan elements must lie in the adjoint representation of the $SU(3)$ global symmetry and have charge $(0,0)$ under it, and thus they are precisely our two chargeless generators from (\ref{a2 gens}):
\begin{equation}\label{cartans a2}
    \begin{split}
        &C_1 = {c^1}_1 \lambda_1 + {c^2}_1 \lambda_2,\\[5pt]
        &C_2 = {c^1}_2 \lambda_1 + {c^2}_2 \lambda_2.\\
    \end{split}
\end{equation}
We can then demand that (\ref{gs charges condition a2}) holds and substitute in (\ref{cartans a2}) for $C_i$, and implement the results of the Poisson brackets we derived in Step $2$ (for example (\ref{zero pbs a2}) and (\ref{eg pbs a2})). Then using the bilinearity, antisymmetry and Jacobi identity of the Poisson bracket, we can solve for the constants ${\alpha^{(i)}}_{abcdefgh}$ and ${c^i}_j$. This is where the computational difficulty comes in: for higher dimensional Coulomb branches, there are simply too many unknown constants of proportionality $\alpha,c$ introduced to solve for. However in this case we can do it, and find that the Cartans are
\begin{equation}
\begin{gathered}
    C_1 = \lambda_1,\\[5pt]
    C_2 = \lambda_1 + \lambda_2,\\
\end{gathered}
\end{equation}
with Poisson brackets between the generators given by
\begin{equation}\label{pbs a2}
    \begin{array}{c|c|c|c|c|c|c|c|c}
         & C_1 & C_2 & V_{-1-1} & V_{-10} & V_{0-1} & V_{01} & V_{10} & V_{11} \\
         \hline
         C_1 & 0 & 0 & -V_{-1-1} & -2V_{-10} & V_{0-1} & -V_{01} & 2V_{10} & V_{11} \\
         C_2 & \cdot & 0 & -V_{-1-1} & V_{-10} & -2V_{0-1} & 2V_{01} & -V_{10} & V_{11} \\
         V_{-1-1} & \cdot & \cdot & 0 & 0 & 0 & V_{-10} & V_{0-1} & C_1+C_2 \\
         V_{-10} & \cdot & \cdot & \cdot & 0 & V_{-1-1} & 0 & C_1 & V_{01} \\
         V_{0-1} & \cdot & \cdot & \cdot & \cdot & 0 & - C_2 & 0 & V_{10} \\
         V_{01} & \cdot & \cdot & \cdot & \cdot & \cdot & 0 & V_{11} & 0 \\
         V_{10} & \cdot & \cdot & \cdot & \cdot & \cdot & \cdot & 0 & 0 \\
         V_{11} & \cdot & \cdot & \cdot & \cdot & \cdot & \cdot & \cdot & 0 \\
    \end{array}.
\end{equation}
All dots in the table (\ref{pbs a2}) are fixed by the negative of their transpose entries due to antisymmetry of the Poisson bracket.

\section{Poisson Brackets for Generating Representations}\label{sec: pbs for cb as reps}
In Sections \ref{sec: free spaces} -- \ref{sec: nilp orbs} we used the \nameref{Alg: MPBA} of Section \ref{sec: pbs for monops} to explicitly compute Poisson bracket relations for Coulomb branch monopole operators, but noted that this method cannot be performed on most quivers. While we do not by any means rule out the existence of more effective methods to perform computations and find the Poisson brackets explicitly for a generic quiver, we would like to conjecture the results in the cases that are currently too difficult to find using the \nameref{Alg: MPBA} of Section \ref{sec: pbs for monops}. In this section, we illustrate how this is possible for families of quivers for which the representation content of the Coulomb branch is known (from the HWG) to low orders: we forget about the monopole construction of the Coulomb branch, viewing it simply as a space of representations, and then conjecture the Poisson bracket relations between the generating representations using purely the known HWG, the antisymmetric property of the Poisson bracket, and representation theory. Section \ref{sec:eg pbs from reps} will see us detail the results for a small number of families of quivers derived from $5$ and $6$ dimensional physics, which have just one or two generating representations other than the adjoint.\\

\noindent The explicit outline of the approach to conjecture the relevant Poisson brackets is as follows:
\begin{enumerate}
    \item Identify the generating representations by taking the plethystic logarithm of the Hilbert series.
    \item Use the tensor/antisymmetric product of the generating representations to constrain the possible result of the Poisson brackets between generators.
    \item Use the representations appearing at the appropriate degree in the HWG (according to (\ref{delta for PB})) to constrain the possible result of the Poisson brackets between generators.
    \item Find the simplest representation common to the constraints from Step $2$ and Step $3$, and find constants to contract relevant indices.
    \item Insert appropriate flavour symmetry invariants to ensure consistency in conformal dimension.
\end{enumerate}

\noindent The reason for Step $2$ is that the Poisson bracket is antisymmetric, hence when taking the Poisson bracket of a representation with itself, the result will lie in the second rank antisymmetric product of this representation. For the Poisson bracket of two different representations, the result will lie in the tensor product of these two representations (see Appendix \ref{app:tens symm antisymm prods} for a more thorough discussion of tensor and antisymmetric products of representations). Step $3$ is there to ensure that the result of the Poisson bracket between two generators actually lies on the particular Coulomb branch variety we are studying. To clarify any uncertainties in Steps $2$, $3$, $4$ and $5$, we turn to an example.

\paragraph{Example} We will use the same example (\ref{u1 2f}) as in Sections \ref{sec: cb ops} -- \ref{sec: explicit pbs from gs charges} to illustrate the success of this more representation theoretic and less monopole-focused approach. We follow the steps above:
\begin{enumerate}
    \item We have already seen before in (\ref{pl hs u1 2f}) that the generators of this Coulomb branch are simply in the adjoint representation $\mu_1^2$ of $SU(2)$ with conformal dimension $\Delta=1$. The $3$ generators lie in the complexification of the global symmetry algebra $\mathfrak{su(2)_{\mathbb{C}}}=\mathfrak{sl(2;\mathbb{C})}$: $a^i$ for $i=1,2,3$. Since the adjoint of $\mathfrak{sl(2;\mathbb{C})}$ is the second rank symmetric of the fundamental, these three generators can be encapsulated in a symmetric $2\times 2$ ``matrix of generators" $a_{\alpha \beta}$ for $\alpha,\beta=1,2$ symmetric indices labelling which generator is which. The only Poisson bracket we need to determine is then
    \begin{equation}\label{u1 2f rep th pb of interest}
        \{a_{\alpha \beta} \, , \, a_{\gamma \delta} \}.
    \end{equation}
    
    \item Then we can ask what possible representations (\ref{u1 2f rep th pb of interest}) could actually be in. Here it is trivial as we know the adjoint representation is closed under the Lie (and hence Poisson) bracket and so the result must also lie in the adjoint representation. This can also be seen by noting that the second rank antisymmetric product of the adjoint representation is just itself:\footnote{\label{footnote:double mu meanign}Note that we will use $\mu$ or $\mu_i$ both as an index labelling generators that takes on certain specified values, and as the highest weight fugacities to denote a representation. Its meaning in a given situation should be clear from context (i.e. whether it lies in an exponent/subscript or not).}
    \begin{equation}\label{AS adj su2}
        \Lambda^2(\mu_1^2) = \mu_1^2,
    \end{equation}
    and since the Poisson structure antisymmetrises comparable arguments, the result of (\ref{u1 2f rep th pb of interest}) must lie in (\ref{AS adj su2}).
    
    \item If (\ref{AS adj su2}) had contained multiple representations, we could have further constrained the representations that (\ref{u1 2f rep th pb of interest}) could lie in by examining the HWG. Using (\ref{delta for PB}), the result must have conformal dimension $1+1-1=1$, i.e. lie at $t^2$ in the Hilbert series. For $\mathcal{C}(\mathcal{Q}_{A_1})$, the HWG is
    \begin{equation}
        HWG(\mathcal{C}(\mathcal{Q}_{A_1})) = PE[\mu_1^2 t^2].
    \end{equation}
    We can therefore see easily that the only representation at $t^2$ on this Coulomb branch is indeed the adjoint, and so this enforces that the Poisson bracket of the adjoint generators lies again in the adjoint:
    \begin{equation}\label{rough pb u1 2f gp th}
        \{a \, , \, a \} \propto a.
    \end{equation}
    We see that in this case either the available Coulomb branch operators or possible representations would be sufficient to deduce (\ref{rough pb u1 2f gp th}), but in more complicated cases they will work in tandem to help postulate the brackets.
    
    \item Finally, we aim to get rid of the proportionality sign in (\ref{rough pb u1 2f gp th}) and make an explicit conjecture. To do this, we note that on the left-hand side of (\ref{rough pb u1 2f gp th}) there will be indices labelling the generators (say for instance $\alpha,\beta,\gamma,\delta$ as in (\ref{u1 2f rep th pb of interest})) which must be matched on the right hand side. Two of these indices will be held by the ``$a$" on the right-hand side, which leaves two remaining indices for the equation to be consistent. In the algebra of $\mathfrak{sl(2;\mathbb{C})}$, the only thing that makes sense to go here is the epsilon invariant (the delta invariant of $\mathfrak{sl(n;\mathbb{C})}$ can be constructed from the epsilon invariant in the $n=2$ case). So we expect (\ref{rough pb u1 2f gp th}) to take form along the lines of $\{ a_{\alpha \beta} \, , \, a_{\gamma \delta} \} \sim A\,( \epsilon_{\alpha \delta} a_{\gamma \beta} +\epsilon_{\beta \delta} a_{\gamma \alpha} + \epsilon_{\alpha \gamma} a_{\delta \beta} + \epsilon_{\beta \gamma} a_{\delta \alpha})$; the relative signs chosen to ensure the result is antisymmetric upon simultaneously exchanging $\alpha \leftrightarrow \gamma$ and $\beta \leftrightarrow \delta$, and symmetric upon exchanging $\alpha \leftrightarrow \beta$ or $\gamma \leftrightarrow \delta$. The overall constant $A$ is fixed by the way in which we identify the $a_{\alpha \beta}$ with the symmetric generators of $\mathfrak{sl(2;\mathbb{C})}$. For example, we could use the representation of the generators of $\mathfrak{sl(2;\mathbb{C})}$ given in (\ref{sl2c gens}). Then if we identify $a_{12} = \lambda$, $a_{11} =2 \, V_{-1}$ and $a_{22} = 2 \, V_{+1}$, a positive overall sign:
    \begin{equation}\label{su2 struc consts}
    \{ a_{\alpha \beta} \, , \, a_{\gamma \delta} \} = \epsilon_{\alpha \delta} a_{\gamma \beta} + \epsilon_{\beta \delta} a_{\gamma \alpha} + \epsilon_{\alpha \gamma} a_{\delta \beta} + \epsilon_{\beta \gamma} a_{\delta \alpha}
    \end{equation}
    for $\alpha, \beta, \gamma, \delta = 1,2$ would recover the appropriate Poisson brackets (\ref{full pb u1 2f}) of (\ref{sl2c gens}).
    
    \item Our result (\ref{su2 struc consts}) is already consistent with conformal dimension: the left hand side has $\Delta=1+1-1=1$, which matches the $\Delta=1$ of the right hand side.
\end{enumerate}

\noindent Although the example above did not illustrate it, non-trivial ammendments can be made in Step $5$, causing the need for a scale to be introduced in the form of some invariant of the global symmetry. In such cases, there is an operator which spontaneously breaks the conformal and $R$-symmetry but preserves the flavour symmetry. We will see this explicitly in the examples of Section \ref{sec:eg pbs from reps}, where we will discuss its physical significance in more detail. \\

\paragraph{Structure constants for $\mathbf{\mathfrak{sl(n>2;\mathbb{C})}}$} Note that, as mentioned in the discussion preceeding (\ref{e matrix mult}), in $\mathfrak{sl(2;\mathbb{C})}$ the fundamental representation is pseudo-real, not complex, and all non-trivial representations are symmetric products of this. Hence all indices are lowered and we can write the structure constants in the special form of (\ref{su2 struc consts}). For $\mathfrak{sl(n>2;\mathbb{C})}$ the fundamental representation is complex, so the structure constants here must be written in a more general form; we cannot generalise (\ref{su2 struc consts}) to all $\mathfrak{sl(n>2;\mathbb{C})}$. The adjoint of $\mathfrak{sl(n>2;\mathbb{C})}$ lies in the tensor product of the fundamental and antifundamental (the complex conjugate and dual of the fundamental) representations, hence the matrix of generators is given by one upper and one lower index ${a^\mu}_{\nu}$ for $\mu,\nu = 1,...,n$ traceless labelling which generator is which. In this case, taking into account the upper and lower indices that appear on the left and right side of (\ref{rough pb u1 2f gp th}), we have two spare indices on the right hand side that do not accompany ``$a$": one upper and one lower. The only invariant of $\mathfrak{sl(n>2;\mathbb{C})}$ with this structure is ${\delta^{\mu}}_{\nu}$, and so we find that the structure constants of $\mathfrak{sl(n;\mathbb{C})}$ are
    \begin{equation}\label{sun struc consts}
        \{ {a^\mu}_\nu \, , \, {a^\rho}_\sigma \} = \,{\delta^\mu}_\sigma {a^\rho}_\nu - {\delta^\rho}_\nu {a^\mu}_\sigma,
    \end{equation}
where analogously to Step $5$ above, the overall sign was determined by ensuring that the result matched the expected result for the $n=2$ case, if we took the generators to be $\mathfrak{sl(n=2;\mathbb{C})}$ matrices. Because here we are viewing $\mathfrak{sl(n=2;\mathbb{C})}$ as the $n=2$ case of $\mathfrak{sl(n;\mathbb{C})}$, we take the generators to be the canonical generators for $2\times 2$ traceless complex matrices (unlike in Step $1$ above, or in (\ref{sl2c gens}), where it was more natural to view them as symmetric complex matrices):
\begin{equation}\label{gens of sl2 trless}
    \tilde{\lambda} =
        \begin{pmatrix}
            1 & 0\\
            0 & -1
        \end{pmatrix}, \ \ \ 
        \tilde{V}_{+1}=
        \begin{pmatrix}
            0 & 1\\
            0 & 0
        \end{pmatrix}, \ \ \
        \tilde{V}_{-1}=
        \begin{pmatrix}
            0 & 0\\
            1 & 0
        \end{pmatrix}.
\end{equation}
The Poisson bracket acts on two such generators by matrix multiplication, using contraction with $\delta$ (rather than $\epsilon$ as we did when we were treating $\mathfrak{sl(2;\mathbb{C})}$ separately from $\mathfrak{sl(n;\mathbb{C})}$, for example in the discussion preceeding (\ref{e matrix mult})). We see that this gives
\begin{equation}\label{sl2c trless pbs}
    \begin{gathered}
        \{\tilde{\lambda},\tilde{V}_{+1}\} = 2 \, \tilde{V}_{+1},\\[5pt]
        \{\tilde{\lambda},\tilde{V}_{-1}\} = -2 \, \tilde{V}_{-1},\\[5pt]
        \{\tilde{V}_{+1},\tilde{V}_{-1}\} = \tilde{\lambda}.\\
    \end{gathered}
\end{equation}
If we identify our ${a^{\mu}}_{\nu}$ generators as
\begin{equation}
    \begin{gathered}
        {a^1}_1-{a^2}_2 = \tilde{\lambda},\\[5pt]
        {a^1}_2 = \tilde{V}_{+1},\\[5pt]
        {a^2}_1 = \tilde{V}_{-1},
    \end{gathered}
\end{equation}
for the $\tilde{\lambda}$, $\tilde{V}_{+1}$, $\tilde{V}_{-1}$ of (\ref{gens of sl2 trless}), then we see that (\ref{sun struc consts}) with its positive overall sign matches (\ref{sl2c trless pbs}). With this identification, we can see that (\ref{sun struc consts}) are the structure constants for $\mathfrak{sl(n>2;\mathbb{C})}$.\\

\noindent We can see that the approach to determine Poisson brackets outlined in this section is slightly more abstract than that detailed in the previous three sections, and in the case of this example is actually somewhat redundant; since the moduli space is solely generated by the adjoint representation at $\Delta=1$ -- a closed algebra -- we could have said without any calculations or thought that the Poisson brackets had to be given by the structure constants. The method outlined in Sections \ref{sec: cb ops} -- \ref{sec: explicit pbs from gs charges} uncovered additional information that we did not know a-priori: in particular, the Poisson bracket relations for the specific monopole operators on the Coulomb branch. That explicit construction is clearly preferable, but is not accessible for many of the quivers for which the approach in this section can be used, hence the utility of this more abstract method. We will now move on to illustrate examples of such cases.\\

\section{\texorpdfstring{Higgs Branches of Certain $5d$ and $6d$ Theories at Infinite Coupling}{Higgs Branches of Certain 5d and 6d Theories at Infinite Coupling}}\label{sec:eg pbs from reps}

In this section we find the Poisson brackets between the Higgs branch generators $\{\mathcal{G}_{\mathcal{H}},\mathcal{G}_{\mathcal{H}}\}$ of certain $5$ and $6$ dimensional theories at infinite coupling by computing $\{\mathcal{G}_{\mathcal{C}},\mathcal{G}_{\mathcal{C}}\}$ for their magnetic quivers \cite{Cabrera:2019izd}.\footnote{Recall that in general, a $3d$ $\mathcal{N}=4$ quiver $Q^\prime$ is called the magnetic quiver to an ``electric" quiver $Q$ if $\mathcal{C}(Q^\prime)=\mathcal{H}(Q)$. In the case of a $5d$ electric theory which is at infinite coupling, since the gauge coupling is a parameter of the theory and not inherently specified by the electric quiver $Q$, we write $\mathcal{C}(Q^\prime)=\mathcal{H}_{\infty}(Q)$. Note the distinction from the terminology of ``Coulomb quiver" used in Sections \ref{sec: pbs for monops} -- \ref{sec: nilp orbs}; this was used in relation to a variety $\mathcal{V}$, as opposed to in relation to the Higgs branch $\mathcal{H}(\mathcal{Q})$ of an electric quiver $Q$.}\\

\noindent This task sees the utility of the method outlined in Section \ref{sec: pbs for cb as reps}; the families we study have one or two generating representations other than the adjoint, hence the Poisson brackets cannot be trivially concluded to be the structure constants a-priori (as they could have been in the example of Section \ref{sec: pbs for cb as reps}). In some cases, Step $4$ of Section \ref{sec: pbs for cb as reps} finds many representations, the set of which we call $R_{\text{mult}}$, that are common to both the tensor/antisymmetric product of the input representations for a given Poisson bracket and the operators on the Coulomb branch at the appropriate conformal dimension. With these constraints alone, we can only say for certain that the Poisson bracket in question would be some undetermined linear combination of $R_{\text{mult}}$, but we \textit{conjecture} that the principle of Occam's razor applies: the coefficients of all representations \textit{other than the simplest} vanish. We call this simplest representation $R_{\text{simp}}\in R_{\text{mult}}$, and in all cases studied it turns out to be either the trivial or the adjoint representation of the global symmetry. The motivation for such a conjecture is that the moduli spaces we study are fairly simple spaces; there is no reason to expect that the symplectic form should take unecessarily elaborate values. To check this conjecture and verify the exact coefficients would require further analysis of the $5$ and $6$ dimensional physics, a task we leave for future work.\\

\paragraph{The breaking of conformality}
In all the $5$ and $6$ dimensional theories studied, we find that the Higgs branch Poisson brackets between representations other than the adjoint include a Casimir to some non-zero power to ensure consistency of conformal dimension. This Casimir takes some numerical value: it is a scale. This tells us that on the Higgs branch of these theories, the conformal symmetry at infinite coupling is spontaneously broken. This sounds like nothing new; any VEV will break the conformal symmetry in the vacuum. However the difference between this scale and other VEVs is that other VEVs also break the Coulomb branch flavour symmetry, whereas this Casimir scale preserves it. This is a powerful statement and is worth unpacking a little, but we should be careful to distinguish between $5$ and $6$ dimensions. In $5d$, the gauge coupling is a parameter of the theory, and so the Poisson brackets we find below in Section \ref{sec: zhenghao quivs} tell us that while conformal symmetry is spontaneously broken in SQCD theories taken at infinite coupling, the flavour symmetry is preserved. In $6d$, the inverse gauge coupling $\frac{1}{g^2}$ is a modulus, not a parameter. If we look at the Higgs branch at finite coupling (i.e. the Higgs branch over a generic point of the tensor branch), $\frac{1}{g^2}$ is an obvious scale and so conformal symmetry is broken. If we look at the Higgs branch at infinite coupling (i.e. the Higgs branch over the origin of the tensor branch), the scale $\frac{1}{g^2}$ disappears, but the Poisson brackets we find in Sections \ref{sec: e8 fam} tell us that another one emerges, $C_2$. That is, as we move towards the origin of the tensor branch the scale which breaks conformal symmetry transitions from $\frac{1}{g^2}$, a scalar in the tensormultiplet, to $C_2$, a scalar in the hypermultiplet which preserves the flavour symmetry. It would be interesting to see how observables (e.g. correlation functions) in the $6d$ theory lying close to the origin of the tensor branch (i.e. when the gauge coupling is very large but not infinite) vary as a function of these two scales $\frac{1}{g^2}$ and $C_2$.\\

\noindent For the first family of theories we consider (the subject of Section \ref{sec: e8 fam}), we go through in detail the method used to obtain the Coulomb branch Poisson brackets $\{\mathcal{G}_{\mathcal{C}},\mathcal{G}_{\mathcal{C}}\} $ of the magnetic quiver. One Poisson bracket on this Coulomb branch -- the bracket of the generating spinor representation with itself -- fits into the categories of brackets discussed above for which the methods of Section \ref{sec: pbs for cb as reps} do not constrain the result to lie in a single representation. We detail the full list of representations it could lie in ($R_{\text{mult}}$), and then end with the conjecture that it actually lies in the simplest ($R_{\text{simp}}$). For the remaining families (the subjects of Sections \ref{sec: e7 fam} - \ref{sec: truck fam}) we simply state the conjectured results (and in particular give only $R_{\text{simp}}$ and not the full list $R_{\text{mult}}$ for any brackets which the methods of Section \ref{sec: pbs for cb as reps} fail to constrain to a single representation), skipping the method. The quivers in Sections \ref{sec: e8 fam} -- \ref{sec: e6 fam 2} are taken from \cite{2018JHEP...07..061F}, and those in Section \ref{sec: zhenghao quivs} from \cite{Cabrera:2018jxt} and an unpublished work by Hanany and Zhong.

\subsection{\texorpdfstring{$E_{8,n}$ family}{E8n family}} \label{sec: e8 fam}
The first family we discuss is that arising from the Higgs branch of the $6d$ theory of $Sp(n)$ gauge group and $2n+8$ flavours \cite{2018JHEP...07..098H}, or equivalently the Higgs branch of the $5d$ theory of $SU(n+2)$ gauge group with Chern Simons level $k=\pm\frac{1}{2}$ and $2n+7$ flavours \cite{2018JHEP...07..061F}, for $n \geq 0$. The magnetic quiver in question for these theories at infinite coupling is \cite{Hanany:2017pdx, 2018JHEP...07..098H, 2018JHEP...07..061F}
\begin{equation}\label{pini e8}
  \begin{tikzpicture}[x=1.2cm,y=.8cm]
    \node (g1) at (0,0) [gauge,label=below:{$1$}] {};
    \node (g2) at (1,0) [gauge,label=below:{$2$}] {};
    \node (g3) at (2,0) {$\cdots$};
    \node (g4) at (3,0) [gauge,label=below:{$2n+6$}] {};
    \node (g7) at (4,0) [gauge,label=below:{$n+4$}] {};
    \node (g8) at (5,0) [gauger,label=below:{$2$}] {};
    \node (g9) at (3,1) [gauge,label=right:{$n+3$}] {};
    \draw (g1)--(g2)--(g3)--(g4)--(g7)--(g8);
    \draw (g4)--(g9);
    \end{tikzpicture}
\end{equation}
The right hand rank $2$ node is coloured red to indicate that it is unbalanced (for $n \neq 0$). Generically, we can read off that this quiver has global symmetry $SO(4n+16)$ \cite{Gledhill:2021cbe}. For $n=-1$ however, (\ref{pini e8}) is the Dynkin quiver of $E_7$ ($\mathsf{D}_{E_7}$, as discussed in Section \ref{sec: Dynkin quivs}) hence the global symmetry is enhanced to $Sp(16) \supset SO(12)$ and the Poisson brackets are those given in Section \ref{sec: free spaces}. For $n=0$, (\ref{pini e8}) is the affine $E_8$ quiver and hence the global symmetry is enhanced to $E_8 \supset SO(16)$, and the Poisson brackets are the structure constants of $E_8$. For all $n \geq 1$ the global symmetry of the Coulomb branch is $SO(4n+16)$ and there are $2$ generating representations, as can be found from the HWG:
\begin{equation}\label{hwg d2kplus8}
    HWG= \frac{1}{(1-t^4)(1-\mu_{2n+8}\, t^{n+2})(1-\mu_{2n+8}\, t^{n+4})\prod_{i=1}^{n+3}(1- \mu_{2i} \, t^{2i})},
\end{equation}
where $\mu_1,...,\mu_{2n+8}$ are highest weight fugacities of the global symmetry $D_{2n+8}=SO(4n+16)$. The Poisson brackets in the $n \geq 1$ case are thus not so trivial; we must proceed with Steps $1-4$ of Section $\ref{sec: pbs for cb as reps}$ to determine them. To illustrate these steps we start by carrying them out for the simplest case of $n=0$, viewing the representations of the global symmetry $E_8$ in terms of the $SO(16)$ subalgebra, and then use this to help us find the Poisson brackets for general $n\geq 1$.

\subsubsection{\texorpdfstring{$n=0$}{n=0}}\label{sec: pini e8 n=1 case}
In this case, (\ref{pini e8}) morphs into the affine Dynkin diagram of $E_8$ and hence its Coulomb branch is the corresponding minimal nilpotent orbit closure, $e_8$. Since here the global symmetry enhances to $E_8$, we could write the HWG \cite{Hanany:2014dia} in terms of representations of $E_8$ -- as $PE[m_7 \, t^2]$ for $m_1,...,m_8$ highest weight fugacities of $E_8$ -- however it will be more useful to write it as the $n=0$ case of (\ref{hwg d2kplus8}) -- i.e. in terms of $SO(16)\subset E_8$ representations -- so that we can use our results to generalise to higher $n$:
\begin{equation}\label{hwg e8}
    HWG = PE[(\mu_2 + \mu_8) t^2 + (1+\mu_4 + \mu_8)t^4 + \mu_6 t^6].
\end{equation}
Recall from Section \ref{sec: pbs for cb as reps} that the first step to finding $\{\mathcal{G}_{\mathcal{C}},\mathcal{G}_{\mathcal{C}}\}$ is to write down the generating representations $\mathcal{G}_{\mathcal{C}}$ themselves. This can be done by converting (\ref{hwg e8}) into a Hilbert series and taking the plethystic logarithm \cite{Feng:2007ur}. It turns out that all generators here lie at order $t^2$ (i.e. have $\Delta=1$), and fall into the adjoint $a_{\mu \nu}$\footnote{In the $16d$ analogy to the Lorentz transformations, $\mu, \nu = 1,...,16$ are antisymmetrised.} ($\mu_2$ in Dynkin labels) and spinor $s_\alpha$\footnote{Each spinor $s_\alpha$ is a $128$ dimensional vector, and $\alpha=1,...,2^{\frac{16}{2}-1}=128$ is a spinor index.} ($\mu_8$ in Dynkin labels) representations of the $SO(16) \subset E_8$ global symmetry. We collect this information on the generators in the following table:
\begin{equation}\label{gens of e8}
    \begin{array}{|c|c|c|}
        \hline
        \text{Generator} & \Delta & SO(16) \text{ representation}\\
        \hline 
        a_{\mu\nu} & 1 & \text{adjoint}\\
         \hline
        s_{\alpha} & 1 & \text{spinor}\\
    \hline
    \end{array}.
\end{equation}
Note that both of these representations are real; we don't need to worry about raising or lowering indices. This is Step $1$ of the method in Section \ref{sec: pbs for cb as reps} accomplished. We then carry out Steps $2 - 4$ for each of the Poisson brackets we need to calculate:
\begin{equation}\label{generic pbs for gens of e8}
    \{a_{\mu \nu},a_{\rho \sigma}\} \, , \ \ \  \{ a_{\mu \nu}, s_\alpha\} \, , \ \ \ \{ s_\alpha, s_\beta \} \,.
\end{equation}

\paragraph{Adjoint with adjoint, $\mathbf{\{a,a\}}$}
\begin{enumerate}
\setcounter{enumi}{1}

    \item $\{a,a\}$ generates a representation in the second rank antisymmetric product of the adjoint. One can calculate this to be
        \begin{equation}
            \Lambda^2(\mu_2) = \mu_2 + \mu_1 \mu_3,
        \end{equation}
    and thus conclude that the Poisson bracket of two $a$'s must lie in one of the following representations of $SO(16)$:
        \begin{equation}\label{poss reps gp theory e8}
            \mu_2\, , \  \mu_1 \mu_3 \, .
        \end{equation}

    \item Using (\ref{delta for PB}) and (\ref{gens of e8}), we see that $\{a,a\}$ must have conformal dimension $\Delta=1$, and therefore lie in a representation in the Hilbert series appearing at $t^2$. This constrains $\{a,a\}$ to lie in one of the following representations:
        \begin{equation}\label{poss reps hs e8}
            \mu_2\, , \ \mu_8 \,,
        \end{equation}
    
    \item The only overlap between (\ref{poss reps gp theory e8}) and (\ref{poss reps hs e8}) is $\mu_2$. That is, under the Poisson bracket the adjoint representation is closed. Schematically, this means
        \begin{equation}
            \{a,a\} \sim a.
        \end{equation}
    To make this rigorous, we need to put the indices in and contract appropriately. Suppose on the left hand side we choose indices as follows:
    \begin{equation}
        \{a_{\mu \nu}, a_{\rho \sigma} \},
    \end{equation}
    On the right hand side all these indices must remain, and since the Poisson bracket is just the Lie bracket on the $SO(16)$ algebra, the result is determined by the structure constants:
        \begin{equation}\label{lorentz struc consts}
            \{a_{\mu \nu},a_{\rho \sigma}\} = \delta_{\nu \rho} a_{\mu \sigma} - \delta_{\mu \rho} a_{\nu \sigma} + \delta_{\mu \sigma} a_{\nu \rho} - \delta_{\nu \sigma} a_{\mu \rho}.
        \end{equation}
        
    \item Our result (\ref{lorentz struc consts}) is already consistent with conformal dimension: the left hand side has $\Delta=1+1-1=1$, which matches the $\Delta=1$ of the right hand side.
\end{enumerate}

\noindent Note the use of $\delta_{\mu \nu}$ as apposed to $\eta_{\mu \nu}$: we are not using representations of Minkowski $SO(15,1)$ spacetime, but rather representations of $SO(16)$ whose indices are Euclidean in nature. (\ref{lorentz struc consts}) are the structure constants for all $SO(2k)$, with all indices ranging from $1,...,2k$. Note how the result is antisymmetric under the three index permutations $(a) \ \mu \leftrightarrow \nu$, $ (b) \ \rho \leftrightarrow \sigma$, and $(c) \ \mu \leftrightarrow \rho, \ \nu \leftrightarrow \sigma$, as expected.\\

\paragraph{Adjoint with spinor, $\{a,s\}$}
\begin{enumerate}
\setcounter{enumi}{1}

    \item The tensor product of $\mu_2$ with $\mu_8$ restricts this Poisson bracket to lie in the following representations:
        \begin{equation}\label{poss reps a with s gp theory e8}
            \mu_8\, , \ \mu_1 \mu_7 \, , \  \mu_2 \mu_8 \, .
        \end{equation}
        
    \item Since $\Delta=1$ for the spinor representation also, again $\{a,s\}$ must lie at $t^2$ in the Hilbert series, and so must lie among (\ref{poss reps hs e8}).
    
    \item This time the only overlap between (\ref{poss reps a with s gp theory e8}) and (\ref{poss reps hs e8}) is $\mu_8$, and so schematically we have
        \begin{equation}
            \{a,s\} \sim s
        \end{equation}
    Again we could have known this a-priori as any representation is an ``Eigenrepresentation" of the adjoint under action by the Lie bracket, with Eigenvalue equal to its weight. To put in the indices, we need something on the right hand side to contract the two vector indices of $a$, leaving just the spinor index of $s$, which transforms compatibly under $SO(16)$. A natural candidate is the generator of the spinor representation: $\gamma_{\mu \nu} = \frac{1}{4} [\gamma_\mu,\gamma_\nu]$,\footnote{This factor of $\frac{1}{4}$ is typical in the literature, to ensure the $\gamma_{\mu \nu}$ satisfy the structure constants of the $SO$ algebra (\ref{lorentz struc consts}). Here it is also convenient to use as it results in no factors of $\frac{1}{4}$ in the results of the Poisson brackets (\ref{summary e8 fam brackets}).} where $\gamma_\mu$ are the Euclidean gamma matrices, satisfying the Clifford algebra $\{\gamma_\mu , \gamma_\nu \} = 2\delta_{\mu \nu}$.\footnote{Note that confusingly, unlike in the rest of the paper, here in the Clifford algebra $\{ \cdot , \cdot \}$ denotes the anticommutator.} Concretely,\footnote{Note that the indices $\mu$ and $\nu$ are labelling which gamma matrix we are referring to of the possible $120=\text{dim}(SO(16))$, and the spinor indices $\alpha$ and $\beta$ label the matrix components of $\gamma_{\mu \nu}$ (spinor indices remind us that $s_\alpha$ transforms as a spinor, not a vector, and that the $(\gamma_{\mu\nu})_{\alpha\beta}$ act on spinors and satisfy the Clifford algebra).}
        \begin{equation}\label{a with s pb for e8}
            \{a_{\mu \nu}, s_{\alpha}\} =  (\gamma_{\mu \nu})_{\alpha \beta} \,  s_\beta.
        \end{equation}
    \item  Our result (\ref{a with s pb for e8}) is already consistent with conformal dimension: the left hand side has $\Delta=1+1-1=1$, which matches the $\Delta=1$ of the right hand side.
\end{enumerate}

\paragraph{Spinor with spinor, $\{s,s\}$}

\begin{enumerate}
\setcounter{enumi}{1}

    \item The Poisson bracket lying in the second rank antisymmetric of the spinor representation $s$ constrains $\{s,s\}$ to lie among
        \begin{equation}\label{poss reps s with s gp theory e8}
            \mu_6\, , \ \mu_2 \, .
        \end{equation}
    
    \item Again, the conformal dimension of $\{s,s\}$ must be $1$, meaning that again it must lie among the representations of (\ref{poss reps hs e8}).
    
    \item The only representation common to (\ref{poss reps s with s gp theory e8}) and (\ref{poss reps hs e8}) is $\mu_2$, so schematically
        \begin{equation}
            \{s,s\} \sim a.
        \end{equation}
    On the left hand side there are two spinor indices, on the right hand side there are two antisymmetrised vector indices. A constant with matching spinor indices must contract these vector indices. Again $\gamma_{\mu \nu}$ is the natural candidate:
        \begin{equation}\label{s with s pb for e8}
            \{ s_\alpha, s_\beta \} = (\gamma_{\mu\nu})_{\alpha \beta} \, a_{\mu \nu}.
        \end{equation}
    
    \item As in the previous two cases, our result (\ref{s with s pb for e8}) is already consistent with conformal dimension.
\end{enumerate}

\subsubsection{\texorpdfstring{$n\geq 0$}{n>=0}}

Now that we have seen the explicit working for the $n=0$ case, it is easy to follow the same logic and arrive at a postulate of the Poisson brackets for the generating operators of the $\mathcal{C}(\ref{pini e8})$ for general $n \geq 0$.\\

\noindent For generic $n$, the generators are a slightly non-trivial generalisation of (\ref{gens of e8}), found in the same way as described in Section \ref{sec: pini e8 n=1 case}:

\begin{equation}\label{gens of e8 fam}
    \begin{array}{|c|c|c|}
        \hline
        \text{Generator} & \Delta & SO(4n+16) \text{ representation}\\
        
        \hline 
        
        a_{\mu\nu} & 1 & \text{adjoint}\\
        
        \hline
         
        s_{\alpha} 
        & 
        \frac{n+2}{2}
        & \text{spinor}\\
        
    \hline
    
    \end{array}.
\end{equation}

\noindent Note that both of these representations are real (or pseudo-real in the case of the spinor representation for odd $n$); we don't need to worry about raising or lowering indices. The form of the $\{a,a\}$ and $\{a,s\}$ Poisson brackets do not change from those discussed in the $n=0$ case of Section \ref{sec: pini e8 n=1 case}: (\ref{lorentz struc consts}) and (\ref{a with s pb for e8}) hold for all $n \geq 0$, with $\mu,\nu$ and $\rho,\sigma$ pairs of antisymmetrised vector indices going from $1,...,4n+16$ and $\alpha,\beta$ spinor indices going from $1,...,2^{\frac{4n+16}{2}-1}$. However, the Poisson bracket between two spinors depends on whether $n$ is odd or even.

\paragraph{\texorpdfstring{Even $n$}{Even n}}
\begin{enumerate}
    \item[2.] In the case of even $n$, the second rank antisymmetric of the spinor is (see Appendix \ref{app:tens symm antisymm prods} for more details)\footnote{Note that this shows that the spinor representation of $SO(4n+16)$ is real for even $n$, as since the singlet doesn't lie in $\Lambda^2(s)$, it must lie in the second rank symmetric product.}
        \begin{equation}
            \Lambda^2(\mu_{2n+8}) = \mu_{2n+6} + \mu_{2n+2} + \cdots + \mu_2,
        \end{equation}
    and so $\{s,s\}$ must lie in some combination of the following representations of $SO(4n+16)$:
        \begin{equation}\label{poss reps s with s gp theory odd k d2kplus8}
            \mu_{2n+6}\, , \ \mu_{2n+2} \, , \  \dots \ , \ \mu_{2}.
        \end{equation}

    \item[3--4.] The spinor appears at $t^{n+2}$ in the Hilbert series, i.e. has $\Delta=\frac{n}{2}+1$, hence (\ref{delta for PB}) tells us that $\{s,s\}$ must have $\Delta=n+1$. The representations with this conformal dimension on the Coulomb branch in question are those which appear at order $t^{2n+2}$ in the HWG. The only overlap between such representations and (\ref{poss reps s with s gp theory odd k d2kplus8}) are 
        \begin{equation}
            R_{mult} = \{ \mu_{2n+2} \, , \ \mu_{2n-2} \, , \,  \dots \, , \ \mu_{2} \},
        \end{equation}
    and so the Poisson brackets must take the form
        \begin{equation}\label{s s d2kplus8 n even pre conj}
            \{s_\alpha,s_\beta\} \sim (\gamma_{\mu \nu})_{\alpha \beta} \, a_{\mu \nu} + \sum_{i=1}^{\frac{n}{2}} \, (\gamma_{\mu_1 \cdots \mu_{4i+2}})_{\alpha \beta} \, b_{\mu_1 \cdots \mu_{4i+2}},
        \end{equation}
    where the indices on $b$ are completely antisymmetrised to reflect the $(4i-2)^{th}$ rank antisymmetric of the vector representation of $SO(4n+16)$, and each term has some coefficient preceeding it. As stated at the start of Section \ref{sec:eg pbs from reps}, to explicitly determine these coefficients would require analysis of the physics in the $6d$ theory from which these states originate. We leave this as a challenge for future work, but we conjecture that the solution should be the simplest: that the coefficients of all terms other than the first of (\ref{s s d2kplus8 n even pre conj}) are zero (i.e. $R_{\text{simp}}=\text{adjoint}$)
    \begin{equation}\label{s s d2kplus8 n even post conj}
        \{s_\alpha,s_\beta\} \sim (\gamma_{\mu \nu})_{\alpha \beta} \, a_{\mu \nu}.
    \end{equation}
    
    \item[5.] As it stands in (\ref{s s d2kplus8 n even post conj}), the conformal dimension of the right hand side does not match what it should do: the left hand side has $\Delta=\frac{n+2}{2}+\frac{n+2}{2}-1=n+1$, but the right hand side only has $\Delta=1$. The right hand side \textit{should} be in the adjoint representation of $SO(4n+16)$ with $\Delta = n+1$: an additional factor which is an $SO(4n+16)$ invariant with $\Delta=n$ must be included. \\
    
    \noindent The HWG tells us that, on this Coulomb branch, the only generating singlet of $SO(4n+16)$ lies at $t^4$ (or equivalently $\Delta=2$). The only invariant with this dimension is the second Casimir:\footnote{Here, as above, $a$ schematically represents any matrix in the adjoint of the $SO(4n+16)$ algebra in question.} 
    \begin{equation}
        C_2=Tr(a^2),
    \end{equation}
    and hence we see that all other $SO(4n+16)$ invariants are powers of $C_2$.\footnote{This is another indicator, along with the Hasse diagram, that this Coulomb branch is ``very close" to being a nilpotent orbit.} Consequently, the only candidate to include in (\ref{s s d2kplus8 n even post conj}) to rectify the current inconsistency in conformal dimension is the $\left(\frac{n}{2}\right)^{\text{th}}$ power of $C_2$:
    \begin{equation}\label{s s d2kplus8 n even post conj post dim check}
        \{s_\alpha,s_\beta\} = {C_2}^{\frac{n}{2}} \, (\gamma_{\mu \nu})_{\alpha \beta} \, a_{\mu \nu}.
    \end{equation}
    
\end{enumerate}

\paragraph{\texorpdfstring{Odd $n$}{Odd n}}
\begin{enumerate}
    \item[2.] For odd $n$, the second rank antisymmetric of the spinor is\footnote{Note that this shows that the spinor representation of $SO(4n+16)$ is pseudo-real for odd $n$, as the singlet lies in the second rank antisymmetric product.}
        \begin{equation}
            \Lambda^2(\mu_{2n+8}) = \mu_{2n+6} + \mu_{2n+2} + \cdots + \mu_4 + 1,
        \end{equation}
    hence $\{s,s\}$ must transform in some combination of the following representations of $SO(4n+16)$:
        \begin{equation}\label{poss reps s with s gp theory even k d2kplus8}
            \mu_{2n+6}\, , \ \mu_{2n+2} \, , \  \dots \ , \ 1 \, .
        \end{equation}

    \item[3--4.] As in the even case, the result lies at $t^{2n+2}$ in the Hilbert series. The representations under the $SO(4n+16)$ global symmetry at this order which overlap with (\ref{poss reps s with s gp theory even k d2kplus8}) are
        \begin{equation}
            R_{\text{mult}}= \{\mu_{2n+2} \, , \  \mu_{2n-2} \, , \ \dots \ , \ 1 \, \},
        \end{equation}
    and so we find that the Poisson bracket must take the form
        \begin{equation}\label{s s d2kplus8 n odd pre conj}
            \{s_\alpha,s_\beta\} \sim  \Omega_{\alpha\beta} + \sum_{i=1}^{\frac{n+1}{2}} \, (\gamma_{\mu_1 \cdots \mu_{4i}})_{\alpha \beta} \, b_{\mu_1 \cdots \mu_{4i}},
        \end{equation}
    for $\Omega_{\alpha \beta}$ the skew-symmetric form of $Sp(k=2^{2n+6})$ as in (\ref{omega spk}). As above the indices of $b$ are antisymmetrised and each term has some coefficient preceeding it, but we conjecture that the coefficients of all terms other than the first of (\ref{s s d2kplus8 n odd pre conj}) are zero ($R_{\text{simp}}=\text{trivial}$):
        \begin{equation}\label{s s d2kplus8 n odd post conj}
            \{s_\alpha,s_\beta\} \sim  \Omega_{\alpha\beta}
        \end{equation}
    
    \item[5.] As above, in (\ref{s s d2kplus8 n odd post conj}) the right hand side should have $\Delta=n+1$, but it currently has $\Delta=0$. To rectify this, an $SO(4n+16)$ invariant factor with $\Delta=n+1$ must be included. Following the same logic as above, the only possibility is ${C_2}^{\frac{n+1}{2}}$. The Poisson brackets are
        \begin{equation}\label{s s d2kplus8 n odd post conj post dim}
            \{s_\alpha,s_\beta\} = {C_2}^{\frac{n+1}{2}} \, \Omega_{\alpha\beta}.
        \end{equation}
    
\end{enumerate}

\subsubsection{Summary of conjectured brackets}
We collect the results derived in this section. We conjecture that the Poisson bracket relations for the Coulomb branch generators of the $E_8$ family (\ref{pini e8}) for $n \geq 0$ are given by:
\begin{equation}\label{summary e8 fam brackets}
    \begin{gathered}
     \{a_{\mu \nu},a_{\rho \sigma}\} = \delta_{\nu \rho} a_{\mu \sigma} - \delta_{\mu \rho} a_{\nu \sigma} + \delta_{\mu \sigma} a_{\nu \rho} - \delta_{\nu \sigma} a_{\mu \rho} \ \ \ \forall  \ n, \\[10pt]
    \{a_{\mu \nu},s_\alpha \} = (\gamma_{\mu \nu})_{\alpha \beta} \, s_\beta \ \ \  \forall \ n, \\[10pt]
    \{s_\alpha, s_\beta\} = 
    \begin{cases}
    {C_2}^{\frac{n}{2}} (\gamma_{\mu \nu})_{\alpha \beta} \, a_{\mu \nu}, & \text{if } n \text{ even}\\[10pt]
    {C_2}^{\frac{n+1}{2}} \, \Omega_{\alpha\beta} & \text{if } n \text{ odd}\\
    \end{cases}
    \end{gathered}
\end{equation}
where $\alpha$ and $\beta$ spinor indices going from $1,...,2^{\frac{4n+16}{2}-1}$; any other indices are vector indices going from $1,...,4n+16$, antisymmetrised with those appearing with them in a given subscript; $\Omega_{\alpha \beta}$ is the skew-symmetric two form of $Sp(k=2^{2n+6})$ (\ref{omega spk}); and $C_2=Tr(a^2)$ is the second Casimir of $SO(4n+16)$, which is normalised to give no numerical coefficient in the final bracket of (\ref{summary e8 fam brackets}).

\subsection{\texorpdfstring{$E_{7,n}$ family}{E7n family}} \label{sec: e7 fam}
The Higgs branch UV fixed point of a $5d$ $SU(n)$ gauge theory with Chern Simons level $0$ and $2n+2$ flavours has magnetic quiver \cite{Hanany:2017pdx, 2018JHEP...07..061F}
\begin{equation}\label{e7 pini}
\begin{tikzpicture}[x=1cm,y=.8cm]
    \node (g1) at (0,0) [gauge,label=below:{$1$}] {};
    \node (g2) at (1,0) {$\cdots$};
    \node (g3) at (2,0) [gauge,label=below:{$n+2$}] {};
    \node (g4) at (3,0) {$\cdots$};
    \node (g5) at (4,0) [gauge,label=below:{$1$}] {};
    \node (g6) at (2,1) [gauger,label=left:{$2$}] {};
    \draw (g1)--(g2)--(g3)--(g4)--(g5);
    \draw (g3)--(g6);
\end{tikzpicture}
\end{equation}
for $n \geq 1$. The rank $2$ node is coloured red to indicate that it is unbalanced (for $n \neq 2$). Generically, we can read off that this quiver has global symmetry $SU(2n+4)$. For $n=1$ however, (\ref{e7 pini}) is the Dynkin quiver of $E_6$ ($\mathsf{D}_{E_6}$, as discussed in Section \ref{sec: Dynkin quivs}) hence the global symmetry is enhanced to $Sp(10) \supset SU(6)$ and the Poisson brackets are those given in Section \ref{sec: free spaces}. For $n=2$, (\ref{e7 pini}) is the affine $E_7$ quiver; the global symmetry is enhanced to $E_7 \supset SU(8)$, and the Poisson brackets are the structure constants of $E_7$. For all $n \geq 3$ the global symmetry of the Coulomb branch is $SU(2n+4)$ and there are $2$ generating representations, as can be found from the HWG:
\begin{equation}\label{hwg e7 pini}
    HWG= \frac{1}{(1-t^4)(1-\mu_{n+2}\, t^{n})(1-\mu_{n+2}\, t^{n+2})\prod_{i=1}^{n+1}(1- \mu_{i} \, \mu_{2n+4-i} \, t^{2i})}.
\end{equation}
where $\mu_1,...,\mu_{2n+3}$ are the highest weight fugacities for the $SU(2n+4)$ global symmetry. Explicitly, the generators are:
\begin{equation}\label{gens of e7 fam}
    \begin{array}{|c|c|c|}
        \hline
        \text{Generator} & \Delta & SU(2n+4) \text{ representation}\\
        
        \hline 
        
        {a^\mu}_{\nu} & 1 & \text{adjoint}\\
        
        \hline
         
        b^{\mu_1 \cdots \mu_{n+2}} 
        & 
        \frac{n}{2}
        & (n+2)^{th} \text{ rank antisymmetric}\\
        
    \hline
    
    \end{array} \ \ ,
\end{equation}
where all indices involving $\mu$ and $\nu$ range from $1,...,2n+4$, and any indices on $b$ are completely antisymmetrised. We conjecture that the Poisson brackets of (\ref{gens of e7 fam}) are given by:
\begin{equation}\label{final e7 brackets}
    \begin{gathered}
     \{{a^\mu}_{\nu},{a^\rho}_{\sigma}\} = {\delta^\mu}_\sigma {a^\rho}_\nu - {\delta^\rho}_\nu {a^\mu}_\sigma, \\[10pt]
    \{{a^\mu}_{\nu}, b^{\mu_1 \cdots \mu_{n+2}} \} = {\delta^{[\mu_1}}_{\nu } \, b^{\mu_2 \cdots \mu_{n+2}]\mu}, \\[10pt]
    \{b^{\mu_1 \cdots \mu_{n+2}} , b^{\nu_1 \cdots \nu_{n+2}} \} = \begin{cases}
    {C_2}^{\frac{n-2}{2}} \left( \epsilon^{\rho \mu_1 \cdots \mu_{n+2} [\nu_2 \cdots \nu_{n+2}} \, {a^{\nu_1]}}_\rho + \epsilon^{\rho \nu_1 \cdots \nu_{n+2} [\mu_2 \cdots \mu_{n+2}} \, {a^{\mu_1]}}_\rho \right) & \text{if } n \text{ even},\\[10pt]
    {C_2}^{\frac{n-1}{2}} \epsilon^{\mu_1 \cdots \mu_{n+2} \nu_1 \cdots \nu_{n+2}} & \text{if } n \text{ odd},\\
    \end{cases}   \end{gathered}
\end{equation}
where $C_2=Tr(a^2)$ is the only non-zero $SU(2n+4)$ Casimir invariant of this moduli space, which is normalised to give no numerical coefficient in the final bracket of (\ref{final e7 brackets}).

\subsection{\texorpdfstring{$E_{6,n}^I$ family}{E6n family I}} \label{sec: e6 fam 1}
The Higgs branch UV fixed point of a $5d$ $SU(n)$ gauge theory with Chern Simons level $\pm \frac{1}{2}$ and $2n+1$ flavours has magnetic quiver \cite{Hanany:2017pdx, 2018JHEP...07..061F}
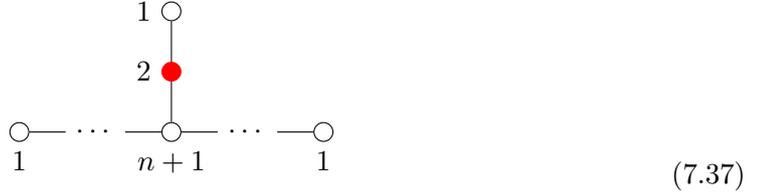
\begin{equation}\label{e6 pini}
\begin{tikzpicture}[x=1cm,y=.8cm]
    \node (g1) at (0,0) [gauge,label=below:{$1$}] {};
    \node (g2) at (1,0) {$\cdots$};
    \node (g3) at (2,0) [gauge,label=below:{$n+1$}] {};
    \node (g4) at (3,0) {$\cdots$};
    \node (g5) at (4,0) [gauge,label=below:{$1$}] {};
    \node (g6) at (2,1) [gauger,label=left:{$2$}] {};
    \node (g7) at (2,2) [gauge,label=left:{$1$}] {};
    \draw (g1)--(g2)--(g3)--(g4)--(g5);
    \draw (g3)--(g6)--(g7);
\end{tikzpicture}
\end{equation}
for $n \geq 1$. The rank $2$ node is coloured red to indicate that it is unbalanced (for $n \neq 2$). Generically, we can read off that this quiver has global symmetry $SU(2n+2) \times SU(2)$. For $n=1$, (\ref{e6 pini}) is the Dynkin quiver of $D_5$ ($\mathsf{D}_{D_5}$, as discussed in Section \ref{sec: Dynkin quivs}) and so the global symmetry is enhanced to $Sp(6)$ and the Poisson brackets are those given in Section \ref{sec: free spaces}. For $n=2$, (\ref{e6 pini}) is the affine $E_6$ quiver; the global symmetry is enhanced to $E_6 \supset SU(6) \times SU(2)$, and the Poisson brackets are the structure constants of $E_6$. For all $n \geq 3$ the global symmetry of the Coulomb branch is $SU(2n+2) \times SU(2)$ and there are $3$ generating representations, as can be found from the HWG:
\begin{equation}\label{hwg e6 pini}
    HWG = \frac{1-\nu^2 \, \mu_{n+1}^2 \, t^{2n+4}}{(1-\nu^2\,t^2)(1-t^4)(1-\nu \, \mu_{n+1} \, t^{n})(1 - \nu \, \mu_{n+1} \, t^{n+2})\prod_{i=1}^{n+1} (1 - \mu_i \, \mu_{2n+2-i} \, t^{2i})},
\end{equation}
where $\mu_1,...,\mu_{2n+1}$ and $\nu$ are the highest weight fugacities for the $SU(2n+2)$ and $SU(2)$ factors in the global symmetry respectively. Explicitly, the generators are:
\begin{equation}\label{gens of e6 I fam}
    \begin{array}{|c|c|c|}
        \hline
        \text{Generator} & \Delta & SU(2n+2) \times SU(2) \text{ representation}\\
        
        \hline 
        
        {A^{\mu}}_{\nu} & 1 & \text{adjoint } \times \text{ trivial}\\
        
        \hline
        
        a_{\alpha \beta} & 1 & \text{trivial } \times \text{ adjoint}\\
        
        \hline
        
        {B^{\mu_1 \cdots \mu_{n+1}}}_{\alpha}
        &
        \frac{n}{2}
        & (n+1)^{th} \text{ rank antisymmetric } \times \text{ fundamental}\\
        
        \hline
    
    \end{array} \ \ ,
\end{equation}
where all indices involving $\mu$ and $\nu$ are $SU(2n+2)$ indices ranging from $1,...,2n+2$ and those that appear in exponent of $B$ are antisymmetrised among themselves; and indices involving $\alpha$ and $\beta$ are $SU(2)$ indices ranging from $1,2$. We conjecture that the non-zero Poisson brackets of (\ref{gens of e6 I fam}) are given by:
\begin{equation}\label{final e61 pbs}
    \begin{gathered}
    \{{A^{\mu_1}}_{\nu_1},{A^{\mu_2}}_{\nu_2}\} = {\delta^{\mu_1}}_{\nu_2} {A^{\mu_2}}_{\nu_1} - {\delta^{\mu_2}}_{\nu_1} {A^{\mu_1}}_{\nu_2}, \\[10pt]
    \{ a_{\alpha_1 \beta_1} , a_{\alpha_2 \beta_2} \} = \epsilon_{\alpha_1 \beta_2} a_{\alpha_2 \beta_1} + \epsilon_{\beta_1 \beta_2} a_{\alpha_2 \alpha_1} + \epsilon_{\alpha_1 \alpha_2} a_{\beta_2 \beta_1} + \epsilon_{\beta_1 \alpha_2} a_{\beta_2 \alpha_1} ,\\[10pt]
    \{{A^{\mu}}_{\nu},{B^{\mu_1 \cdots \mu_{n+1}}}_{\alpha}\} = {\delta^{[\mu_1}}_{\nu } {B^{\mu_2  \cdots \mu_{n+1}] \, \mu}}_{\alpha}, \\[10pt]
    \{a_{\alpha \beta} ,{B^{\mu_1 \cdots \mu_{n+1}}}_{\alpha_1}\} =  {B^{\mu_1 \cdots \mu_{n+1}}}_{(\beta} \, \epsilon_{\alpha) \alpha_1}, \\[10pt]
    \{{B^{\mu_1 \cdots \mu_{n+1}}}_{\alpha},{B^{\nu_1 \cdots \nu_{n+1}}}_{\beta}\} =
    \begin{dcases}
        \ \ \begin{multlined}
            {C_2}^{\frac{n-2}{2}} \, \Big( \epsilon^{\mu_1 \cdots \mu_{n+1} \nu_1 \cdots \nu_{n+1}} \, \epsilon_{\alpha_1 \, [\alpha} \, a_{\beta] \, \alpha_1} + \epsilon_{\alpha \beta} \\
            \Big( \epsilon^{\nu \nu_1 \cdots \nu_{n+1} [ \mu_2 \cdots \mu_{n+1}} \, {A^{\mu_1]}}_{\nu}
            + \epsilon^{\nu \mu_1 \cdots \mu_{n+1} [\nu_2 \cdots \nu_{n+1}} \, {A^{\nu_1]}}_{\nu} \Big) \Big)
        \end{multlined}  
        & 
        \text{if } n \text{ even},\\[10pt]
        \ \ {C_2}^{\frac{n-1}{2}} \, \epsilon_{\alpha \beta} \, \epsilon^{\mu_1 \cdots \mu_{n+1} \nu_1 \cdots \nu_{n+1}}
        &
        \text{if } n \text{ odd},\\[10pt]
    \end{dcases}\\[10pt]
    \end{gathered}
\end{equation}
where $(\cdot)$/$[\cdot]$ indicates a symmetrisation/antisymmetrisation over the enclosed indices with no numerical prefactor; and $C_2$ is the second and only non-zero Casimir for $SU(2n+2)\times SU(2)$ on this Coulomb branch, which is normalised to give no numerical coefficient in the final bracket of (\ref{final e61 pbs}). The second Casimir $C_2$ of the product group $SU(2n+2)\times SU(2)$ is proportional to the second Casimir of each individual group ($SU(2n+2)$ and $SU(2)$ respectively).\footnote{The $2^{\text{nd}}$ Casimirs of $SU(2n+2)$ and $SU(2)$, $C_{2,SU(2n+2)}=Tr(A^2)$ and $C_{2,SU(2)}=Tr(a^2)$ respectively, both have conformal dimension $2$. The HWG tells us that there is just a single invariant of the Coulomb branch global symmetry, and that it has $\Delta=2$. This means that some linear combination of the individual Casimirs $C_{2,SU(2n+2)}$ and $C_{2,SU(2)}$, say $\beta_1 \, C_{2,SU(2n+2)} + \beta_2 \, C_{2,SU(2)}$, must vanish, and all Casimirs of the flavour symmetry $SU(2n+2)\times SU(2)$ on this Coulomb branch must be proportiinal to an orthogonal linear combination, say $C_2 = \alpha_1 \, C_{2,SU(2n+2)}+\alpha_2 \, C_{2,SU(2)}$. The vanishing linear combination tells us that $C_{2,SU(2n+2)}$ and $C_{2,SU(2)}$ are proportional, and so we can see that the second Casimir of the product group $C_2$ is proportional to either of the second Casimirs of the individual groups.} All other Poisson brackets between (\ref{gens of e6 I fam}) that are not listed in (\ref{final e61 pbs}) vanish.

\subsection{\texorpdfstring{$E_{6,n}^{II}$ family}{E6n family II}} \label{sec: e6 fam 2}
The Higgs branch UV fixed point of a $5d$ $SU(n)$ gauge theory with Chern Simons level $\pm \frac{3}{2}$ and $2n+1$ flavours has magnetic quiver \cite{Hanany:2017pdx, 2018JHEP...07..061F}
\begin{equation}\label{e6 pini II}
\begin{tikzpicture}[x=1cm,y=.8cm]
    \node (g1) at (0,0) [gauge,label=below:{$1$}] {};
    \node (g2) at (1,0) {$\cdots$};
    \node (g3) at (2,0) [gauge,label=below:{$2n-1$}] {};
    \node (g4) at (3,0) [gauge,label=below:{$n$}] {};
    \node (g5) at (4,0) [gauger,label=below:{$1$}] {};
    \node (g6) at (2,1) [gauge,label=left:{$n$}] {};
    \node (g7) at (2,2) [gauger,label=left:{$1$}] {};
    \draw (g1)--(g2)--(g3)--(g4)--(g5);
    \draw (g3)--(g6)--(g7);
\end{tikzpicture}
\end{equation}
for $n \geq 1$. The two rank $1$ nodes are coloured red to indicate that they are unbalanced (for $n \neq 2$). Generically, we can read off that this quiver has global symmetry $SO(4n+2) \times U(1)$. For $n=1$, (\ref{e6 pini II}) is Dynkin quiver of $A_5$ ($\mathsf{D}_{A_5}$, as discussed in Section \ref{sec: Dynkin quivs}) and hence the global symmetry is enhanced to $Sp(4) \supset SO(6) \times U(1)$, and the Poisson brackets are given in Section \ref{sec: free spaces}. For $n=2$, (\ref{e6 pini II}) is the affine $E_6$ quiver; the global symmetry is enhanced to $E_6 \supset SO(10) \times U(1)$, and the Poisson brackets are the structure constants of $E_6$. For all $n \geq 3$ the global symmetry of the Coulomb branch is $SO(4n+2) \times U(1)$ and there are $4$ generating representations, as can be found from the HWG:
\begin{equation}\label{hwg e6 pini II}
    HWG = \frac{1}{(1-t^2)(1- \frac{\mu_{2n+1} \, t^{n}}{q})(1- q \, \mu_{2n} \, t^{n})\prod_{i=1}^{n-1} (1 - \mu_{2i} \, t^{2i})},
\end{equation}
where $\mu_1,...,\mu_{2n+1}$ are the highest weight fugacities for $SO(4n+2)$ and $q$ is the fugacity for the $U(1)$ charge. Explicitly, the generators are:
\begin{equation}\label{gens of e6 II fam}
    \begin{array}{|c|c|c|}
        \hline
        \text{Generator} & \Delta & SO(4n+2) \times U(1) \text{ representation}\\
        
        \hline 
        
        a_{\mu \nu} & 1 & \text{adjoint } \times (0)\\
        
        \hline
        
        C_1 & 1 & \text{trivial } \times (0)\\
        
        \hline
        
        s^{\alpha}
        &
        \frac{n}{2}
        & \text{left spinor } \times (+1)\\
        
        \hline
        
        s_\alpha
        &
        \frac{n}{2}
        & \text{right spinor } \times (-1)\\
        
        \hline
    
    \end{array} \ \ ,
\end{equation}
where $\alpha$ and $\beta$ are spinor indices going from $1,...,2^{2n}$; and any other indices are vector indices going from $1,...,4n+2$, antisymmetrised with those appearing alongside them in a given subscript. $C_1$ is the invariant of the $U(1)$ factor in the flavour symmetry, so-called to make connection with the notation for the flavour symmetry invariants in the Coulomb branches of Sections \ref{sec: e8 fam} -- \ref{sec: e6 fam 1}. This $U(1)$ is formed from a linear combination of the two $U(1)$ symmetries in the classical (finite coupling) Higgs branch of the $5d$ theory in question: the $U(1)_B\subset U(2n+1)$ baryon symmetry associated to the trace of the meson matrix $M$, and the $U(1)_I$ instanton symmetry associated to the gaugino bilinear $S$ one can construct.\footnote{By associated, we mean that $M$ and $S$ are the scalar superpartners of the conserved currents associated to these symmetries. We use this association because, for study of the moduli space, it is the scalars which we are interested in.} The orthogonal linear combination of these $U(1)$ symmetries is combined with the $SU(2n+1)$ factor in the classical flavour symmetry, enhancing it to the $SO(4n+2)$ we see at infinite coupling. The left and right spinors of $SO(4n+2)$ are complex representations which are conjugate to one another, hence the raised and lowered indices respectively. This makes matching the indices in the Poisson brackets (Step $4$ of Section \ref{sec: pbs for cb as reps}) slightly more complicated in this case. The matrices $\delta$ and $\gamma_{\mu\nu}$ are spinor valued (i.e. their entries are labelled by two spinor indices $\alpha$ and $\beta$), and lie in the trivial ($1$) and adjoint ($\mu_2$) representations of $SO(4n+2)$ respectively. To see what form their spinor indices take, it is therefore important to ascertain whether these representations lie in the second rank symmetric/antisymmetric of one of the spinors or in the tensor product of the two conjugate spinors. It turns out that both $1$ and $\mu_2$ lie in the tensor product $\mu_{2n} \otimes \mu_{2n+1}$, and hence have one upper and one lower spinor index labelling their matrix entries: ${\delta^{\alpha}}_{\beta}$ and ${(\gamma_{\mu\nu})^{\alpha}}_{\beta}$.\\

\noindent We conjecture that the non-zero Poisson brackets of (\ref{gens of e6 II fam}) are given by:
\begin{equation}\label{final e62 pbs}
    \begin{gathered}
     \{a_{\mu \nu},a_{\rho \sigma}\} = \delta_{\nu \rho} a_{\mu \sigma} - \delta_{\mu \rho} a_{\nu \sigma} + \delta_{\mu \sigma} a_{\nu \rho} - \delta_{\nu \sigma} a_{\mu \rho}, \\[10pt]
    \{a_{\mu \nu},s^{\alpha}\} = {(\gamma_{\mu \nu})^{\alpha}}_{\beta} \, s^{\beta},  \\[10pt]
    \{a_{\mu \nu},s_\alpha\} = s_\beta \, {(\gamma_{\mu \nu})^{\beta}}_{\alpha}, \\[10pt]
    \{C_1,s^{\alpha}\}=+ \, s^{\alpha}\\[10pt]
    \{C_1,s_\alpha\}=- \, s_\alpha\\[10pt]
    \{s^{\alpha},s_\beta\} = {C_1}^{n-1} \, {\delta^{\alpha}}_{\beta},\\[10pt]
    \end{gathered}
\end{equation}
where $C_1$ is normalised such that there are no additional numerical coefficients in (\ref{final e62 pbs}). All other Poisson brackets between (\ref{gens of e6 II fam}) that are not listed in (\ref{final e62 pbs}) vanish.

\subsection{\texorpdfstring{Magnetic quivers for $5d$ $\mathcal{N}=1$ SQCD}{Magentic quivers for 5d N=1 SQCD}}\label{sec: zhenghao quivs}
The following four sections are based on magnetic quivers \cite{Cabrera:2019izd} found for certain cones of the Higgs branch at infinite coupling of $5d$ $\mathcal{N}=1$ special unitary SQCD theories with UV fixed points \cite{Cabrera:2018jxt}. Such a $5d$ theory is specified by three parameters: the number of colours $N_c$ of the gauge group; the number of flavours $N_f$ transforming in its fundamental representation; and the Chern-Simons level $k$;
\begin{equation}\label{sunc sqcd}
\begin{tikzpicture}[x=1cm,y=0.8cm]
    \node (g1) at (0,0) [gauge,label=left:{$SU(N_c)_{\pm |k|}$}] {};
    \node (g2) at (1,0) [flavor,label=right:{$N_f$}] {};
    \node (g4) at (1.9,-0.1) {$.$};
    \draw (g1)--(g2);
\end{tikzpicture}
\end{equation}
To have a UV fixed point, these three parameters must satisfy the following inequality \cite{Bergman:2014kza}:
\begin{equation}\label{uv fixed ineq}
    |k| \leq N_c - \frac{N_f}{2} +2.
\end{equation}
In these instances, we can investigate the case of infinite coupling. \\

\noindent In \cite{Cabrera:2018jxt}, the authors split (\ref{uv fixed ineq}) into four regions and computed the magnetic quivers -- plural as each Higgs branch is the union of various cones -- for the various Higgs branches in each region. In an unpublished work by Zhenghao Zhong and Amihay Hanany, the HWG for each of these magnetic quivers was computed. We will use these HWGs to derive the Poisson brackets for the Coulomb branch generators of these magnetic quivers. Many of the quivers appearing in \cite{Cabrera:2018jxt} either only have generators in the adjoint representation or have already appeared in Sections \ref{sec: e8 fam} -- \ref{sec: e6 fam 2}: we will omit the Poisson brackets for these quivers as they have been covered in previous sections. In each section a brief comment will be given regarding such quivers. The remaining quivers (bar certain exceptional cases of the Chern Simons coupling in regions $2$ and $4$) fall into four families: the so-called trapezium, pyramid, kite and truck families. The first two such families are specified by three parameters, $N_f-1$, $n$ and $\sigma$, and the latter two families by just $n$ and $\sigma$. Each of these parameters is a function of the parameters of the $5d$ theory, $N_f$, $N_c$ and $k$, and one can check the original paper \cite{Cabrera:2018jxt} for the specific relationship (the new parameters $n$ and $\sigma$ have been introduced to encapsulate multiple cones by the same quiver: the only difference between cones being the dependence of $n$ and $\sigma$ on $N_f$, $N_c$ and $k$). The structure of each family of quivers restricts the values that $n$ and $\sigma$ can take (for example, for the Trapezium family (\ref{trapez fam}) can only have $n=1$ for $N_f=2$ or $3$). In Sections \ref{sec: trapez fam} -- \ref{sec: truck fam} we will explore the Poisson brackets for the Coulomb branch generators of each of these four families in turn.\\

\subsubsection{Trapezium family} \label{sec: trapez fam}
The trapezium family encapsulates the magnetic quivers for two cones which make up the Higgs branch of $5d$ $\mathcal{N}=1$ SQCD with parameters satisfying the following subsets of (\ref{uv fixed ineq}), ``region $1$" and ``region $3$" of \cite{Cabrera:2018jxt} respectively:
\begin{equation}\label{region 1 and 3}
        |k| < N_c - \frac{N_f}{2} \ \ \ \ \text{ and } \ \ \ \ |k| = N_c - \frac{N_f}{2} + 1.
\end{equation}
Note that there is an additional cone comprising the Higgs branch of each of these regions (\ref{region 1 and 3}) with a magnetic quiver that does not conform to the structure of (\ref{trapez fam}), but they are closures nilpotent orbits.\footnote{For the exceptional cases of $k=\frac{1}{2}$ and $k=0$ in region $3$, the magnetic quiver does not take the form of (\ref{trapez fam}), and instead takes the form of (\ref{e6 pini}) for $N_f=2n+1$ and (\ref{e7 pini}) for $N_f=2n+2$ respectively (where this $n$ is that used in (\ref{e6 pini}) and (\ref{e7 pini}) respectively, not to be confused with the $n$ of (\ref{trapez fam}).} The family is given by
\begin{equation}\label{trapez fam}
\begin{tikzpicture}[x=1cm,y=0.8cm]
    \node (g1) at (0,0) [gauge,label=below:{$1$}] {};
    \node (g2) at (1,0) [gauge,label=below:{$2$}] {};
    \node (g3) at (2,0) {$\cdots$};
    \node (g4) at (3,0) [gauge,label=below:{$n$}] {};
    \node (g5) at (4,0) {$\cdots$};
    \node (g6) at (5,0) [gauge,label=below:{$n$}] {};
    \node (g7) at (6,0) {$\cdots$};
    \node (g8) at (7,0) [gauge,label=below:{$2$}] {};
    \node (g9) at (8,0) [gauge,label=below:{$1$}] {};
    \node (g10) at (3.5,1) [gauger,label=above:{$1$}] {};
    \node (g11) at (4.5,1) [gauger,label=above:{$1$}] {};
    \draw (g1)--(g2)--(g3)--(g4)--(g5)--(g6)--(g7)--(g8)--(g9);
    \draw (g4)--(g10);
    \draw (g6)--(g11);
    \draw[transform canvas = {yshift=1.5pt}] (g10) -- (g11) node [midway, above] {$\sigma$};
    \draw[transform canvas = {yshift=-1.5pt}] (g10)--(g11) ;
    \draw [decorate,decoration={brace,mirror,amplitude=6pt}] (0,-0.8) --node[below=6pt] {$N_f - 1$} (8,-0.8);
\end{tikzpicture}
\end{equation}
for $N_f \geq 2$, where a red node indicates that it is unbalanced, and $\sigma$ indicates the multiplicity of the hypermultiplet linking the two unbalanced $U(1)$ nodes. Generically, we can read off that this quiver has global symmetry $SU(N_f) \times U(1)$. Note that this is less than in the classical case. In the case where $n=\sigma=1$ however, the two red nodes are actually balanced and this quiver becomes the affine $A_{N_f}$ quiver; the global symmetry is enhanced to $SU(N_f+1) \supset SU(N_f) \times U(1)$, and the Poisson brackets are the structure constants of $SU(N_f + 1)$. From the structure of the quiver, one can see that we must necessarily have $1 \leq n \leq \lfloor \frac{N_f}{2} \rfloor$. When we don't have $n=\sigma=1$, there are $4$ generating representations, as can be found from the HWG:
\begin{equation}
    HWG = \frac{1 - \mu_n \, \mu_{N_f-n} \, t^{2(n+\sigma)}}{(1-t^2)(1-\mu_n \, q \, t^{n+\sigma})(1- \frac{\mu_n}{q} \, t^{n+\sigma})\prod_{i=1}^{n}(1-\mu_i \, \mu_{N_f-i} \, t^{2i})}.
\end{equation}
where $\mu_1,...,\mu_{N_f-1}$ are the highest weight fugacities for $SU(N_f)$, and $q$ is the fugacity for the $U(1)$ charge. Explicitly, the generators are:\footnote{The $(N_f-n)^{\text{th}}$ rank antisymmetric could be written with $N_f-n$ upper antisymmetrised indices, but the invariance of $\epsilon^{\mu_1 \cdots \mu_{N_f}}$ means we can also write it with $n$ lower antisymmetrised indices instead. This choice of notation makes clear the fact that it is the conjugate to the $n^{\text{th}}$ rank antisymmetric representation.}
\begin{equation}\label{gens of trapez fam}
    \begin{array}{|c|c|c|}
        \hline
        \text{Generator} & \Delta & SU(N_f) \times U(1) \text{ representation}\\
        
        \hline 
        
        {a^{\mu}}_{\nu} & 1 & \text{adjoint } \times (0)\\
        
        \hline
        
        C_1 & 1 & \text{trivial } \times (0)\\
        
        \hline
        
        b^{\mu_1 \cdots \mu_n}
        &
        \frac{n+\sigma}{2}
        & n^{th} \text{ rank antisymmetric } \times (+1)\\
        
        \hline
        d_{\mu_1 \cdots \mu_{n}}
        &
        \frac{n+\sigma}{2}
        & (N_f-n)^{th} \text{ rank antisymmetric } \times (-1)\\
        
        \hline
    
    \end{array} \ \ ,
\end{equation}
where all indices involving $\mu$ and $\nu$ range from $1,...,N_f$; those on $b$ are completely antisymmetrised; and $C_1$ is the invariant of the $U(1)$ factor in the flavour symmetry (see Section \ref{sec: e6 fam 2} for more details). The parameter $n$ takes different values with regards to the parameters $N_c$, $N_f$ and $k$ on each of the Higgs branch cones whose magnetic quiver is encapsulated by (\ref{trapez fam}). The specific expression for $n$ will dictate the physical states of $b$ and $d$, hence their generic names.\\

\noindent We conjecture that the non-zero Poisson brackets of (\ref{gens of trapez fam}) are given by:
\begin{equation}\label{final trapez pbs}
    \begin{gathered}
     \{{a^\mu}_{\nu},{a^\rho}_{\sigma}\} = {\delta^\mu}_\sigma {a^\rho}_\nu - {\delta^\rho}_\nu {a^\mu}_\sigma, \\[10pt]
    \{{a^{\mu}}_{\nu}, b^{\mu_1 \cdots \mu_n} \} = {\delta^{[\mu_1}}_{\nu } \, b^{\mu_2 \cdots \mu_n] \, \mu} \\[10pt]
    \{{a^{\mu}}_{\nu}, d_{\mu_1 \cdots \mu_{n}}\} = 
    {\delta^{\mu}}_{[\mu_1} \, d_{\mu_2  \cdots \mu_{n}] \, \nu},\\[10pt]
    \{C_1,b^{\mu_1 \cdots \mu_n}\}=+\, b^{\mu_1 \cdots \mu_n},\\[10pt]
    \{C_1,d_{\mu_1 \cdots \mu_{n}}\}=- \, d_{\mu_1 \cdots \mu_{n}},\\[10pt]
    \{b^{\mu_1 \cdots \mu_n}, d_{\nu_1 \cdots \nu_{n}} \} = {C_1}^{n+\sigma-1} \, {\delta^{[\mu_1}}_{[\nu_1} \, \cdots \, {\delta^{\mu_n]}}_{\nu_n]}. \\[10pt]
    \end{gathered}
\end{equation}
where $C_1$ is normalised such that there are no additional numerical coefficients in (\ref{final trapez pbs}). In one of the cones $n=N_f-N_c$, which renders $b$ and $d$ the baryons which we would expect to show up (due to $C_1$ acting as the baryon number: the $U(1)_B$ baryon symmetry is preserved). In the other cone, we expect $b$ and $d$ to be instanton operators. All other Poisson brackets between (\ref{gens of trapez fam}) that are not listed in (\ref{final trapez pbs}) vanish.

\subsubsection{Pyramid family} \label{sec: pyramid fam}
The pyramid family are magnetic quivers for one cone which makes up the Higgs branch of $5d$ $\mathcal{N}=1$ SQCD with parameters satisfying the following subset of (\ref{uv fixed ineq}), ``region $2$" of \cite{Cabrera:2018jxt}:
\begin{equation}\label{region 2}
        |k| = N_c - \frac{N_f}{2}.
\end{equation}
Note that there is an additional Higgs branch cone with a magnetic quiver that does not conform to the structure of (\ref{pyramid fam}), but it is the closure of a nilpotent orbit or a product thereof (depending on the values of the parameters).\footnote{Note also that for the exceptional case of $k=0$ in (\ref{region 2}), the magnetic quiver does not take the form of (\ref{pyramid fam}).} The family is given by
\begin{equation}\label{pyramid fam}
\begin{tikzpicture}[x=1cm,y=0.8cm]
    \node (g1) at (0,0) [gauge,label=below:{$1$}] {};
    \node (g2) at (1,0) [gauge,label=below:{$2$}] {};
    \node (g3) at (2,0) {$\cdots$};
    \node (g4) at (3,0) [gauge,label=below:{$n$}] {};
    \node (g5) at (4,0) {$\cdots$};
    \node (g6) at (5,0) [gauge,label=below:{$n$}] {};
    \node (g7) at (6,0) {$\cdots$};
    \node (g8) at (7,0) [gauge,label=below:{$2$}] {};
    \node (g9) at (8,0) [gauge,label=below:{$1$}] {};
    \node (g10) at (3.5,1) [gauger,label=above:{$1$}] {};
    \node (g11) at (4.5,1) [gauger,label=above:{$1$}] {};
    \node (g12) at (4,2) [gauge,label=above:{$1$}] {};
    \draw (g1)--(g2)--(g3)--(g4)--(g5)--(g6)--(g7)--(g8)--(g9);
    \draw (g4)--(g10);
    \draw (g6)--(g11);
    \draw (g10)--(g12)--(g11);
    \draw[transform canvas = {yshift=1.5pt}] (g10) -- (g11) ;
    \draw[transform canvas = {yshift=-1.5pt}] (g10)--(g11) node [midway, below] {$\sigma$};
    \draw [decorate,decoration={brace,mirror,amplitude=6pt}] (0,-0.8) --node[below=6pt] {$N_f - 1$} (8,-0.8);
\end{tikzpicture}
\end{equation}
for $N_f \geq 2$, where a red node indicates that it is unbalanced, and $\sigma$ indicates the multiplicity of the hypermultiplet stretching between the two unbalanced $U(1)$ nodes. Generically, we can read off that this quiver has global symmetry $SU(N_f) \times SU(2) \times U(1)$. The structure of the quiver means that we must necessarily have $1 \leq n \leq \lfloor \frac{N_f}{2} \rfloor$. There are $5$ generating representations, as can be found from the HWG:
\begin{equation}
    HWG = \frac{1 - \nu^2 \, \mu_n \, \mu_{N_f-n} \, t^{2(n+\sigma+1)}}{(1-t^2)(1-\nu^2 \, t^2)(1-\nu \, \mu_n \, q \, t^{n+\sigma+1})(1- \frac{\nu \, \mu_n}{q} \, t^{n+\sigma+1})\prod_{i=1}^{n}(1-\mu_i \, \mu_{N_f-i} \, t^{2i})},
\end{equation}
where $\nu$ and $\mu_1,...,\mu_{N_f-1}$ are the highest weight fugacities for $SU(2)$ and $SU(N_f)$ respectively, and $q$ is the fugacity for the $U(1)$ charge. Explicitly, the generators are:
\begin{equation}\label{gens of pyramid fam}
    \begin{array}{|c|c|c|}
        \hline
        \text{Generator} & \Delta & SU(N_f) \times SU(2) \times U(1) \text{ representation}\\
        
        \hline 
        
        {A^{\mu}}_{\nu} & 1 & \text{adjoint } \times \text{ trivial} \times (0)\\
        
        \hline
        
        a_{\alpha \beta} & 1 & \text{trivial } \times \text{ adjoint} \times (0)\\
        
        \hline
        
        C_1 & 1 & \text{trivial } \times \text{ trivial} \times (0)\\
        
        \hline
        
        {B^{\mu_1 \cdots \mu_n}}_{\alpha}
        &
        \frac{n+\sigma+1}{2}
        & n^{th} \text{ rank antisymmetric } \times \text{ fundamental} \times (+1)\\
        
        \hline
        
        D_{\mu_1 \cdots \mu_{n} \, \alpha}
        &
        \frac{n+\sigma+1}{2}
        & (N_f-n)^{th} \text{ rank antisymmetric } \times \text{ fundamental} \times (-1)\\
        
        \hline
    
    \end{array} \ \ ,
\end{equation}
where all indices involving $\mu$ and $\nu$ are $SU(N_f)$ indices ranging from $1,...,N_f$, and those that appear in exponents or subscripts of $B$ or $D$ are antisymmetrised among themselves; indices involving $\alpha$ and $\beta$ are $SU(2)$ indices ranging from $1,2$; and $C_1$ is the invariant of the $U(1)$ factor in the flavour symmetry (see Section \ref{sec: e6 fam 2} for more details). We conjecture that the non-zero Poisson brackets of (\ref{gens of pyramid fam}) are given by:
\begin{equation}\label{final pyramid pbs}
    \begin{gathered}
     \{{A^{\mu_1}}_{\nu_1},{A^{\mu_2}}_{\nu_2}\} = {\delta^{\mu_1}}_{\nu_2} \,  {A^{\mu_2}}_{\nu_1} - {\delta^{\mu_2}}_{\nu_1} \, {A^{\mu_1}}_{\nu_2},  \\[10pt]
    \{ a_{\alpha_1 \beta_1} , a_{\alpha_2 \beta_2} \} = \epsilon_{\alpha_1 \beta_2} a_{\alpha_2 \beta_1} + \epsilon_{\beta_1 \beta_2} a_{\alpha_2 \alpha_1} + \epsilon_{\alpha_1 \alpha_2} a_{\beta_2 \beta_1} + \epsilon_{\beta_1 \alpha_2} a_{\beta_2 \alpha_1} ,\\[10pt]
    \{{A^{\mu}}_{\nu}, {B^{\mu_1 \cdots \mu_n}}_{\alpha} \} = {\delta^{[\mu_1}}_{\nu } {B^{\mu_2 \cdots \mu_n] \, \mu}}_{\alpha}, \\[10pt]
    \{{A^{\mu}}_{\nu}, D_{\mu_1 \cdots \mu_{n} \, \alpha} \} = {\delta^{\mu}}_{[\mu_1 } D_{\mu_2 \cdots \mu_n] \, \nu \, \alpha }, \\[10pt]
    \{a_{\alpha \beta}, {B^{\mu_1 \cdots \mu_n}}_{\alpha_1} \} =  {B^{\mu_1 \cdots \mu_n}}_{(\alpha} \, \epsilon_{\beta) \alpha_1} , \\[10pt]
    \{a_{\alpha \beta}, D_{\mu_1 \cdots \mu_{n} \, \alpha_1} \} = D_{\mu_1 \cdots \mu_{n} (\beta} \, \epsilon_{\alpha) \alpha_1}, \\[10pt]
    \{C_1,{B^{\mu_1 \cdots \mu_n}}_{\alpha} \}=+ \, {B^{\mu_1 \cdots \mu_n}}_{\alpha},\\[10pt]
    \{C_1,D_{\mu_1 \cdots \mu_{n} \alpha}\}=-D_{\mu_1 \cdots \mu_{n} \alpha},\\[10pt]
    \{{B^{\mu_1 \cdots \mu_n}}_{\alpha} , D_{\nu_1 \cdots \nu_{n} \beta} \} = {C_1}^{n+\sigma} \, {{\delta}^{[\mu_1}}_{[\nu_1} \ \cdots \ {{\delta}^{\mu_n]}}_{\nu_{n}]} \, \epsilon_{\alpha \beta}, \\[10pt]
    \end{gathered}
\end{equation}
where $(\cdot)$/$[\cdot]$ indicates a symmetrisation/antisymmetrisation over the enclosed indices with no numerical prefactor; and $C_1$ is normalised such that there are no additional numerical coefficients in (\ref{final pyramid pbs}). All other Poisson brackets between (\ref{gens of pyramid fam}) that are not listed in (\ref{final  pyramid pbs}) vanish.

\subsubsection{Kite family} \label{sec: kite fam}
The kite family are magnetic quivers for one cone in the Higgs branch of $5d$ $\mathcal{N}=1$ SQCD with parameters satisfying the following subset of (\ref{uv fixed ineq}), ``region $4$" of \cite{Cabrera:2018jxt}:
\begin{equation}\label{region 4}
        2<|k| = N_c - \frac{N_f}{2} + 2,
\end{equation}
with $N_f$ even. There is an additional Higgs branch cone with a magnetic quiver that does not conform to the structure of (\ref{kite fam}) for $N_f\geq 2$, but it is the closure of a nilpotent orbit.\footnote{Note also that for the exceptional case of $k=1$, the magnetic quiver does not take the form of (\ref{kite fam}).} The family is given by
\begin{equation}\label{kite fam}
\begin{tikzpicture}[x=1cm,y=0.8cm]
    \node (g1) at (0,0) [gauge,label=below:{$1$}] {};
    \node (g2) at (1,0) [gauge,label=below:{$2$}] {};
    \node (g3) at (2,0) {$\cdots$};
    \node (g4) at (3,0) [gauge,label=below:{$2n-2$}] {};
    \node (g5) at (4,0) [gauge,label=below:{$n$}] {};
    \node (g6) at (4.8,0.8) [gauger,label=right:{$1$}] {};
    \node (g7) at (4.8,-0.8) [gauger,label=right:{$1$}] {};
    \node (g8) at (3,1) [gauge,label=above:{$n-1$}] {};
    \draw (g1)--(g2)--(g3)--(g4)--(g5)--(g6);
    \draw (g5)--(g7);
    \draw (g4)--(g8);
    \draw[transform canvas = {xshift=1.5pt}] (g6)--(g7) node [midway, right] {$\sigma$};
    \draw[transform canvas = {xshift=-1.5pt}] (g6)--(g7);
\end{tikzpicture}
\end{equation}
for $n \geq 2$, where a red node indicates that it is unbalanced, and $\sigma$ indicates the multiplicity of the hypermultiplet stretching between the two unbalanced $U(1)$ nodes. Generically, we can read off that this quiver has global symmetry $SO(4n) \times U(1)$. There are $4$ generating representations, as can be found from the HWG:
\begin{equation}
    HWG = \frac{1 - \mu_{2n}^2 \, t^{2(n+\sigma)}}{(1-t^2)(1-\mu_{2n}^2 \, t^{2n})(1- \mu_{2n} \, q \, t^{n+\sigma})(1- \frac{\mu_{2n}}{q} \, t^{n+\sigma})\prod_{i=1}^{n-1}(1-\mu_{2i} \, t^{2i})},
\end{equation}
where $\mu_1,...,\mu_{2n}$ are the highest weight fugacities for the $SO(4n)$ factor in the global symmetry, and $q$ is the fugacity for the $U(1)$ charge. Explicitly, the generators are:
\begin{equation}\label{gens of kite fam}
    \begin{array}{|c|c|c|}
        \hline
        \text{Generator} & \Delta & SO(4n) \times U(1) \text{ representation}\\
        
        \hline
        
        a_{\mu \nu} & 1 & \text{adjoint } (0)\\
        
        \hline
        
        C_1 & 1 & \text{trivial } \times (0)\\
        
        \hline
        
        s^+_\alpha
        &
        \frac{n+\sigma}{2}
        & \text{spinor } \times (+1)\\
        
        \hline
        
        s^-_\alpha
        &
        \frac{n+\sigma}{2}
        & \text{spinor } \times (-1) \\
        
        \hline
    
    \end{array} \ \ ,
\end{equation}
where $\alpha$ and $\beta$ spinor indices going from $1,...,2^{2n-1}$; any other indices are vector indices going from $1,...,4n$, antisymmetrised with those appearing alongside them in a given subscript; and $C_1$ is the invariant of the $U(1)$ factor in the flavour symmetry (see Section \ref{sec: e6 fam 2} for more details). The spinor representation of $SO(4n)$ is real/pseudo-real for even/odd $n$. We conjecture that the non-zero Poisson brackets of (\ref{gens of kite fam}) are given by:
\begin{equation}\label{final kite pbs}
    \begin{gathered}
     \{a_{\mu \nu},a_{\rho \sigma}\} = \delta_{\nu \rho} a_{\mu \sigma} - \delta_{\mu \rho} a_{\nu \sigma} + \delta_{\mu \sigma} a_{\nu \rho} - \delta_{\nu \sigma} a_{\mu \rho}, \\[10pt]
    \{a_{\mu \nu}, s^{\pm}_\alpha\} =  (\gamma_{\mu \nu})_{\alpha \beta} \, s^{\pm}_\beta,  \\[10pt]
    \{C_1,s^{\pm}_\alpha\}= \pm \,  s^{\pm}_\alpha,\\[10pt]
    \{s^{+}_\alpha,  s^{-}_\beta\} = 
    \begin{cases}
    {C_1}^{n + \sigma - 2} \, (\gamma_{\mu \nu})_{\alpha \beta} \, a_{\mu \nu} & \text{ if } n \text{ even},\\[10pt]
    {C_1}^{n + \sigma - 1} \, \Omega_{\alpha \beta} & \text{ if } n \text{ odd},\\[10pt]
    \end{cases}\\[10pt]
    \end{gathered}
\end{equation}
where $C_1$ is normalised such that there are no additional numerical coefficients in (\ref{final kite pbs}); and $\Omega_{\alpha \beta}$ is the invariant skew-symmetric two form of $Sp(k=2^{2n-2})$ (\ref{omega spk}). All other Poisson brackets between (\ref{gens of kite fam}) that are not listed in (\ref{final kite pbs}) vanish.

\subsubsection{Truck family} \label{sec: truck fam}
The truck family is the magnetic quiver for the Higgs branch of $5d$ $\mathcal{N}=1$ SQCD with parameters satisfying the same subset of (\ref{uv fixed ineq}) as the kite family, (\ref{region 4}) (``region $4$" of \cite{Cabrera:2018jxt}), but with $N_f$ odd.\footnote{Note that for $k=\frac{1}{2}$, the magnetic quiver does not take the form (\ref{truck fam}), and instead takes the form of (\ref{pini e8}) for $N_f=2n+7$ (where this $n$ is that used in (\ref{pini e8}), not to be confused with the $n$ of (\ref{truck fam}).} The family is given by
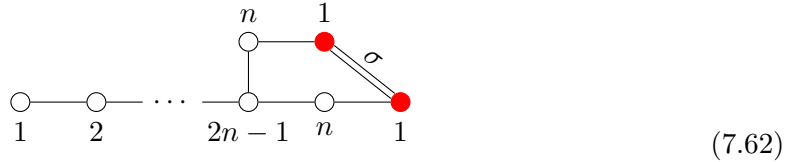
\begin{equation}\label{truck fam}
\begin{tikzpicture}[x=1cm,y=0.8cm]
    \node (g1) at (0,0) [gauge,label=below:{$1$}] {};
    \node (g2) at (1,0) [gauge,label=below:{$2$}] {};
    \node (g3) at (2,0) {$\cdots$};
    \node (g4) at (3,0) [gauge,label=below:{$2n-1$}] {};
    \node (g5) at (4,0) [gauge,label=below:{$n$}] {};
    \node (g6) at (5,0) [gauger,label=below:{$1$}] {};
    \node (g7) at (3,1) [gauge,label=above:{$n$}] {};
    \node (g8) at (4,1) [gauger,label=above:{$1$}] {};
    \draw (g1)--(g2)--(g3)--(g4)--(g5)--(g6);
    \draw[double distance = 2pt] (g6)--(g8) node [midway, sloped, above] {$\sigma$};
    \draw (g4)--(g7)--(g8);
\end{tikzpicture}
\end{equation}
for $n \geq 2$, where a red node indicates that it is unbalanced, and $\sigma$ indicates the multiplicity of the hypermultiplet stretching between the two unbalanced $U(1)$ nodes. Generically, we can read off that this quiver has global symmetry $SO(4n+2) \times U(1)$. There are $4$ generating representations, as can be found from the HWG:
\begin{equation}
    HWG = \frac{1 - \mu_{2n} \, \mu_{2n+1} \, t^{2(n+\sigma)}}{(1-t^2)(1-\mu_{2n} \, \mu_{2n+1} \, t^{2n})(1- \mu_{2n} \, q \, t^{n+\sigma})(1- \frac{\mu_{2n+1}}{q} \, t^{n+\sigma})\prod_{i=1}^{n-1}(1-\mu_{2i} \, t^{2i})},
\end{equation}
where $\mu_1,...,\mu_{2n+1}$ are the highest weight fugacities for the $SO(4n+2)$ factor in the global symmetry, and $q$ is the fugacity for the $U(1)$ charge. Explicitly, the generators are:
\begin{equation}\label{gens of truck fam}
    \begin{array}{|c|c|c|}
        \hline
        \text{Generator} & \Delta & SO(4n+2) \times U(1) \text{ representation}\\
        
        \hline
        
        a_{\mu \nu} & 1 & \text{adjoint } (0)\\
        
        \hline
        
        C_1 & 1 & \text{trivial } \times (0)\\
        
        \hline
        
        s^{\alpha}
        &
        \frac{n+\sigma}{2}
        & \text{left spinor } \times (+1)\\
        
        \hline
        
        s_{\alpha}
        &
        \frac{n+\sigma}{2}
        & \text{right spinor } \times (-1) \\
        
        \hline
    
    \end{array} \ \ ,
\end{equation}
where $\alpha$ and $\beta$ spinor indices going from $1,...,2^{2n}$; any other indices are vector indices going from $1,...,4n+2$, antisymmetrised with those appearing alongside them in a given subscript; and $C_1$ is the invariant of the $U(1)$ factor in the flavour symmetry (see Section \ref{sec: e6 fam 2} for more details). The left and right spinors of $SO(4n+2)$ are complex representations, and conjugate to one another, hence the raised and lower indices respectively. We conjecture that the non-zero Poisson brackets of (\ref{gens of truck fam}) are given by:
\begin{equation}\label{final truck pbs}
    \begin{gathered}
     \{a_{\mu \nu},a_{\rho \sigma}\} = \delta_{\nu \rho} a_{\mu \sigma} - \delta_{\mu \rho} a_{\nu \sigma} + \delta_{\mu \sigma} a_{\nu \rho} - \delta_{\nu \sigma} a_{\mu \rho}, \\[10pt]
    \{a_{\mu \nu}, s^{\alpha} \} =  {(\gamma_{\mu \nu})^{\alpha}}_{\beta} \, s^{\beta},  \\[10pt]
    \{a_{\mu \nu}, s_{\alpha} \} = s_{\beta} \, {(\gamma_{\mu \nu})^{\beta}}_{\alpha} , \\[10pt]
    \{C_1,s^{\alpha}\}=+ \, s^{\alpha},\\[10pt]
    \{C_1,s_{\alpha}\}=- \, s_{\alpha},\\[10pt]
    \{ s^{\alpha}, s_{\beta} \} = 
    {C_1}^{n+\sigma-1} \, {\delta^{\alpha}}_{\beta}, \\[10pt]
    \end{gathered}
\end{equation}
where the spinor indices on the $\gamma_{\mu\nu}$ and $\delta$ are determined as in Section \ref{sec: e6 fam 2}; and $C_1$ is normalised such that there are no additional numerical coefficients in (\ref{final truck pbs}). All other Poisson brackets between (\ref{gens of truck fam}) that are not listed in (\ref{final truck pbs}) vanish.

\clearpage

\section*{Acknowledgements}
    We would like to thank Zhenghao Zhong for sharing the results of his work calculating the HWG for the quivers of Section \ref{sec: zhenghao quivs}; Julius Grimminger for insights on the topic of this note; and Matthew Bullimore for discussions regarding the consistency of Poisson brackets, and the results derived in \cite{Bullimore:2015lsa}. The work of KG is supported by STFC DTP research studentship grant ST/V506734/1, and AH and KG are both supported by STFC grant ST/P000762/1 and ST/T000791/1.

\appendix

\section{Symplectic Singularities and the Poisson Bracket}\label{app: ss}
\noindent The Coulomb branch $\mathcal{C}$ of the quivers we consider are all symplectic singularities in the sense of Beauville \cite{2000InMat.139..541B}, which in our languauge essentially means an algebraic variety on which there exists a $2$-form which degenerates at zero or more points. This $2$-form is called the symplectic form $\omega$, and defines a pairing on the tangent space at any point on the variety. Since $\omega$ maps two tangent vectors at a point $p$ on $\mathcal{C}$ to $\mathbb{C}$, it can be thought of as a map from a tangent space at $p$ to the cotangent space at $p$,
\begin{equation}
    \omega : T_p \mathcal{C} \rightarrow T_p^* \mathcal{C}.
\end{equation}
This induces a correspondence between one-forms on $\mathcal{C}$ and vector fields. Recall that we are interested in investigating the chiral operators lying on $\mathcal{C}$, which are precisely the holomorphic functions over it. The Hilbert series counts such functions on the Coulomb branch, but another thing we can investigate, the purpose of this note, is the Poisson brackets between them.\\

\noindent A Poisson bracket $\{ \cdot , \cdot \}$ is simply a bilinear map which obeys anticommutativity, Leibniz and the Jacobi identity. It is a natural thing to consider on the Coulomb branches we study, because any symplectic singularity automatically has a Poisson structure induced by the symplectic form. In particular, holomorphic functions $f$ are differentiable and so one can consider their differential $df$ (a one form). The symplectic form, which is bijective everywhere but at the singularities, then says that away from singularities this is equivalent to a vector field $X_f$. Thus we can define the Poisson bracket of two chiral operators i.e. holomorphic functions on the Coulomb branch, $f$ and $g$, to be
\begin{equation}
    \{f,g\} \equiv \omega(X_f,X_g).
\end{equation}
 
\noindent The properties of the Poisson bracket match those of a Lie bracket, and this fact can be exploited in both directions. For example, consider the generators of a given Coulomb branch that lie in the adjoint representation of the topological symmetry algebra. If the structure constants of this algebra are known, we can use them to conclude the Poisson brackets. If the structure constants are unknown and we can compute the Poisson brackets, then we can use these to conclude the structure constants.\\

\noindent We are interested in the Poisson brackets on our moduli space as it encodes the symplectic form: the defining property of a symplectic singularity. In principle, this symplectic form can be used to compute the metric on the Coulomb branch, but as far as we are aware, a general method for doing this is not known (although some progress has been made for certain theories, see for instance \cite{Bullimore:2015lsa}). Understanding how to do this would be very fruitful, but we expect the task to be challenging.

\section{Tensor, Symmetric and Antisymmetric Products}
\label{app:tens symm antisymm prods}
In Sections \ref{sec: pbs for cb as reps} - \ref{sec:eg pbs from reps}, we use the antisymmetric property of the Poisson bracket to constrain the representations that the output can lie in. In this appendix, we outline how this works in practise.\\

\noindent Suppose we want to find the Poisson bracket between two generators $g_1$ and $g_2$, which lie in the representations $R_1$ and $R_2$ of some group $G$ respectively. Appropriate indices on $g_1$ and $g_2$ will reflect the components of the respective representations. The Poisson bracket in some sense intertwines $g_1$ and $g_2$, and so the result must lie in the tensor product representation $R_1 \otimes R_2$ in all cases. If $R_1=R_2$, the antisymmetry of the Poisson bracket means that the result $\{g_1,g_2\}$ not only lies in the tensor product $R_1 \otimes R_1$, but in fact in the second rank antisymmetric product of $R_1$: $\Lambda^2(R_1)=\Lambda(R_1 \otimes R_1) \subset R_1 \otimes R_1$. But how do we compute this?\\

\noindent In general $R_1 \otimes R_2$ is not an irreducible representation. The weights lying in $R_1 \otimes R_2$ are those obtained by adding all weights of $R_2$ to each weight of $R_1$. These weights form several irreducible representations of $G$, and this is the tensor product decomposition of $R_1 \otimes R_2$. That is, if the weights of the representation $R$ are denoted by $w_i^R$ for $i=1,...,d_R=\text{dim}(R)$, then 
\begin{equation}
    \left\{w_i^{R_1 \otimes R_2} \ \Big\vert \ i=1,...,d_{R_1}\, d_{R_2} \right\} = \left\{w_j^{R_1} + w_k^{R_2} \ \Big\vert \ j=1,...,d_{R_1}, \ k=1,...,d_{R_2} \right\}
\end{equation}
and we write
\begin{equation}
    R_1 \otimes R_2 = \mathop{\oplus}_{i} c_i \, \tilde{R}_i
\end{equation}
where $c_i$ are non-negative integer coefficients, to illustrate how the tensor product representation decomposes into the irreducible representations $\tilde{R}_i$ of $G$. Sometimes the direct sum symbol is dropped and we just write
\begin{equation}
    R_1 \otimes R_2 = \sum_{i} c_i \,R_i.
\end{equation}
For example, for $G=SU(2)$, consider $R_1=\mu$\footnote{Here we use the Dynkin label notation for a representation: the index on $\mu$ is equal to the highest weight of the representation. For groups of higher rank $r$, we introduce $r$ fugacities e.g. $\mu_1,...,\mu_r$ whose powers give the highest weight in a particular representation.} whose weights are $\{w_i^{\mu}\}=\{\, (1) \, , \, (-1) \, \}$, and $R_2=\mu^2$ whose weights are $\{w_i^{\mu^2}\}=\{ \, (2) \, , \, (0) \, , \, (-2) \, \}$. The tensor product $\mu \otimes \mu^2$ then contains the weights
\begin{equation}
\begin{split}
    \left\{w_i^{\mu \, \otimes \, \mu^2} \ \Big\vert \ i=1,...,6 \right\} &= \left\{w_j^{\mu} + w_k^{\mu^2} \ \Big\vert \ j=1,2, \ k=1,2,3 \right\}\\[5pt]
    &= \left\{\, (3) \, , \, (1) \, , \, (-1) \, , \, (1) \, , \, (-1) \, , \, (-3) \,\right\},
\end{split}
\end{equation}
and thus we see that
\begin{equation}
    \mu \otimes \mu^2 = \mu^3 + \mu.
\end{equation}

\noindent In the case where $R_1=R_2=R$, we can consider the symmetric and antisymmetric parts of such a product. We can see that in this case the weights of the tensor product can be decomposed as follows:
\begin{equation}
    \begin{split}
        \left\{w_j^{R} + w_k^{R} \ \Big\vert \ j,k=1,...,d_{R} \right\} 
        & = \left\{w_j^{R} + w_k^{R} \ \Big\vert \ j\leq k=1,...,d_{R} \right\} + \left\{w_j^{R} + w_k^{R} \ \Big\vert \ j>k=1,...,d_{R} \right\}\\[10pt]
        &\equiv \left\{w_i^{S^2(R)} \ \Big\vert \ i=1,...,d_{S^2(R)} \right\} + \left\{w_i^{\Lambda^2(R)} \ \Big\vert \ i=1,...,d_{\Lambda^2(R)} \right\},
    \end{split}
\end{equation}
and under these definitions of the second rank symmetric product $S^2(R)$ and the second rank antisymmetric product $\Lambda^2(R)$ that
\begin{equation}
    R \otimes R = S^2(R) + \Lambda^2(R).
\end{equation}
One can see that the dimensions of the second rank symmetric and antisymmetric products are
\begin{equation}
\begin{gathered}
    d_{S^2(R)}=\frac{d_R(d_R+1)}{2},\\[5pt]
    d_{\Lambda^2(R)}=\frac{d_R(d_R-1)}{2}.\\
\end{gathered}
\end{equation}
\noindent For example, for $G=SU(2)$ and $R=\mu^2$, if we label the weights $w_j^{\mu^2}$ for $j=1,2,3$ without loss of generality as
\begin{equation}
    w_1^{\mu^2} = (2), \ \ \ w_2^{\mu^2} = (0), \ \ \ w_3^{\mu^2} = (-2),
\end{equation}
then the second rank symmetric product is specified by the weights 
\begin{equation}
    w_{ij}^{S^2(\mu^2)}\equiv w_i^{\mu^2} + w_j^{\mu^2}
\end{equation} 
for 
\begin{equation}
    (i,j)=\{ \, (1,1) \, , \, (1,2) \, , \, (1,3) \, , \, (2,2) \, , \, (2,3) \, , \, (3,3) \, \},
\end{equation}
giving 
\begin{equation}\label{S2mu2 weights}
    w_{ij}^{S^2(\mu^2)} = \{ \, (4) \, , \, (2) \, , \, (0) \, , \, (0) \, , \, (-2) \, , \, (-4) \, \}.
\end{equation}
Similarly the second rank antisymmetric product is specified by the weights 
\begin{equation}
    w_{ij}^{S^2(\mu^2)}\equiv w_i^{\mu^2} + w_j^{\mu^2}
\end{equation} 
for 
\begin{equation}
    (i,j)=\{ \, (2,1) \, , \, (3,1) \, , \, (3,2) \, \},
\end{equation}
giving 
\begin{equation}\label{L2mu2 weights}
    w_{ij}^{\Lambda^2(\mu^2)} = \{ \, (2) \, , \, (0) \, , \, (-2) \, \}.
\end{equation}
Thus we conclude from (\ref{S2mu2 weights}) and (\ref{L2mu2 weights}) that
\begin{equation}
    \begin{gathered}
        S^2(\mu^2) = \mu^4 + 1,\\[5pt]
        \Lambda^2(\mu^2) = \mu^2.\\
    \end{gathered}
\end{equation}

\bibliographystyle{JHEP}
\bibliography{bibli.bib}

\end{document}